%% file: paper.tex
\documentclass[11pt]{article}
\usepackage{fullpage}
\usepackage{amsmath, amsfonts, amssymb, amsthm, amsbsy, amscd, bm,
  bbm} 
\usepackage{array} 
\usepackage{graphicx}
\usepackage[small,bf]{caption}
\usepackage{subcaption}
\usepackage{color}
\usepackage[colorlinks,citecolor={black},urlcolor={black},linkcolor={black}]{hyperref}
\usepackage{url}
\usepackage{algorithm, algorithmic}
\usepackage{multirow}

\usepackage{mdframed}

\definecolor{darkred}{rgb}{0.8,0,0}
\definecolor{lightgray}{rgb}{0.94,0.94,0.94}
\newmdenv[linewidth=1pt,linecolor=black,backgroundcolor=lightgray,roundcorner=0pt]{shadedbox}

\def\st{\textsc{ST}}                
\def\1i{\imath}                     

\def\minim{\mathop{\hbox{minimize}}}
\def\maxim{\mathop{\hbox{maximize}}}
\def\minimize#1{\displaystyle\minim_{#1}}
\def\maximize#1{\displaystyle\maxim_{#1}}
\def\st{\mathop{\hbox{subject to}}}

\def\ol#1#2{J_{#2}(#1)}
\def\olplain{J}
\def\olconj#1#2{J^*_{#2}(#1)}
\def\olconjplain{J^*}

\newtheorem{theorem}{Theorem}[section]
\newtheorem{lemma}[theorem]{Lemma}

\newtheorem{proposition}[theorem]{Proposition}

\newtheorem{assumption}[theorem]{Assumption}

\newcommand{\R}{\mathbb{R}}

\renewcommand{\>}{\rangle}

\newcommand{\goto}{\rightarrow}
\newcommand{\sgn}{\textrm{sgn}}
\renewcommand{\P}{\operatorname{\mathbb{P}}}
\newcommand{\E}{\operatorname{\mathbb{E}}}
\newcommand{\norm}[1]{{\left\lVert{#1}\right\rVert}}

\renewcommand{\i}{\imath}

\newcommand{\half}{{\textstyle\frac{1}{2}}}

\def \endprf{\hfill {\vrule height6pt width6pt depth0pt}\medskip}

\renewenvironment{proof}{\noindent {\bf Proof} }{\endprf\par}

\definecolor{eac}{RGB}{200,50,50}
\definecolor{gosia}{RGB}{50,200,50}
\definecolor{ejc}{RGB}{255,0,0}
\definecolor{wjs}{RGB}{100,100,0}
\definecolor{vdb}{RGB}{0,0,255}
\definecolor{srb}{RGB}{200,10,255}

\newcommand{\LBH}{\lambda_{\text{BH}}}
\newcommand{\LBHc}{\lambda_{\text{BHc}}}
\newcommand{\LBHthm}{\lambda_{\text{\em BH}}}

\graphicspath{{./figspaper/}}

\numberwithin{equation}{section}

\allowdisplaybreaks

\title{Statistical Estimation and Testing via the Sorted $\ell_1$ Norm}

\author{ Ma{\l}gorzata Bogdan\textsuperscript{a} \, Ewout van den
  Berg\textsuperscript{b} \, Weijie Su\textsuperscript{c} \, Emmanuel
  J.~Cand\`es\textsuperscript{c,d}\thanks{Corresponding author}}
\date{}

\begin{document}
\maketitle

{\centering
\vspace*{-0.3cm}
\noindent\textsuperscript{a} Departments of Mathematics and Computer Science,
Wroc{\l}aw University of Technology and Jan D{\l}ugosz University,
Poland\\
\textsuperscript{b} IBM T.J. Watson Research Center,
Yorktown Heights, NY 10598, U.S.A.\\
\textsuperscript{c} Department of Statistics, Stanford
  University, Stanford, CA 94305, U.S.A.\\
  \textsuperscript{d} Department of Mathematics, Stanford University,
  Stanford, CA 94305, U.S.A.\par\bigskip 
\date{October 2013}\par
}






\begin{abstract}
  We introduce a novel method for sparse regression and variable
  selection, which is inspired by modern ideas in multiple
  testing. Imagine we have observations from the linear model $y = X
  \beta + z$, then we suggest estimating the regression coefficients
  by means of a new estimator called SLOPE, which is the solution to
\[
\underset{b}{\text{minimize}} \quad \half \|y - Xb\|_{\ell_2}^2 + \lambda_1
|b|_{(1)} + \lambda_2 |b|_{(2)} + \ldots + \lambda_p |b|_{(p)};
\]
here, $\lambda_1 \ge \lambda_2 \ge \ldots \ge \lambda_p$ and
$|b|_{(1)} \ge |b|_{(2)} \ge \ldots \ge |b|_{(p)}$ is the order
statistics of the magnitudes of $b$. In short, the regularizer is a
sorted $\ell_1$ norm which penalizes the regression coefficients
according to their rank: the higher the rank---the closer to the
top---the larger the penalty. This is similar to the famous
Benjamini-Hochberg procedure (BHq) \cite{BH95}, which compares the
value of a test statistic taken from a family to a critical threshold
that depends on its rank in the family. SLOPE is a convex program and
we demonstrate an efficient algorithm for computing the solution. We
prove that for orthogonal designs with $p$ variables, taking
$\lambda_i = F^{-1}(1-q_i)$ ($F$ is the cumulative distribution
function of the errors), $q_i = iq/(2p)$, controls the false discovery
rate (FDR) for variable selection. This holds under the assumption
that the errors are i.i.d.~symmetric and continuous random
variables. When the design matrix is nonorthogonal there are inherent
limitations on the FDR level and the power which can be obtained with
model selection methods based on $\ell_1$-like penalties. However,
whenever the columns of the design matrix are not strongly correlated,
we demonstrate empirically that it is possible to select the
parameters $\lambda_i$ as to obtain FDR control at a reasonable level
as long as the number of nonzero coefficients is not too large. At the
same time, the procedure exhibits increased power over the lasso,
which treats all coefficients equally.  The paper illustrates further
estimation properties of the new selection rule through comprehensive
simulation studies.
\end{abstract}

{\bf Keywords.}  Sparse regression, variable selection, false
discovery rate, lasso, sorted $\ell_1$ penalized estimation (SLOPE),
prox operator.

\input{intro}

\input{algo}
\input{fdr_ortho}
\input{fdr_general}
\input{numerical}
\input{discussion}

{\small
\subsection*{Acknowledgements}
E.~C.~is partially supported by AFOSR under grant FA9550-09-1-0643, by
ONR under grant N00014-09-1-0258 and by a gift from the Broadcom
Foundation.  M.~B.~was supported by the Fulbright Scholarship and NSF
grant NSF 1043204. E.~v.d.B.~was supported by National Science
Foundation Grant DMS 0906812 (American Reinvestment and Recovery
Act). W.~S.~is supported by a General Wang Yaowu SGF
Fellowship. E.~C.~would like to thank Stephen Becker for all his help
in integrating the sorted $\ell_1$ norm software into TFOCS, and
Chiara Sabatti for fruitful discussions about this
project. M.~B.~would like to thank David Donoho and David Siegmund for
encouragement and Hatef Monajemi for helpful discussions. We are
very grateful to Lucas Janson for suggesting the acronym SLOPE, and to
Rina Foygel Barber for useful comments about an early version of the
manuscript.

\bibliographystyle{plain}
\bibliography{paper}
}

\appendix

\input{appendix}

\end{document}

%% file: intro.tex
\section{Introduction}
\label{sec:introduction}


\subsection{The model selection problem}

This paper is concerned with estimation and/or testing in the (high-dimensional) statistical linear model in which we observe
\begin{equation}
  \label{eq:linear}
  y = X \beta + z;
\end{equation}
as usual, $y \in \R^n$ is the vector of responses, $X \in \R^{n\times
  p}$ is the design matrix, $\beta \in \R^p$ is the unknown parameter
of interest, and $z \in \R^n$ is a vector of stochastic errors (unless
specified otherwise, we shall assume that the errors are
i.i.d.~zero-mean normal variables). In an estimation problem, we
typically wish to predict the response variable as accurately as
possible or to estimate the parameter $\beta$ as best as we can. In
the testing problem, the usual goal is to identify those variables for
which the corresponding regression coefficient is nonzero. In the
spirit of the recent wave of works on multiple testing where we
collect many measurements on a single unit, we may have reasons to
believe that only a small fraction of the many variables in the study
are associated with the response. That is to say, we have few effects
of interest in a sea of mostly irrelevant variables.

Such problems are known under the name of model selection and have
been the subject of considerable studies since the linear model has
been in widespread use. Canonical model selection procedures find
estimates $\hat \beta$ by solving
\begin{equation}
  \label{eq:l0}
  \min_{b \in \R^p} \,  
  \|y - Xb\|_{\ell_2}^2 + \lambda \|b\|_{\ell_0}, 
\end{equation}
where $\|b\|_{\ell_0}$ is the number of nonzero components in $b$. The
idea behind such procedures is of course to achieve the best possible
trade-off between the goodness of fit and the number of variables
included in the model. Popular selection procedures such as AIC and
$C_p$ \cite{AIC,Cp} are of this form: when the errors are
i.i.d.~$\mathcal{N}(0,\sigma^2)$, AIC and $C_p$ take $\lambda = 2
\sigma^2$. In the high-dimensional regime, such a choice typically
leads to including very many irrelevant variables in the model
yielding rather poor predictive power in sparse settings (when the
true regression coefficient sequence is sparse). In part to remedy
this problem, Foster and George \cite{RIC} developed the risk
inflation criterion (RIC) in a breakthrough paper. They proposed using
a larger value of $\lambda$ effectively proportional to $2 \sigma^2
\log p $, where we recall that $p$ is the total number of variables in
the study. 
Under orthogonal designs, it can be
shown that this yields control of the familywise error rate (FWER);
that is to say, the probability of including a single irrelevant
variable is very low. Unfortunately, this procedure is also rather
conservative (it is similar to a Bonferroni-style procedure in
multiple testing) and RIC may not have much power in detecting those
variables with nonvanishing regression coefficients unless they are
very large.

The above dichotomy has been recognized for some time now and several
researchers have proposed more adaptive strategies. One frequently
discussed idea in the literature is to let the parameter $\lambda$ in
\eqref{eq:l0} decrease as the number of included variables
increases. For instance, penalties with appealing information- and
decision-theoretic properties are roughly of the form
\begin{equation}
  \label{eq:pen_adpative1}
  P(b) = p(\|b\|_{\ell_0}), \quad p(k) = 2\sigma^2 k \log(p/k)
\end{equation}
(the fitted coefficients minimize the residual sum of squares (RSS)
plus the penalty) or with 
\begin{equation}
  \label{eq:pen_adpative2}
p(k) =   2\sigma^2 \sum_{1 \le j \le k} \log(p/j). 
\end{equation}
Among others, we refer the interested reader to
\cite{FosterStine,BirgeMassart} and to \cite{TibshiraniKnight} for a
related approach. When $\|b\|_{\ell_0} = 1$, these penalties are
similar to RIC but become more liberal as the fitted coefficients
become denser.

The problem with the selection strategies above is that they are
computationally intractable. Solving \eqref{eq:l0} or the variations
\eqref{eq:pen_adpative1}--\eqref{eq:pen_adpative2} would involve a
brute-force search essentially requiring to fit least-squares
estimates for {\em all} possible subsets of variables. This is not
practical for even moderate values of $p$---for $p > 60$, say---and is the
main reason why computationally manageable alternatives have attracted
considerable attention in applied and theoretical communities. In
statistics, the most popular alternative is by and large the lasso
\cite{lasso}, which operates by substituting the nonconvex $\ell_0$
norm by the $\ell_1$ norm---its convex surrogate---yielding
\begin{equation}
  \label{eq:lasso}
  \min_{b \in \R^p} \, \half \|y - Xb\|^2_{\ell_2} + 
\lambda \,  \|b\|_{\ell_1}.
\end{equation}
We have the same dichotomy as before: on the one hand, if the selected
value of $\lambda$ is too small, then the lasso would select very many
irrelevant variables (thus compromising its predicting
performance). On the other hand, a large value of $\lambda$ would
yield little power as well as a large bias.

\subsection{The sorted $\ell_1$ norm}

This paper introduces a new variable selection procedure, which is
computationally tractable and adaptive in the sense that the
`effective penalization' is adaptive to the sparsity level (the number
of nonzero coefficients in $\beta \in \R^p$). This method relies on
the {\em sorted $\ell_1$ norm}: letting $\lambda$ be a nonincreasing
sequence of nonnegative scalars,
\begin{equation}
  \label{eq:lambda}
  \lambda_1 \ge \lambda_2 \ge \ldots \ge \lambda_p \ge 0,
\end{equation}
with $\lambda_1 > 0$, we define the sorted $\ell_1$ norm of a vector
$b \in \R^p$ as
\begin{equation}
  \label{eq:orderedl1}
  \ol{b}{\lambda} = \lambda_1 |b|_{(1)} + \lambda_2 |b|_{(2)} + \ldots + \lambda_p |b|_{(p)}.
\end{equation}
Here, $|b|_{(1)} \ge |b|_{(2)} \ge \ldots \ge |b|_{(p)}$ is the order
statistic of the magnitudes of $b$, namely, the absolute values ranked
in decreasing order. For instance, if $b = (-2.1,-0.5,3.2)$, we would
have $|b|_{(1)} = 3.2$, $|b|_{(2)} = 2.1$ and $|b|_{(3)} =
0.5$. Expressed differently, the sorted $\ell_1$ norm of $b$ is thus
$\lambda_1$ times the largest entry of $b$ (in magnitude), plus
$\lambda_2$ times the second largest entry, plus $\lambda_3$ times the
third largest entry, and so on.  As the name suggests,
$\ol{\cdot}{\lambda}$ is a norm and is, therefore,
convex.\footnote{Observe that when all the $\lambda_i$'s take on an
  identical positive value, the sorted $\ell_1$ norm reduces to the
  usual $\ell_1$ norm. Also, when $\lambda_1 > 0$ and $\lambda_2 =
  \ldots = \lambda_p = 0$, the sorted $\ell_1$ norm reduces to the
  $\ell_\infty$ norm.}
\begin{proposition}
  \label{prop:orderedl1}
  The functional $\eqref{eq:orderedl1}$ is a norm provided
  \eqref{eq:lambda} holds and $\lambda \neq 0$. In fact, the sorted
  $\ell_1$ norm can be characterized as
  \begin{equation}
    \label{eq:dual1}
    \ol{b}{\lambda} = \sup_{w \in C_\lambda} \<w, b\>,
\end{equation}
where $w$ is in the convex set $C_\lambda$ if and only if for all $i =
1, \ldots, p$,
\begin{equation}
  \label{eq:dual2} 
  \sum_{j \le i} |w|_{(j)}
  \le \sum_{j \le i} \lambda_j.
  \end{equation}
\end{proposition}
The proof of this proposition is in the Appendix. For now, observe
that $C_\lambda = \cap_{i = 1}^p C^{(i)}_\lambda$, where
\[
C^{(i)}_\lambda = \Bigl\{w : f_i(w) \le \sum_{j \le i} \lambda_j\Bigr\}, \qquad 
f_i(w) = \sum_{j \le i}
|w|_{(j)}.
\]
For each $i$, $f_i$ is convex so that $C^{(i)}_\lambda$ is
convex. Hence, $C_\lambda$ is the intersection of convex sets and is,
therefore, convex.  The convexity of $f_i$ follows from its
representation as a supremum of linear functions, namely,
\[
f_i(w) = \sup \,\, \<z, w\>, 
\]
where the supremum is over all $z$ obeying $\|z\|_{\ell_0} \le i$ and
$\|z\|_{\ell_\infty} \le 1$ (the supremum of convex functions is
convex \cite{Boyd04}). In summary, the characterization via
\eqref{eq:dual1}--\eqref{eq:dual2} asserts that $J_\lambda(\cdot)$ is
a norm and that $C_\lambda$ is the unit ball of its dual norm.

\subsection{SLOPE}

The idea is to use the sorted $\ell_1$ norm for variable selection and
in particular, we suggest a penalized estimator of the form 
\begin{equation}
  \label{eq:ol}
 \min_{b \in \R^p} \,  \half \|y - X b\|_{\ell_2}^2 + \sum_{i = 1}^p \lambda_i |b|_{(i)}. 
\end{equation}
We call this Sorted L-One Penalized Estimation (SLOPE). SLOPE is
convex and, hence, tractable. As a matter of fact, we shall see in
Section \ref{sec:algo} that the computational cost for solving this
problem is roughly the same as that for solving the plain lasso. This
formulation is rather different from the lasso, however, and achieves
the adaptivity we discussed earlier: indeed, because the $\lambda_i$'s
are decreasing or sloping down, we see that the cost of including new
variables decreases as more variables are added to the model.

\subsection{Connection with the Benjamini-Hochberg procedure}
\label{sec:BHq}

Our methodology is inspired by the Benjamini-Hochberg (BHq) procedure
for controlling the false discovery rate (FDR) in multiple testing
\cite{BH95}. To make this connection explicit, suppose we are in the
{\em orthogonal design} in which the columns of $X$ have unit norm and
are perpendicular to each other (note that this implies $p \le
n$). Suppose further that the errors in \eqref{eq:linear} are
i.i.d.~$\mathcal{N}(0,1)$. In this setting, we have
\[
\tilde y = X' y \sim \mathcal{N}(\beta, I_p)
\]
where, here and below, $I_p$ is the $p \times p$ identity matrix. For
testing the $p$ hypotheses $H_i: \beta_i = 0$, the BHq {\em step-up}
procedure proceeds as follows:

\newcommand{\iSU}{i_{\text{SU}}}
\newcommand{\iSD}{i_{\text{SD}}}
\newcommand{\iOL}{i_{\text{SLOPE}}}

\begin{shadedbox}
\begin{itemize}
\item[(1)] Sort the entries of $\tilde y$ in decreasing order of magnitude,
  $|\tilde y|_{(1)} \ge |\tilde y|_{(2)} \ge \ldots \ge |\tilde
  y|_{(p)}$ (this yields corresponding ordered hypotheses $H_{(1)},
  \ldots, H_{(p)}$). 
\item[(2)] Find the largest index $i$ such that
\begin{equation}
  \label{eq:stepup}
  |\tilde y|_{(i)} >  \Phi^{-1}(1- q_i), \quad q_i = q\frac{i}{2p},
\end{equation}
where $\Phi^{-1}(\alpha)$ is the $\alpha$th quantile of the standard
normal distribution and $q$ is a parameter in $[0,1]$. Call this index
$i_{\text{SU}}$. (For completeness, the BHq procedure is traditionally
expressed via the inequality $|\tilde y|_{(i)} \ge \Phi^{-1}(1- q_i)$
but this does not change anything since $\tilde y$ is a continuous
random variable.)
\item[(3)] Reject all $H_{(i)}$'s for which $i \le \iSU$ (if there is no
  $i$ such that the inequality in \eqref{eq:stepup} holds, then make
  no rejection). 
\end{itemize}
\end{shadedbox}
This procedure is adaptive in the sense that a hypothesis is rejected
if and only if its $z$-value is above a data-dependent threshold.  In
their seminal paper \cite{BH95}, Benjamini and Hochberg proved that
this procedure controls the FDR. Letting $V$ (resp.~$R$) be the total
number of false rejections (resp.~total number of rejections), we have
\begin{equation}
  \label{eq:BH}
  \text{FDR} = \E \left[ \frac{V}{R \vee 1} \right] = q \frac{p_0}{p},  
\end{equation}
where $p_0$ is the number of true null hypotheses, $p_0 = |\{i:
\beta_i = 0\}|$, so that $p = p_0 + \|\beta\|_{\ell_0}$. This always
holds; that is, no matter the value of the mean vector $\beta$.

One can also run the procedure in a {\em step-down} fashion in which
case the last two steps are as follows:
\begin{shadedbox}
\begin{itemize}
\item[(2)] Find the smallest index $i$ such that
\begin{equation}
  \label{eq:stepdown}
  |\tilde y|_{(i)} \le  \Phi^{-1}(1- q_i)
\end{equation}
and call it $\iSD + 1$. 
\item[(3)] Reject all $H_{(i)}$'s for which $i \le \iSD$ (if 
  there is no $i$ such that the inequality in \eqref{eq:stepdown}
  holds, then reject all the hypotheses).
\end{itemize}
\end{shadedbox}
This procedure is also adaptive. Since we clearly have $\iSD \le
\iSU$, we see that the step-down procedure is more conservative than
the step-up.  The step-down variant also controls the FDR, and obeys
\begin{equation*}
  \label{eq:BH2}
  \text{FDR} = \E \left[ \frac{V}{R \vee 1} \right] \le q \frac{p_0}{p}
\end{equation*}
(note the inequality instead of the equality in \eqref{eq:BH}). We
omit the proof of this fact.

To relate our approach with BHq, we can use SLOPE \eqref{eq:ol} as a
multiple comparison procedure: (1) select weights $\lambda_i$, (2)
compute the solution $\hat \beta$ to \eqref{eq:ol}, and (3) reject
those hypotheses for which $\hat \beta_i \neq 0$.  Now in the
orthogonal design, SLOPE \eqref{eq:ol} reduces to
\begin{equation}
  \label{eq:ol2}
  \min_{b \in \R^p} \,  
\half \|\tilde y - b\|_{\ell_2}^2 + \sum_{i = 1}^p \lambda_i |b|_{(i)} 
\end{equation}
(recall $\tilde y = X'y$).  The connection with the BHq procedure is
as follows:
\begin{proposition}
  \label{prop:BHq-ol}
  Assume an orthogonal design and set $\lambda_i = \LBHthm(i) :=
  \Phi^{-1}(1-q_i)$. Then the {\em SLOPE} procedure rejects $H_{(i)}$
  for $i \le i^\star$ where $i^\star$ obeys\footnote{For completeness,
    suppose without loss of generality that the $\tilde y_i$'s are
    ordered, namely, $|\tilde y_1| \ge |\tilde y_2| \ge \ldots \ge
    |\tilde y_p|$. Then the solution $\hat \beta$ is ordered in the
    same way, i.e.~$|\hat \beta_1| \ge |\hat \beta_2| \ge \ldots \ge
    |\hat \beta_p|$ and we have $|\hat \beta_1| \ge \ldots \ge |\hat
    \beta_{i^\star}| > 0$, $\hat \beta_{i^\star+1} = \ldots = \hat
    \beta_p = 0$.}
\begin{equation}
  \label{eq:BHq-ol}
  i_{\text{\em SD}} \le i^\star  \le i_{\text{\em SU}}.
\end{equation}
This extends to arbitrary sequences $\lambda_1 \ge \lambda_2 \ge
\ldots \ge \lambda_n > 0$. 
\end{proposition}
The proof is also in the Appendix.  In words, SLOPE is at least as
conservative as the step-up procedure and as liberal or more than the
step-down procedure. It has been noted in \cite{ABDJ} that in most
problems, the step-down and step-up points coincide, namely, $\iSD =
\iSU$. Whenever this occurs, all these procedures produce the same
output.

An important question is of course whether the SLOPE procedure
controls the FDR in the orthogonal design. In Section
\ref{sec:fdr_orth}, we prove that this is the case.
\begin{theorem}
\label{teo:fdr_control1}
Assume an orthogonal design with i.i.d.~$\mathcal{N}(0,1)$ errors, and
set $\lambda_i = \LBHthm(i) = \Phi^{-1}(1-iq/2p)$. Then the FDR of the
{\em SLOPE} procedure obeys
\begin{equation}
  \label{eq:fdr_control1}
  \text{\rm FDR} = \E \left[\frac{V}{R \vee 1} \right] \le q \frac{p_0}{p}.   
\end{equation}
Again $p$ is the number of hypotheses being tested and $p_0$ the total
number of nulls.
\end{theorem}
We emphasize that this result is not a consequence of the bracketing
\eqref{eq:BHq-ol}. In fact, the argument appears nontrivial.

\subsection{Connection with FDR thresholding}

When $X$ is the identity matrix or, equivalently, when $y \sim
\mathcal{N}(\beta,\sigma^2 I)$, there exist FDR thresholding
procedures for estimating the mean vector, which also adapts to the
sparsity level. Such a procedure was developed by Abramovich and
Benjamini \cite{AbramovichBenjamini95} in the context of wavelet
estimation (see also \cite{AbramovichBenjamini96}) and works as
follows. We rank the magnitudes of $y$ as before, and let $\iSU$ be
the largest index for which $|y|_{(i)} > \Phi^{-1}(1-q_i)$ as in the
step-up procedure. Letting $t_{\text{FDR}} = \Phi^{-1}(1- q_{\iSU})$,
set
\begin{equation}
\label{eq:FDRthresh}
\hat \beta_i = \begin{cases} y_i & |y_i| \ge t_{\text{FDR}}\\
0 & |y_i| < t_{\text{FDR}}.
\end{cases}
\end{equation}
This is a hard-thresholding estimate but with a data-dependent
threshold: the threshold decreases as more components are judged to be
statistically significant.  It has been shown that this simple
estimate is asymptotically minimax throughout a range of sparsity
classes \cite{ABDJ}.

Our method is similar in the sense that it also chooses an adaptive
threshold reflecting the BHq procedure as we have seen in Section
\ref{sec:BHq}. However, it does not produce a hard-thresholding
estimate. Rather, owing to nature of the sorted $\ell_1$ norm, it
outputs a sort of soft-thresholding estimate. Another difference is
that it is not clear at all how one would extend \eqref{eq:FDRthresh}
to nonorthogonal designs whereas the SLOPE formulation \eqref{eq:ol}
is straightforward. Having said this, it is also not obvious a priori
how one should choose the weights $\lambda_i$ in \eqref{eq:ol} in a
non-orthogonal setting as to control a form of Type I error such as
the FDR.

\subsection{FDR control under random designs}
\label{sec:fdr_gauss_design}

We hope to have made clear that the lasso with a fixed $\lambda$ is
akin to a Bonferroni procedure where each observation is compared to a
fixed value, irrespective of its rank, while SLOPE is adaptive and
akin to a BHq-style procedure. Moving to nonorthogonal designs, we
would like to see whether any of these procedures have a chance to
control the FDR in a general setting.

To begin with, at the very minimum we would like to control the FDR
under the global null, that is when $\beta = 0$ and, therefore, $y =
z$ (this is called weak family-wise error rate (FWER) control in the
literature). In other words, we want to keep the probability that
$\hat \beta \neq 0$ low whenever $\beta = 0$. For the lasso, $\hat
\beta = 0$ if and only if $\|X'z\|_{\ell_\infty} \le \lambda$. Hence,
we would need $\lambda$ to be an upper quantile of the random variable
$\|X'z\|_{\ell_\infty}$. If $X$ is a random matrix with
i.i.d.~$\mathcal{N}(0,1/n)$ (this is in some sense the nicest
nonorthogonal design), then simple calculations would show that
$\lambda$ would need to be selected around $\LBH(1)$, where $\LBH$ is
as in Theorem \ref{teo:fdr_control1}. The problem is that neither this
value nor a substantially higher fixed value (leading to an even more
conservative procedure) is guaranteed to control the FDR in
general. Consequently, the more liberal SLOPE procedure also
cannot be expected to control the FDR in full generality.

\begin{figure}[h!]
  \centering
        \begin{subfigure}[b]{0.495\textwidth}
                \centering
                \includegraphics[width=\textwidth]{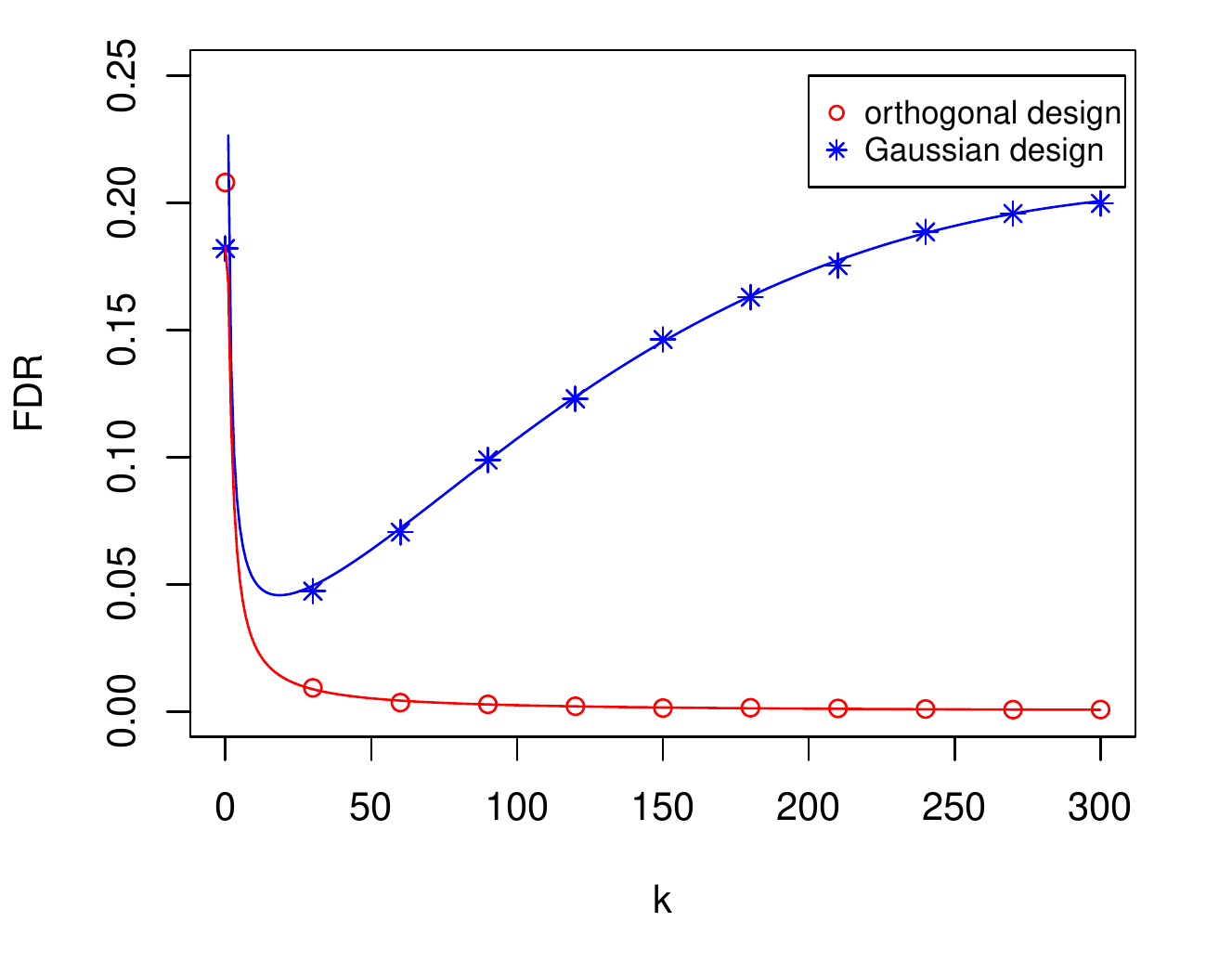}
                \label{fig:Fig1a}
        \end{subfigure}
        \begin{subfigure}[b]{0.495\textwidth}
                \centering
                \includegraphics[width=\textwidth]{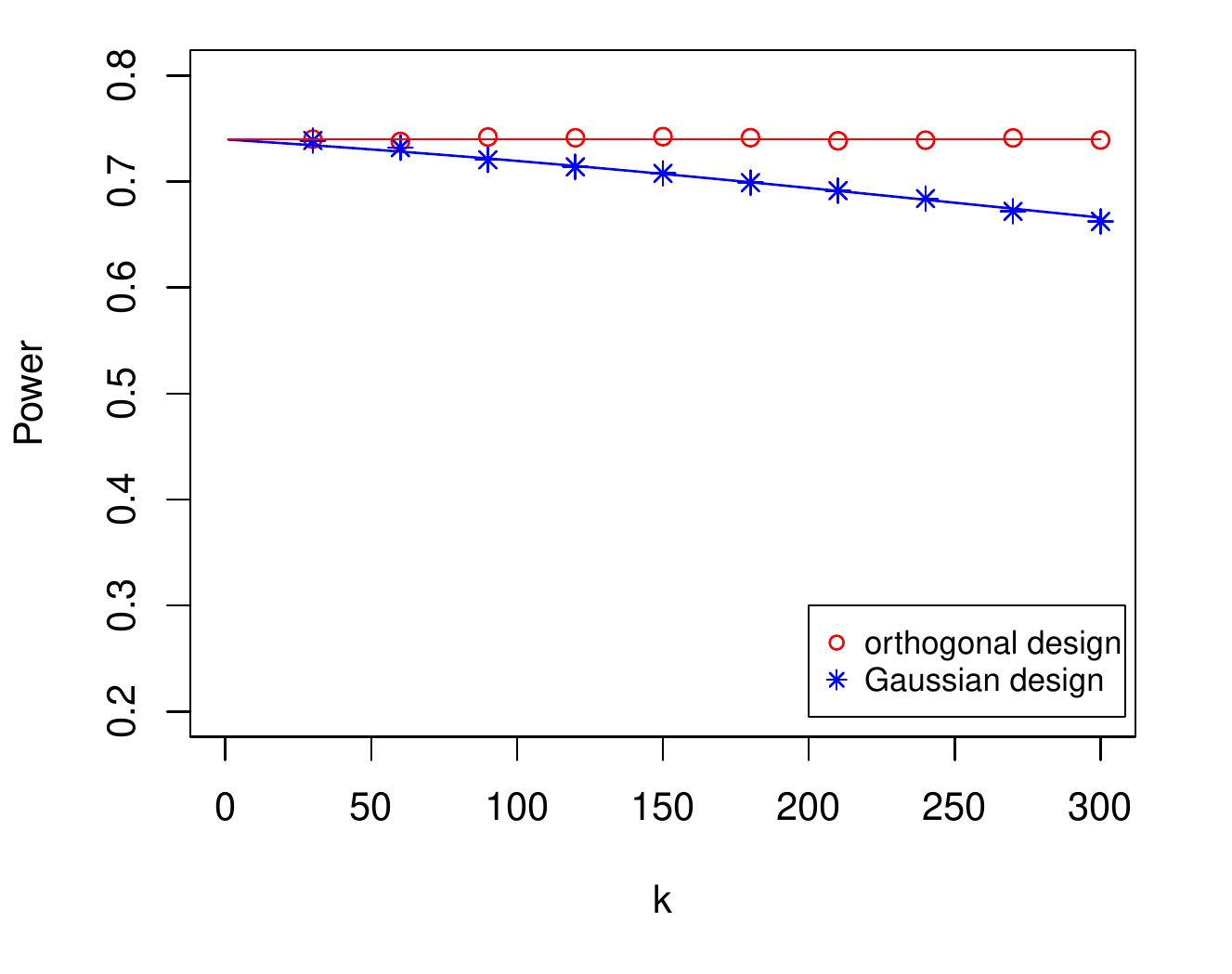}
                \label{fig:Fig1b}
        \end{subfigure}
\vspace*{-1cm}
\caption{Estimated and predicted FDR and power of the lasso with
  $\lambda=3.717$ as a function of $k=\|\beta\|_{\ell_0}$.  The
  entries of the Gaussian design are i.i.d.~$\mathcal{N}(0,1/n)$. In
  both graphs, $n=p=1000$ and $\beta_1,\ldots,\beta_k$ are independent
  normal random variables with mean $0$ and standard deviation equal
  to $3\lambda$. Each data point is a value estimated from 500
  replicated experiments. Solid lines represent predicted curves. The
  predictions for the Gaussian design are based on the extension of
  the results in \cite{BM12}, see Appendix \ref{sec:Montanari}.}
\label{fig:Fig1}
\end{figure}

Figure \ref{fig:Fig1} plots the FDR and power of the lasso with
$\lambda=3.717$, corresponding to $\lambda = \LBH(1)$ for $q \approx
0.207$ and $p=1,000$. True effects were simulated from the normal
distribution with mean zero and standard deviation equal to
$3\lambda$. Our choice of $\lambda$ guarantees that under the
orthogonal design the probability of at least one false rejection is
not larger than about 0.183. Also, the power (average fraction of
properly identified true effects) of the lasso under the orthogonal
design does not depend on $k$ and is equal to 0.74 in our setting.
When the design is orthogonal, the FDR of the lasso quickly decreases
to zero as the number $k = \|\beta\|_{\ell_0}$ of non-nulls
increases. The situation dramatically changes when the design matrix
is a random Gaussian matrix as discussed above. After an initial
decrease, the FDR rapidly increases with $k$. Simultaneously, the
power slowly decreases with $k$. A consequence of the increased FDR is
this: the probability of at least one false rejection drastically
increases with $k$, and for $k=150$ one actually observes an average
of 15.53 false discoveries.

Section \ref{sec:fdr_general} develops heuristic arguments to conclude
on a perhaps surprising and disappointing note: if the columns of the
design matrix are realizations of independent random variables, then
independently on how large we select $\lambda$ nonadaptively, one can
never be sure that the FDR of the lasso solution is controlled below
some $q^\star=q^\star(n/p)>0$.  The inability to control the FDR at a
prescribed level is intimately connected to the shrinkage of the
regression estimates, see Section~\ref{sec:FDR_LASSO}.



\begin{figure}
\centering
          \includegraphics[width=0.6\textwidth]{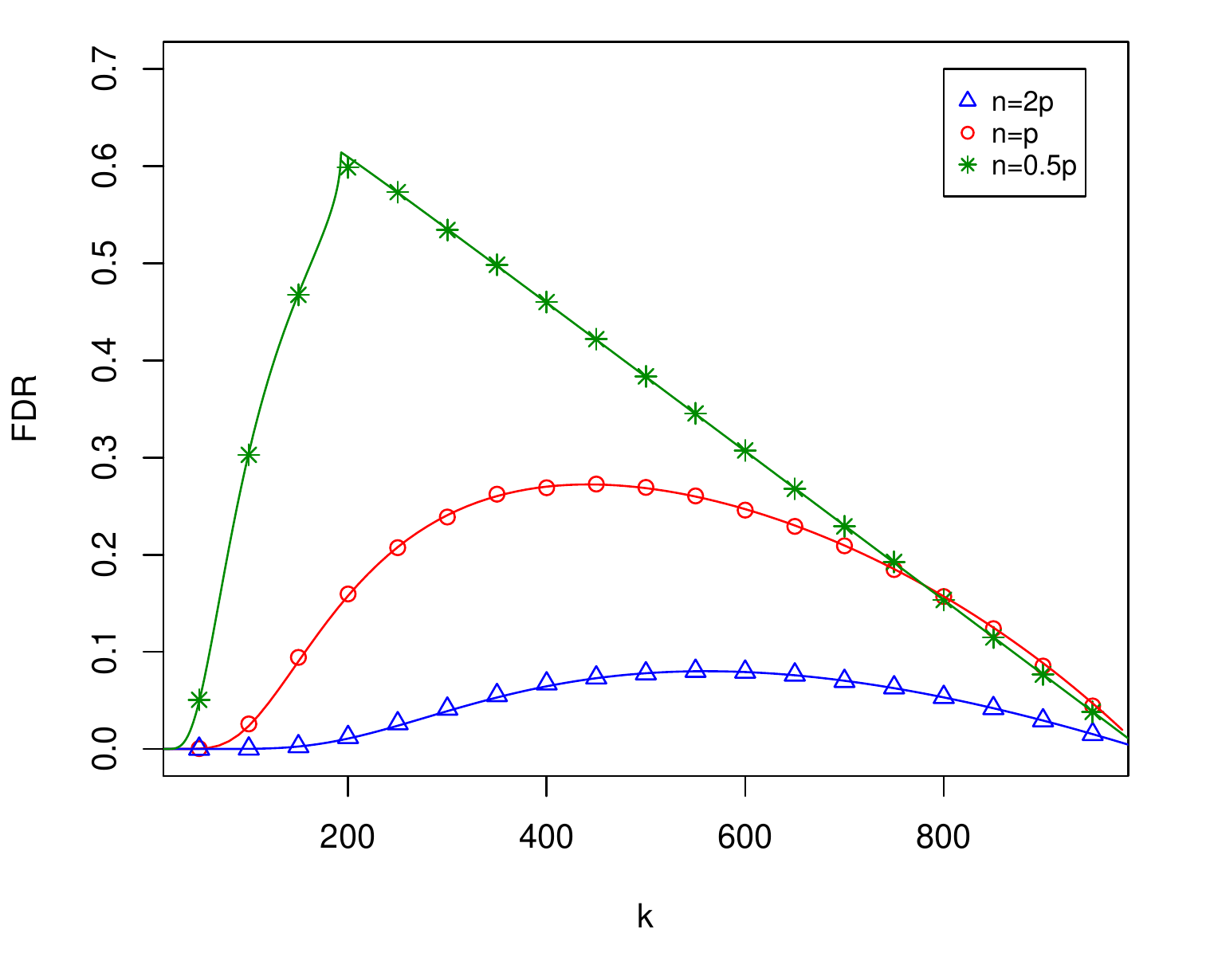}
          \caption{Simulated (markers) and predicted (lines) FDR in
            the high signal-to-noise regime as a function of the
            number of nonzero regression coefficients. Here, the
            design is $n \times p$ Gaussian with $p= 1,000$. The
            formulas for the predicted curves are in Appendix
            \ref{sec:Montanari}.}
\label{fig:minimax}
\end{figure}

Under a Gaussian design, we can in fact predict the asymptotic FDR of
the lasso in the {\em high signal-to-noise} regime in which the
magnitudes of the nonzero regression coefficients lie far above
$\lambda$. Letting $\text{FDR}_{\text{lasso}}(\beta, \lambda)$ be the
FDR of the lasso by employing $\lambda$ as a regularization parameter,
a heuristic inspired by results in \cite{BM12}
gives 
\begin{equation}
  \label{eq:minimax}
  \lim_{p, n \goto \infty} \,\,  \inf_{\lambda} \,\, \sup_{\beta: \|\beta\|_{\ell_0} \le k}  \text{FDR}_{\text{lasso}}(\beta, \lambda) \ge q^\star(\epsilon, \delta),  
\end{equation}
where in the limit, $n/p \goto \delta > 0$ and $k/p \goto
\epsilon>0$. In Appendix \ref{sec:Montanari}, we give explicit
formulas for the lower limit $q^\star(\epsilon,\delta)$. Before
continuing, we emphasize that we do not prove \eqref{eq:minimax}
rigorously, and only provide a heuristic justification. Now the
accuracy of the prediction $q^\star$ is illustrated in Figure
\ref{fig:minimax}, where it is compared with FDR estimates from a
simulation study with a Gaussian design as before, and a value of the
nonzero coefficients set to $1,000 \lambda$ and $\lambda = 300$.  The
figure shows excellent agreement between predicted and observed
behaviors. According to Figure 2, when $n/p=2,1$, or $0.5$, then
independently on how large $\lambda$ is used one cannot be sure that
the FDR is controlled below the values $0.08, 0.27$, or $0.6$,
respectively. We also note a singularity in the predicted curve for
$n/p = 1/2$, which occurs exactly at the point of the classical
phase-transition curve in compressive sensing (or weak-threshold) of
\cite{DonTan05}. All of this is explained in Appendix
\ref{sec:Montanari}.

%
%
%
%

\subsection{Contributions and outline}

This paper introduces a novel method for sparse regression and
variable selection, which is inspired by powerful adaptive ideas in
multiple testing. We conclude this introduction by summarizing our
contributions as well as outlining the rest of the paper. In Section
\ref{sec:algo}, we demonstrate an efficient algorithm for computing
the SLOPE solution. This algorithm is based on a linear time algorithm
for computing the prox to the sorted $\ell_1$ norm (after sorting),
and is new. In Section \ref{sec:fdr_orth}, we prove that SLOPE with
the sequence $\LBH$ controls the FDR under orthogonal
designs. 
In Section \ref{sec:fdr_general}, we detail the inherent limitations
on the FDR level and the power which can be obtained with model
selection methods based on $\ell_1$-like penalties. In this section,
we also derive the predictions introduced in Section
\ref{sec:fdr_gauss_design}. A positive consequence of our
understanding of FDR control is that it leads to an adjusted sequence
$\{\lambda_i\}$ of parameters for use in SLOPE; see Section
\ref{sec:lambda}. Furthermore, in Section \ref{sec:numerical}, we
report on empirical findings demonstrating the following properties:
\begin{itemize}
\item First, when the number of non-nulls (the number of nonzero
  coefficients in the regression model) is not too large, SLOPE
  controls the FDR at a reasonable level, see Section
  \ref{sec:testing}. This holds provided that the columns of the
  design matrix are not strongly correlated. At the same time, the
  method has more power in detecting true regressors than the lasso
  and other competing methods. Therefore, just as the BH procedure,
  SLOPE controls a Type I error while enjoying greater power---albeit
  in a restricted setting.

\item Second, in Section \ref{sec:estimation}, we change our point of
  view and regard SLOPE purely as an estimation procedure. There we
  use SLOPE to select a subset of variables, and then obtain
  coefficient estimates by regressing the response on this subset. We
  demonstrate that with the adjusted weights from Section
  \ref{sec:lambda}, this two-step procedure has good estimation
  properties even in situations when the input is not sparse at
  all. These good properties come from FDR control together with power
  in detecting nonzero coefficients. Indeed, a procedure selecting too
  many irrelevant variables would result in a large variance and
  distort the coefficient estimates of those variables in the
  model. At the same time, we need power as we would otherwise suffer
  from a large bias. Finally in Section \ref{sec:estimation}, we shall
  see that the performance of SLOPE is not very sensitive to the
  choice of the parameter $q$ specifying the nominal FDR level.
\end{itemize}
Section \ref{sec:discussion} concludes the paper with a short
discussion and questions we leave open for future research.  

Finally, it goes without saying that our methods apply to multiple
testing with correlated test statistics. Assume that $n \le p$. Then
multiplying the equation $y = X \beta + z$ on both sides by the pseudo
inverse $X^\dagger$ of $X$ yields
\[
y' = (X^\dagger)'y \sim \mathcal{N}(\beta, \Sigma),
\]
where $\Sigma = (X'X)^{-1}$ whenever $X$ has full column rank. Hence,
procedures for controlling the FDR in the linear model translate into
procedures for testing the means of a multivariate Gaussian
distribution.  Our simulations in Section \ref{sec:testing} illustrate
that under sparse scenarios SLOPE has better properties than the BH
procedure applied to marginal test statistics---with or without
adjustment for correlation---in this context.

As a last remark, SLOPE is looking for a trade-off between the
residual sum of squares and the sorted $\ell_1$ norm but we could
equally contemplate using the sorted $\ell_1$ norm in other penalized
estimation problems. Consider the Dantzig selector \cite{DS} which,
assuming that the columns of $X$ are normalized, takes the form
\begin{equation}
  \label{eq:ds}
  \min_{b \in \R^p} \, \|b\|_{\ell_1} \quad \text{s.~t.}  \quad \|X'(y-  Xb)\|_{\ell_\infty} \le \lambda.  
\end{equation}  
One can naturally use the ideas presented in this paper to tighten
this in several ways, and one proposal is this: take a sequence
$\lambda_1 \ge \lambda_2 \ge \ldots \ge \lambda_p \ge 0$ and consider
\begin{equation}
  \label{eq:ordered_ds}
  \min_{b \in \R^p} \, \|b\|_{\ell_1} \quad \text{s.~t.}  \quad X'(y-  Xb) \in C_{\lambda}, 
\end{equation}  
where $C_\lambda$ is as in Proposition \ref{prop:orderedl1}, see
\eqref{eq:dual2}. With $\lambda_1 = \lambda$, the constraint on the
residual vector in \eqref{eq:ordered_ds} is tighter than that in
\eqref{eq:ds}. Indeed, setting $w = X'(y-Xb)$, the feasible set in
\eqref{eq:ds} is of the form $|w|_{(1)} \le \lambda_1$ while that in
the sorted version is $|w|_{(1)} \le \lambda_1$, $|w|_{(1)} +
|w|_{(2)} \le \lambda_1 + \lambda_2$, $|w|_{(1)} + |w|_{(2)} +
|w|_{(3)} \le \lambda_1 + \lambda_2 + \lambda_3$ and so on. Hence, the
sorted version appears to shrink less and is more liberal.




%% file: algo.tex
\section{Algorithms}
\label{sec:algo}

In this section, we present effective algorithms for computing the
solution to SLOPE \eqref{eq:ol}, which rely on the numerical
evaluation of the proximity operator (prox) to the sorted $\ell_1$
norm. Hence, we first develop a fast algorithm for computing the prox.

\subsection{Preliminaries}
\label{sec:prelim}

\newcommand{\proxh}{\operatorname{prox}_\lambda}

Given $y \in \mathbb{R}^n$ and $\lambda_1 \geq \lambda_2 \geq \cdots
\geq \lambda_n \geq 0$, the prox to the sorted $\ell_1$ norm is the
unique solution\footnote{Unicity follows from the strong convexity of
  the function we minimize.} to
\begin{equation}\label{eq:prox}
  \proxh(y) =  \operatorname{argmin}_{x \in \R^n} \,\, 
  \half \|y-x\|_{\ell_2}^2 + \sum_{i=1}^n \lambda_i |x|_{(i)}. 
\end{equation}

Without loss of generality we can make the following assumption:
\begin{assumption}\label{Ass:SortedY}
  The vector $y$ obeys $y_1 \geq y_2 \geq \cdots \geq y_n \geq 0$.
\end{assumption}

At the solution to \eqref{eq:prox}, the sign of each $x_i \neq 0$ will
match that of $y_i$. It therefore suffices to solve the problem for
$|y|$ and restore the signs in a post-processing step, if
needed. Likewise, note that applying any permutation $P$ to $y$
results in a solution $Px$. We can thus choose a permutation that
sorts the entries in $y$ and apply its inverse to obtain the desired
solution.

\begin{proposition}
Under Assumption~\ref{Ass:SortedY}, the solution $x$ to
\eqref{eq:prox} satisfies
$x_1 \geq x_2 \geq \cdots \geq x_n \geq 0$.
\end{proposition}
\begin{proof}
  Suppose that $x_i < x_j$ for $i < j$ (and $y_i > y_j$), and form a
  copy $x'$ of $x$ with entries $i$ and $j$ exchanged. Letting $f$ be
  the objective functional in \eqref{eq:prox}, we have
\[
  f(x) - f(x')  = \half (y_i - x_i)^2 + \half (y_j - x_j)^2 - \half (y_i - x_j)^2 - \half (y_j - x_i)^2. 
\]
This follows from the fact that the sorted $\ell_1$ norm takes on the
same value at $x$ and $x'$ and that all the quadratic terms cancel but
those for $i$ and $j$. This gives 
\begin{equation*}
  f(x) - f(x')  = x_jy_i - x_iy_i + x_iy_j - x_jy_j =
  (x_j-x_i)(y_i-y_j) > 0,
\end{equation*}
which shows that the objective $x'$ is strictly smaller, thereby
contradicting optimality of $x$.
\end{proof}

Under Assumption~\ref{Ass:SortedY} we can reformulate
\eqref{eq:prox} as
\begin{equation}\label{Eq:ProxFunSorted}
\begin{array}{ll}
  \text{minimize} & \quad \half \|y - x\|_{\ell_2}^2 + \sum_{i=1}^n \lambda_i
  x_i\\
  \text{subject to} & \quad x_1 \geq x_2\geq \cdots\geq x_n \geq 0.
\end{array}
\end{equation}
In other words, the prox is the solution to a quadratic program
(QP). However, we do not suggest performing the prox calculation by
calling a standard QP solver, rather we introduce a dedicated $O(n)$
algorithm we present next. For further reference, we record the
Karush-Kuhn-Tucker (KKT) optimality conditions for this QP.
\begin{description}
\item {\em Primal feasibility}: $x_1 \geq x_2\geq \cdots\geq x_n \geq
  0$.
\item {\em Dual feasibility}: $\mu \in \R^n$ obeys $\mu \ge 0$.
\item {\em Complementary slackness}: $\mu_i(x_i - x_{i+1}) = 0$ for
  all $i = 1, \ldots, n$ (with the convention that $x_{n+1} = 0$).
\item {\em Stationarity of the Lagrangian}:
\[
x_i - y_i + \lambda_i - (\mu_i - \mu_{i-1}) = 0 
\]
with the convention that $\mu_0 = 0$). 
\end{description}

\subsection{A fast prox algorithm} 

\newcommand{\nameproxsp}{FastProxSL1}
\newcommand{\nameprox}{FastProxSL1}

\begin{lemma}
  \label{lem:monotone}
  Suppose $(y - \lambda)_+$ is nonincreasing, then the solution to
  \eqref{eq:prox} obeys
\[
\proxh(y) = (y -\lambda)_+. 
\]
\end{lemma}
\begin{proof} Set $x = (y - \lambda)_+$, which by assumption is primal
  feasible, and let $i_0$ be the last index such that $y_i -
  \lambda_i > 0$.  Set $\mu_1 = \mu_2 = \ldots = \mu_{i_0} = 0$ and
  for $j > i_0$, recursively define
\[
\mu_{j} = \mu_{j-1} - (y_j - \lambda_j) \ge 0. 
\]
Then it is straightforward to check that the pair $(x, \mu)$ obeys the
KKT optimality conditions from Section \ref{sec:prelim}
\end{proof}

We now introduce the \nameproxsp{} algorithm (Algorithm
\ref{alg:fastprox}) for computing the prox: for pedagogical reasons we
introduce it in its simplest form before presenting in Section
\ref{sec:stack} a stack implementation running in $O(n)$ flops.
\begin{algorithm}[h]
\caption{\nameprox} 
\label{alg:fastprox}
\begin{algorithmic}
  \STATE \textbf{input:} Nonnegative and nonincreasing sequences $y$
  and $\lambda$.
%
%
\WHILE{$y - \lambda$ is not nonincreasing}
\STATE Identify strictly increasing subsequences, i.e.~segments $i:j$
    such that
    \begin{equation}
      \label{eq:subseq}
y_i - \lambda_i < y_{i+1} - \lambda_{i+1} < \ldots < y_j - \lambda_j.
    \end{equation}
    \STATE Replace the values of $y$ over such segments by their
    average value: for $k \in \{i, i+1, \ldots, j\}$
\[
y_k \gets  \frac{1}{j-i+1} \sum_{i \le k \le j} y_k.
\]
\STATE Replace the values of $\lambda$ over such segments by their
    average value: for $k \in \{i, i+1, \ldots, j\}$
\[
\lambda_k \gets \frac{1}{j-i+1} \sum_{i \le k \le j} \lambda_k.
\]
\ENDWHILE
\STATE \textbf{output:} $x = (y - \lambda)_+$.
\end{algorithmic}
\end{algorithm}

This algorithm, which obviously terminates in at most $n$ steps, is
simple to understand: we simply keep on averaging until the
monotonicity property holds, at which point the solution is known in
closed form thanks to Lemma \ref{lem:monotone}. The key point
establishing the correctness of the algorithm is that the update does
not change the value of the prox. This is formalized below. 

\begin{lemma} 
  \label{lem:key} Let $(y^+, \lambda^+)$ be the updated value of
  $(y,\lambda)$ after one pass in Algorithm \ref{alg:fastprox}. Then
\[
\proxh(y) = \operatorname{prox}_{\lambda^+}(y^+). 
\]
\end{lemma}
\begin{proof} We first claim that the prox has to be constant over any
  monotone segment of the form
\[
y_i - \lambda_i \le y_{i+1} -
  \lambda_{i+1} \le \ldots \le y_j - \lambda_j.
\]
To see why this is true, set $x = \proxh(y)$ and suppose the contrary:
then over a segment as above, there is $k \in \{i, i+1, \ldots, j-1\}$
such that $x_k > x_{k+1}$ (we cannot have a strict inequality in the
other direction since $x$ has to be primal feasible). By complementary
slackness, $\mu_k = 0$. This gives
\begin{align*}
  x_k & = y_k - \lambda_k - \mu_{k-1}\\
  x_{k+1} & = y_{k+1} - \lambda_{k+1} + \mu_{k+1}.
\end{align*}
Since $y_{k+1} - \lambda_{k+1} \ge y_k - \lambda_{k}$ and $\mu \ge 0$,
we have $x_k \le x_{k+1}$, which is a contradiction.


Now an update replaces an increasing segment as in \eqref{eq:subseq}
with a constant segment and we have just seen that both proxes must be
constant over such segments. Now consider the cost function associated
with the prox with parameter $\lambda$ and input $y$ over an
increasing segment as in \eqref{eq:subseq},
\begin{equation}
  \label{eq:blocksum}
\sum_{i \le k \le j} \Bigl\{\half (y_k - x_k)^2 + \lambda_k x_k\Bigr\}. 
\end{equation}
Since all the variables $x_k$ must be equal to some value $z$ over
this block, this cost is equal to
\begin{align*}
  \sum_{i \le k \le j} \Bigl\{\half (y_k - z)^2 + \lambda_k z\Bigr\}
 & = \sum_k \half (y_k - \bar y)^2 + \sum_{i \le k \le j}
  \Bigl\{\half (\bar y - z)^2 + \bar \lambda z\Bigr\}\\
& = \sum_k \half (y_k - \bar y)^2 + \sum_{i \le k \le j}
  \Bigl\{\half (y_k^+ - z)^2 + \bar \lambda_k^+ z\Bigr\},
\end{align*}
where $\bar y$ and $\bar \lambda$ are block averages. The second term
in the right-hand side is the cost function associated with the prox
with parameter $\lambda^+$ and input $y^+$ over the same segment since
all the variables over this segment must also take on the same value.
Therefore, it follows that replacing each appearance of block sums as
in \eqref{eq:blocksum} in the cost function yields the same
minimizer. This proves the claim.
\end{proof}

In summary, the \nameproxsp{} algorithm finds the solution to
\eqref{eq:prox} in a finite number of steps.

\subsection{Stack-based algorithm for \nameprox}
\label{sec:stack}

As stated earlier, it is possible to obtain an $O(n)$ implementation
of \nameprox.  Below we present a stack-based approach. We use tuple
notation $(a,b)_i = (c,d)$ to denote $a_i = c$, $b_i = d$.

\begin{algorithm}[H]
\caption{Stack-based algorithm for \nameprox.}
\label{Alg:StackAlgorithm}
\begin{algorithmic}[1]
  \STATE \textbf{input:} Nonnegative and nonincreasing sequences $y$
  and $\lambda$.
\STATE {\it \# Find optimal group levels}
\STATE $t \gets 0$
\FOR{$k = 1$ to $n$}
\STATE $t \gets t + 1$
\STATE   $(i,j,s,w)_t = (k,\ k,\ y_i - \lambda_i,\ (y_i - \lambda_i)_+)$
\WHILE{$(t > 1)$ and $(w_{t-1} \leq w_{t})$}
\STATE   $(i,j,s,w)_{t-1} \gets (i_{t-1},\ j_t,\ s_{t-1}+s_{t},
   (\frac{j_{t-1} - i_{t-1} + 1}{j_{t} - i_{t-1} +1}\cdot s_{t-1}
   + \frac{j_t - i_t + 1}{j_{t} - i_{i-1}+1}\cdot s_{t})_+$)
\STATE  Delete $(i,j,s,w)_t$, $t \gets t - 1$
\ENDWHILE
\ENDFOR

\STATE {\it \# Set entries in $x$ for each block}
\FOR{$\ell = 1$ to $t$}
\FOR{$k = i_{\ell}$ to $j_{\ell}$}
\STATE $x_k \gets w_{\ell}$
\ENDFOR
\ENDFOR
\end{algorithmic}
\end{algorithm}

For the complexity of the algorithm note that we create a
total of $n$ new tuples. Each of these tuple is merged into a previous
tuple at most once. Since the merge takes a constant amount of time
the algorithm has the desired ${O}(n)$ complexity.

With this paper, we are making available a C, a Matlab, and an R
implementation of the stack-based algorithm at
\url{http://www-stat.stanford.edu/~candes/SortedL1}.
The algorithm is also included in the current version of the TFOCS
package available here \url{http://cvxr.com}, see \cite{tfocs}. To
give an idea of the speed, we applied the code to a series of vectors
with fixed length and varying sparsity levels. The average runtimes
measured on a MacBook Pro equipped with a 2.66 GHz Intel Core i7 are
reported in Table~\ref{Table:ProxRuntime}.

\begin{table}
\centering
\input{./tables/TableProxRuntime.tex}
\caption{Average runtimes of the stack-based prox implementation 
  with normalization steps (sorting and sign changes) included,
  respectively excluded.}\label{Table:ProxRuntime}
\end{table}



\subsection{Proximal algorithms for SLOPE}
\label{sec:algol}

With a rapidly computable algorithm for evaluating the prox, efficient
methods for computing the SLOPE solution \eqref{eq:ol} are now a
stone's throw away. Indeed, we can entertain a proximal gradient
method which goes as in Algorithm \ref{alg:ista}.
\begin{algorithm}[H]
\caption{Proximal gradient algorithm for SLOPE \eqref{eq:ol}}
\label{alg:ista}
\begin{algorithmic}[1]
\REQUIRE $b^{0} \in \R^p$
\FOR {$k=0,1,\ldots $}
\STATE $b^{k+1} = \proxh(b^{k} - t_k X'(Xb^k - y))$
\ENDFOR
\end{algorithmic}
\end{algorithm}
It is well known that the algorithm converges (in the sense that
$f(b^k)$, where $f$ is the objective functional, converges to the
optimal value) under some conditions on the sequence of step sizes
$\{t_k\}$.  Valid choices include step sizes obeying $t_k < 2/\|X\|^2$
and step sizes obtained by backtracking line search, see
\cite{tfocs,fista}.

Many variants are of course possible and one may entertain accelerated
proximal gradient methods in the spirit of FISTA, see \cite{fista} and
\cite{Nesbook,nes:07}. The scheme below is adapted from \cite{fista}.
\begin{algorithm}[H]
  \caption{Accelerated proximal gradient algorithm for SLOPE \eqref{eq:ol}}
\label{alg:fista}
\begin{algorithmic}[1]
\REQUIRE $b^{0} \in \R^p$, and set $a^{0} = b^{0}$ and $\theta_0=1$ 
\FOR {$k=0,1,\ldots $}
\STATE $b^{k+1} =  \proxh(a^{k} - t_k X'(Xa^k - y))$ 
\STATE $\theta_{k+1}^{-1} = \frac12 (1 + \sqrt{1 + 4/\theta_k^2})$
\STATE $a^{k+1} = b^{k+1} + \theta_{k+1}(\theta_k^{-1} - 1)(b^{k+1} - b^k)$
\ENDFOR
\end{algorithmic}
\end{algorithm} 

The code used for the numerical experiments uses a straightforward
implementation of the standard FISTA algorithm, along with
problem-specific stopping criteria.  Standalone Matlab and R
implementations of the algorithm are available at the website listed
in Section \ref{sec:stack}. TFOCS implements Algorithms \ref{alg:ista}
and \ref{alg:fista} as well as many variants;  
for instance, the Matlab code below prepares the prox and then solves
the SLOPE problem,
\begin{verbatim}
prox = prox_Sl1(lambda);
beta = tfocs( smooth_quad, { X, -y }, prox, beta0, opts );
\end{verbatim}
Here \verb|beta0| is an initial guess (which can be omitted) and
\verb|opts| are options specifying the methods and parameters one
would want to use, please see \cite{tfocs} for details. There is
also a one-liner with default options which goes like this:
\begin{verbatim}
beta  = solver_SLOPE( X, y, lambda);
\end{verbatim}


\subsection{Duality-based stopping criteria}

To derive the dual of \eqref{eq:ol} we first rewrite it as
\[
\minimize{b,r}\quad \half r'r + \ol{b}{\lambda}\quad \st\quad Xb + r = y.
\]
The dual is then given by
\[
\maximize{w}\quad \mathcal{L}(b,r,w),
\]
where
\begin{eqnarray*}
\mathcal{L}(b,r,w) &:=& \inf_{b,r}\ \{\half r'r + \ol{b}{\lambda} -
w'(Xb + r - y)\}\\
& = & w'y -\sup_{r}\ \{w'r - \half r'r\} - \sup_{b}\ \{(X'w)'b - \ol{b}{\lambda}\}. 
\end{eqnarray*}
The first supremum term evaluates to $\half w'w$ by choosing $r =
w$. The second term is the conjugate function $\olconjplain$ of $\olplain$ evaluated
at $v = X'w$, which can be shown to reduce to
\[
\olconj{v}{\lambda} := \sup_{b}\ \{v'b - \ol{b}{\lambda}\} = \begin{cases}
0 & v \in C_\lambda,\\
+\infty & \mbox{otherwise},
\end{cases}
\]
where the set $C_\lambda$ is given by \eqref{eq:dual2}. The dual problem
is thus given by
\[
\maximize{w}\quad w'y - \half w'w\quad \st\quad w \in C_\lambda.
\]
The dual formulation can be used to derive appropriate stopping
criteria. At the solution we have $w = r$, which motivates estimating
a dual point by setting $\hat{w} = r =: y - Xb$. At this point the
primal-dual gap at $b$ is the difference between the primal and dual
objective:
\[
\delta(b) = (Xb)'(Xb-y) +
\ol{b}{\lambda}. 
\]
However, $\hat w$ is not guaranteed to be feasible, i.e.,~we may not
have $\hat w \in C_\lambda$. Therefore we also need to compute a level
of infeasibility of $\hat{w}$, for example
\[
\mbox{infeasi}(\hat w) = \max\Big\{0,\ \max_{i} \sum_{j \leq i} (\vert
\hat w\vert_{(j)} - \lambda_j)\Big\}.
\]
The algorithm used in the numerical experiments terminates whenever
both the infeasibility and primal-dual gap are sufficiently small. In
addition, it imposes a limit on the total number of iterations to
ensure termination.


%% file: tables/TableProxRuntime.tex
\begin{tabular}{lrrr}
\hline\hline
 & $p=10^{5}$ & $p=10^{6}$ & $p=10^{7}$\\
\hline
Total prox time (sec.)  &  9.82e-03 &  1.11e-01 &  1.20e+00\\
Prox time after normalization (sec.) &  6.57e-05 &  4.96e-05 &  5.21e-05\\
\hline
\hline
\end{tabular}

%% file: fdr_ortho.tex
\section{FDR Control Under Orthogonal Designs}
\label{sec:fdr_orth}

\newcommand{\one}[1]{\mathbbm{1}_{\left\{ {#1} \right\} }}

In this section, we prove FDR control in the orthogonal design,
namely, Theorem \ref{teo:fdr_control1}. As we have seen in Section
\ref{sec:introduction}, the SLOPE solution reduces to
\[
 \min_{b \in \R^p} \,  
\half \|\tilde y - b\|_{\ell_2}^2 + \sum_{i = 1}^p \lambda_i |b|_{(i)},
\]
where $\tilde y = X'y \sim \mathcal{N}(\beta,I_p)$. From this, it is clear
that it suffices to consider the setting in which $y \sim
\mathcal{N}(\beta, I_n)$, which we assume from now on.

We are thus testing the $n$ hypotheses $H_i: \beta_i = 0$, $i = 1,
\ldots, n$ and set things up so that the first $n_0$ hypotheses are
null, i.e.~$\beta_i = 0$ for $i \le n_0$. The SLOPE solution
is
\begin{equation}
\label{eq:Id}
\hat \beta = \text{arg min} \, \half \|y - b\|_{\ell_2}^2 + \sum_{i = 1}^n \lambda_i |b|_{(i)}
\end{equation}
with $\lambda_i = \Phi^{-1}(1-iq/2n)$. We reject $H_i$ if and only if
$\hat \beta_i \neq 0$. Letting $V$ (resp.~$R$) be the number of false
rejections (resp.~the number of rejections) or, equivalently, the
number of indices in $\{1, \ldots, n_0\}$ (resp.~in $\{1, \ldots,
n\}$) for which $\hat \beta_i \neq 0$, we have
\begin{equation}
\label{eq:fdr-expression}
  \text{FDR} = \E \left[ \frac{V}{R \vee 1} \right] = \sum_{r = 1}^n
  \E \left[ \frac{V}{r} \one{R =r} \right]
  = \sum_{r = 1}^n \frac{1}{r} \E \left[ \sum_{i =1}^{n_0} \one{H_i \text{ is rejected}} \one{R = r} \right].
\end{equation}
The proof of Theorem \ref{teo:fdr_control1} now follows from the two
key lemmas below.
\begin{lemma}
\label{lem:critical0}
Let $H_i$ be a null hypothesis and let $r \ge 1$. Then  
\[
\{\text{\rm $y$: $H_i$ is rejected and $R = r$}\} = \{\text{\rm $y$:
  $|y_i| > \lambda_r$ and $R = r$}\}.
\]
\end{lemma}
\begin{lemma}
\label{lem:critical}
Consider applying the {\em SLOPE} procedure to $\tilde y = (y_1,
\ldots, y_{i-1}, y_{i+1}, \ldots, y_n)$ with weights $\tilde \lambda =
(\lambda_2, \ldots, \lambda_n)$ and let $\tilde R$ be the number of
rejections this procedure makes. Then with $r \ge 1$,
\[
\{\text{\rm $y$: $|y_i| > \lambda_r$
  and $R = r$}\} \subset
  \{y: \text{\rm $|y_i| > \lambda_r$ and $\tilde R = r-1$}\}.
\]
\end{lemma}

To see why these intermediate results give Theorem
\ref{teo:fdr_control1}, observe that
\begin{align*}
  \P(\text{$H_i$ rejected and $R = r$}) & \le \P(\text{$|y_i|
    \ge
    \lambda_r$ and $\tilde R = r-1$})\\
  & = \P(|y_i| \ge
  \lambda_r)  \P(\tilde R = r-1) \\
  & = \frac{qr}{n} \P(\tilde R = r-1),
\end{align*}
where the inequality is a consequence of the lemmas above and the
first equality follows from the independence between $y_i$ and $\tilde
y$.  Plugging this inequality into \eqref{eq:fdr-expression} gives
\[
\text{FDR} = \sum_{r = 1}^n \frac{1}{r} \sum_{i = 1}^{n_0}
\P(\text{$H_i$ rejected and $R = r$}) \le \sum_{r \ge 1}
\frac{qn_0}{n} \P(\tilde R = r-1) = \frac{qn_0}{n},
\]
which finishes the proof. 

\subsection{Proof of Lemma \ref{lem:critical0}}

We begin with a lemma we shall use more than once. 
\begin{lemma}
\label{lem:gosia}
Consider a pair of nonincreasing and nonnegative sequences $y_1 \ge
y_2 \ge \ldots \ge y_n \ge 0$, $\lambda_1 \ge \lambda_2 \ge \ldots \ge
\lambda_n \ge 0$, and let $\hat b$ be the solution to
\[
\begin{array}{ll}
  \text{\rm minimize} & \quad f(b) = \, \half \|y - b\|_{\ell_2}^2 + \sum_{i = 1}^n \lambda_i b_i \\
  \text{\rm subject to} & \quad b_1 \ge b_2 \ge \ldots \ge b_n \ge 0.  
\end{array}
\]
If $\hat b_r > 0$ and $\hat b_{r+1} = 0$, then for every $j \le r$, it
holds that
\begin{equation}
  \label{eq:gosia1}
\sum_{i  = j}^r (y_i - \lambda_i) > 0
\end{equation}
and for every $j \ge r+1$, 
\begin{equation}
  \label{eq:gosia2}
\sum_{i  = r+1}^j (y_i - \lambda_i) \le 0.
\end{equation}
\end{lemma}
\begin{proof}
  To prove \eqref{eq:gosia1}, consider a new feasible sequence $b$,
  which differs from $\hat b$ only by subtracting a small positive
  scalar $h < \hat b_r$ from $\hat b_j, \ldots, \hat b_r$. Now 
\[
f(b) - f(\hat b) = h \sum_{i = j}^r (y_i - \lambda_i - \hat b_i) + h^2 \sum_{i = j}^r \half.  
\]
Taking the limit as $h$ goes to zero, the optimality of $\hat b$
implies that $\sum_{i = j}^r (y_i - \lambda_i - \hat b_i) \ge 0$,
which gives
\[
\sum_{i = j}^r (y_i - \lambda_i) \ge \sum_{i = j}^r \hat b_i > 0. 
\]

For the second claim \eqref{eq:gosia2}, consider a new sequence $b$,
which differs from $\hat b$ by replacing $\hat b_{r+1}, \ldots, \hat
b_j$ with a positive scalar $0 < h < \hat b_r$. Now observe that
\[
f(b) - f(\hat b) = -h \sum_{i = r+1}^j (y_i - \lambda_i) + h^2 \sum_{i
  = r+1}^j \half.
\]
The claim follows from the optimality of $\hat b$.
\end{proof}

It is now straightforward so see how these simple relationships give
Lemma \ref{lem:critical0}. Observe that when $R = r$, we must have
$|y|_{(r)} > \lambda_r$ and $|y|_{(r+1)} \le \lambda_{r+1}$. Hence, if
$H_1$ is rejected, it must hold that $|y_1| \ge |y|_{(r)} > 
\lambda_r$. This shows that $\{\text{$H_1$ is rejected and $R = r$}\}
\subset \{\text{$|y_1| > \lambda_r$ and $R = r$}\}$. Conversely,
assume that $|y_1| > \lambda_r$ and $R = r$. Then $H_1$ must be
rejected since $|y_1| > |y|_{(r+1)}$. This shows that $\{\text{$H_1$
  is rejected and $R = r$}\} \supset \{\text{$|y_1| > \lambda_r$ and
  $R = r$}\}$.

\subsection{Proof of Lemma \ref{lem:critical}}

We assume without loss of generality that $y \ge 0$ (the extension to
arbitrary signs is trivial). By assumption the solution to
\eqref{eq:Id} with $\lambda_i = \Phi^{-1}(1-iq/2n)$ has exactly $r$ strictly
positive entries, and we need to show that when $y_1$ is rejected, the
solution to
\begin{equation}
\label{eq:reduced}
\text{min} \, J(\tilde b) := \sum_{i = 1}^{n-1} \half (\tilde{y}_i -
\tilde{b}_i)^2 + \sum_{i=1}^{n-1} \tilde \lambda_{i} |\tilde b|_{(i)} 
\end{equation}
in which $\tilde \lambda_i = \lambda_{i+1}$ has exactly $r-1$ nonzero
entries. We prove this in two steps: 
\begin{itemize}
\item[(i)] The optimal solution $\hat b$ to \eqref{eq:reduced} has at
  least $r-1$ nonzero entries.
\item[(ii)] The optimal solution $\hat b$ to \eqref{eq:reduced} has at
  most $r-1$ nonzero entries.
\end{itemize}

\subsubsection{Proof of (i)} 

Suppose by contradiction that $\hat b$ has fewer than $r-1$ entries;
i.e.,~$\hat b$ has $j - 1$ nonzero entries with $j < r$. Letting $I$
be those indices for which the rank of $\tilde y_i$ is between $j$ and
$r-1$, consider a feasible point $b$ as in the proof of Lemma
\ref{lem:gosia} defined as
\[
b_i = \begin{cases} h & i \in I,\\
\hat b_i & \text{otherwise}; 
\end{cases}
\]
here, the positive scalar $h$ obeys $0 < h <
b_{(j-1)}$. By definition, 
\[
J(b) - J(\hat b) = -h \sum_{i = j}^{r-1} (\tilde y_{(i)} - \tilde
\lambda_i) + h^2 \sum_{i = j}^{r-1} \half.
\]
Now 
\[
\sum_{j \le i \le r-1} \tilde y_{(i)} -
\tilde \lambda_i =  \sum_{j+1 \le i \le r} \tilde y_{(i-1)} -
\lambda_i \ge \sum_{j+1 \le i \le r} y_{(i)} -
\lambda_i > 0.  
\]
The first equality follows from $\tilde \lambda_i = \lambda_{i+1}$,
the first inequality from $y_{(i)} \le \tilde y_{(i-1)}$ and the last
from \eqref{eq:gosia1}. By selecting $h$ small enough, this gives
$J(b) < J(\hat b)$, which contradicts the optimality of $\hat b$.

\subsubsection{ Proof of (ii)} 

The proof is similar to that of (i).  Suppose by contradiction that
$\hat b$ has more than $r-1$ entries; i.e.~$\hat b$ has $j$ nonzero
entries with $j \ge r$.  Letting $I$ be those indices for which the
rank of $\tilde y_i$ is between $r$ and $j$, consider a feasible point
$b$ as in the proof of Lemma \ref{lem:gosia} defined as
\[
b_i = \begin{cases} b_i - h & i \in I\\
\hat b_i & \text{otherwise}; 
\end{cases}
\]
here, the positive scalar $h$ obeys $0 < h <
b_{(j)}$. By definition, 
\[
J(b) - J(\hat b) = h \sum_{i = r}^j (\tilde y_{(i)} - \tilde \lambda_i - \hat b_{(i)}) + h^2 \sum_{i = r}^j \half.
\]
Now 
\[
\sum_{r \le i \le j} (\tilde y_{(i)} -
\tilde \lambda_i) =  \sum_{r+1 \le i \le j+1} (y_{(i)} -
\lambda_i) \le 0. 
\]
The equality follows from the definition and the inequality from
\eqref{eq:gosia2}. By selecting $h$ small enough, this gives
$J(b) < J(\hat b)$, which contradicts the optimality of $\hat b$.

%% file: fdr_general.tex
\section{FDR Control Under General Designs}
\label{sec:fdr_general}

\subsection{The notion of FDR in the linear model}

We consider the multiple testing problem in which we wish to decide
whether each of the $p$ regression coefficients in the linear model
\eqref{eq:linear} is zero or not. Now the notion of FDR control in
this situation is delicate and it is best to present the precise
context in which our results hold as to avoid any kind of
misunderstanding. We work with a generative model 
\[
y = X \beta + z
\]
for the data in which the errors are i.i.d.~zero-mean Gaussian random
variables. In other words, we assume that the (full) linear model is
correct and wish to know which of the coefficients $\beta_i$ are
nonzero. For example, in a medical imaging application, $\beta$ may be
the concentrations of hydrogen at various locations in the body
(different tissues are characterized by different concentration
levels) and while we cannot observe such concentrations directly, they
can be measured indirectly by magnetic resonance. In this case, the
linear model above holds, the object of inference is a well-defined
physical quantity, and it makes sense to test whether there are
locations whose concentration exceeds a prescribed level. Another
example may concern gene mapping studies in which we wish to identify
which of the many genes are associated with a given phenotype. Here,
there are mutations that affect the phenotype and others that do not
so that---assuming the correctness of the linear model---there are
true (and false) discoveries to be made. We are thus concerned with a
true linear model in which the notion of a true regressor along with
the value of its regression coefficient has a clear meaning, and can
be explained in the language of the respective science. In contrast,
we are not concerned with the use of the linear model as means of an
interpretative model, which, while being inexact, may be used to
summarize a phenomenon of interest or predict future outcomes since it
may not allow true and false discoveries.

Furthermore, we also wish to stay away from designs where the columns
of $X$ (the variables) are highly correlated as in this case, the
stringent definition we adopt for FDR (expected proportion of
incorrectly selected variables) may not be the right notion of Type I
error, see the discussion in \cite{SiegmundFDR}.  Hence, we shall
present the performance of different selection methods on seemingly
simple examples, where the columns of the design matrix are only
slightly correlated or where they are generated as independent random
variables.  As we shall see, while small correlations can already lead
to many variable selection problems, model selection procedures based
on the sorted $\ell_1$ norm can work well under sparsity, i.e.,~when
the number of true regressors is comparably small.

\subsection{Loss of FDR control due to shrinkage}
\label{sec:FDR_LASSO}

As briefly mentioned in the Introduction, methods using convex
surrogates for sparsity, such as the $\ell_1$ or sorted $\ell_1$
norms, can have major difficulties in controlling the FDR. This
phenomenon has to do with the shrinkage of regression coefficients, as
we now explain.  

We consider the lasso \eqref{eq:lasso} for simplicity and assume that
the columns of $X$ have unit norm. The optimality conditions for the
lasso solution $\hat \beta$ take the form
\begin{equation}
  \label{eq:KKT}
  \hat \beta = \eta_\lambda(\hat \beta - X'(X\hat \beta - y)), 
\end{equation}
where $\eta_\lambda$ is the soft-thresholding operator,
$\eta_\lambda(t) = \sgn(t) (|t| - \lambda)_+$, applied componentwise; 
see \cite[page 150]{BoydProx}. Setting
\begin{equation}
\label{eq:v}
v_i = \<X_i, \sum_{j \neq i} X_j (\beta_j - \hat \beta_j)\>, 
\end{equation}
we can rewrite the optimality condition as 
\begin{equation}
  \label{eq:KKT2}
  \hat \beta_i = \eta_\lambda(\beta_i + X_i' z  + v_i).
\end{equation}
Observe that conditional on $X$, $X_i' z \sim \mathcal{N}(0,1)$.  In
an orthogonal design, $v_i = 0$ for all $i$, and taking $\lambda$ at
the appropriate level (e.g.,~$\Phi^{-1}(1-q/2p)$) controls the FDR and
even the FWER. Imagine now that $X$ has correlated columns and that
$k$ regression coefficients are large. Then the lasso estimates of
these coefficients will be shrunk towards zero by an amount roughly
proportional to $\lambda$. The effect of this is that $v_i$ now looks
like a noise term whose size is roughly proportional to $\lambda$
times the square root of $k$. In other words, there is an inflation of
the `noise variance' so that we should not be thresholding at
$\lambda$ but at a higher value. The consequence is that the lasso
procedure is too liberal and selects too many variables for which
$\beta_i = 0$; hence, the FDR becomes just too large.  One can imagine
using a higher value of $\lambda$. The problem, however, is that the
if the coefficients are really large, the estimation bias is of size
$\lambda$ so that the size of $v_i$ scales like $\lambda$ so there
does not appear to be an easy way out (at least, as long as $\lambda$
is selected non-adaptively).

\begin{figure}[h!]
        \centering
        \begin{subfigure}[b]{0.495\textwidth}
                \centering
                \includegraphics[width=\textwidth]{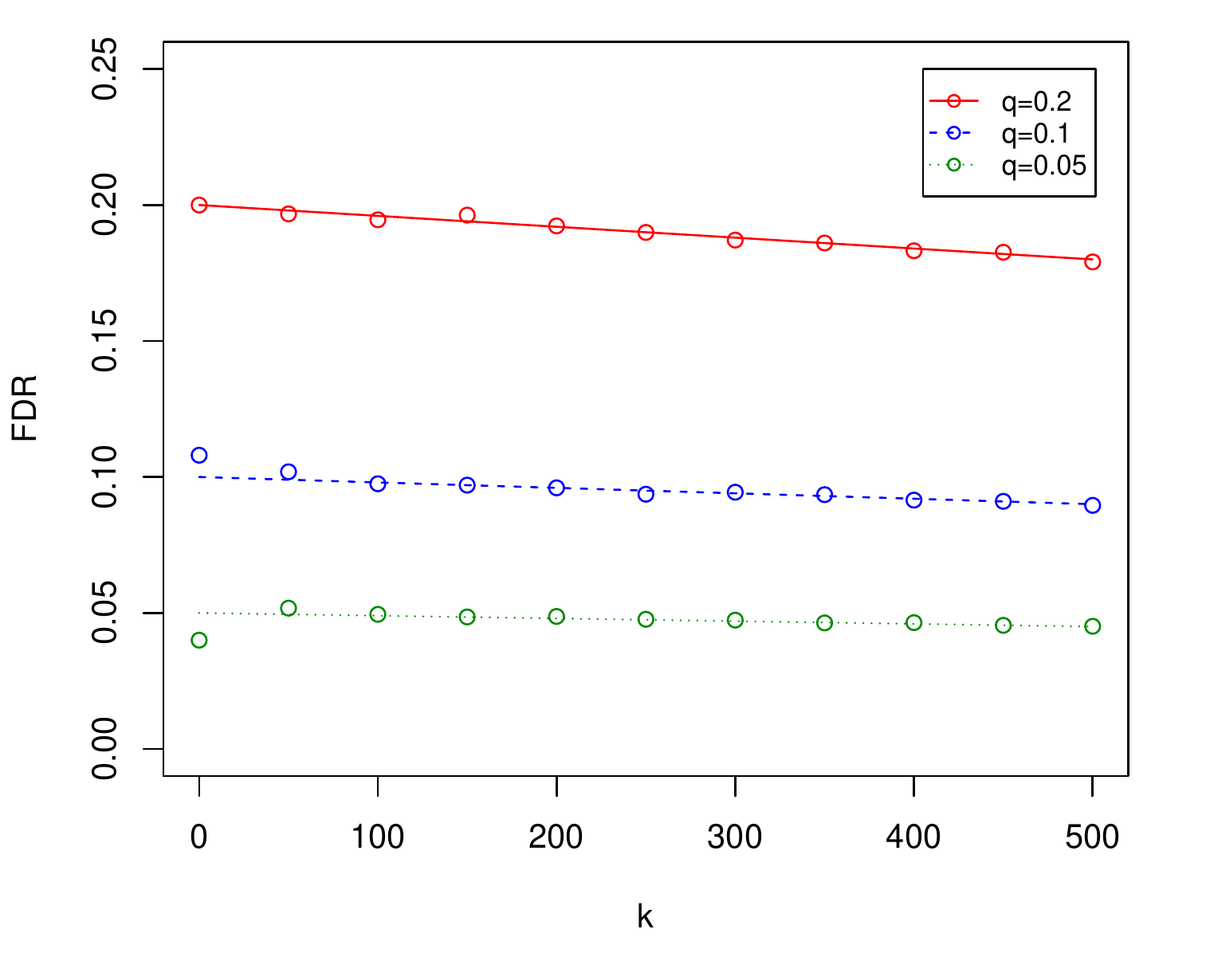}
                \caption{Orthogonal design.}
                \label{fig:Fig3a}
        \end{subfigure}
        \begin{subfigure}[b]{0.495\textwidth}
                \centering
                \includegraphics[width=\textwidth]{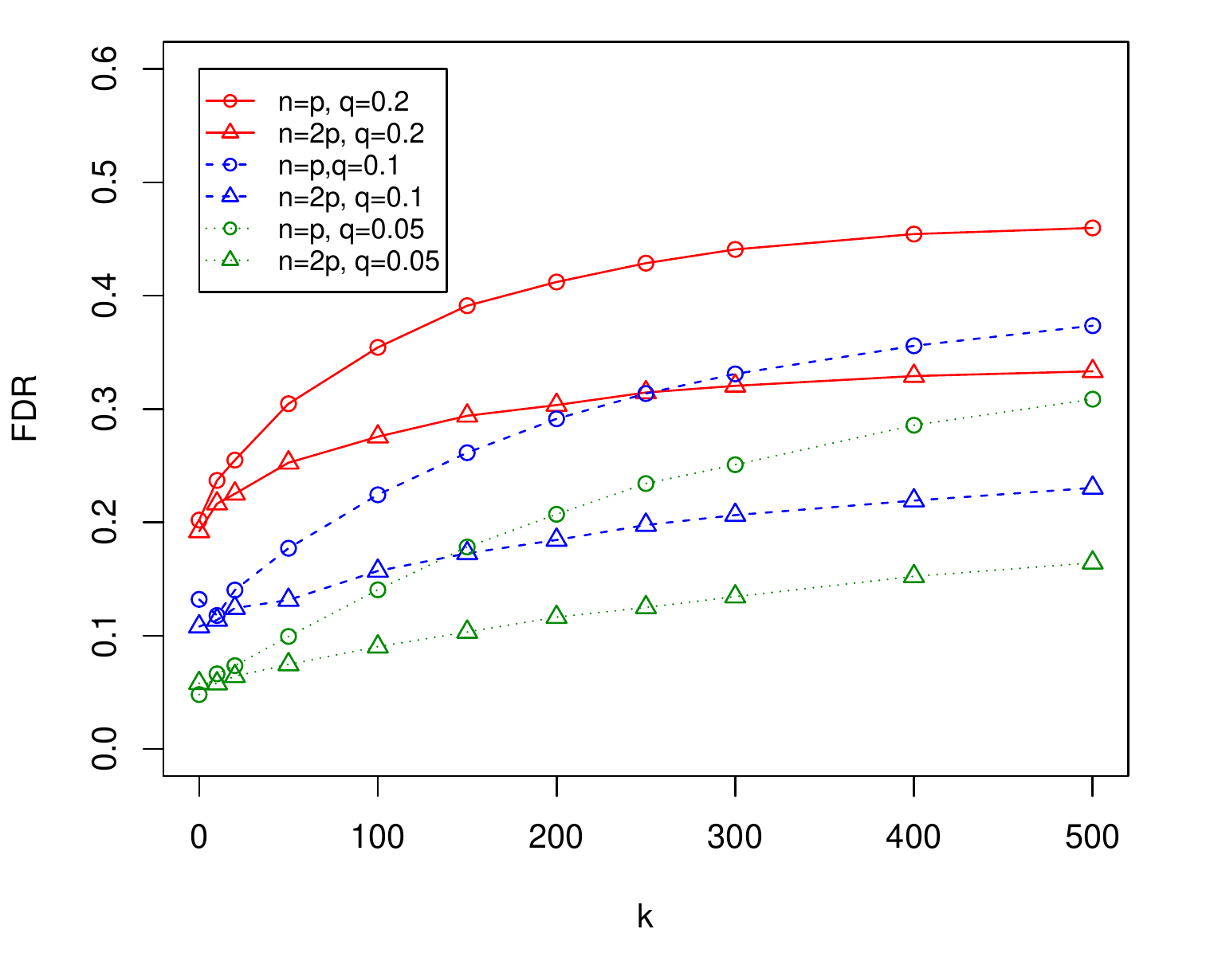}
\caption{Gaussian design.}
\label{fig:Fig3b}
        \end{subfigure}
        \caption{Observed FDR of SLOPE with $\lambda=\LBH$ for
          $p=5,000$. In (a), the straight lines indicate the nominal
          level of the BHq procedure, namely, $q(1-k/p)$.}
          \label{fig:Fig3}
\end{figure}

Following this line of thought, one can make two qualitative
predictions in the model with a few large regressors: (1) the problem
tends to be more severe as the number of regressors increases, at
least initially (when most of the variables are in the model, it gets
harder to make false rejections), and (2) the problem tends to be more
severe when the columns tend to be more correlated.  Obviously, the
problems with FDR control also apply to SLOPE.  Figure 3
presents the estimated FDR of SLOPE both with an
orthogonal design and a random Gaussian design in which $X$ has
i.i.d.~$\mathcal{N}(0,1/n)$ entries so that the columns nearly have
unit norm. The sequence $\lambda$ is set to be $\LBH$.  In the first
case, we work with $n = p = 5,000$ and in the second with $n = p =
5,000$ and $n = 2p = 10,000$. The value of the nonzero regression
coefficients is set to $5\sqrt{2 \log p}$. Figure~\ref{fig:Fig3b}
shows that SLOPE with a sequence of BH values no longer
controls the FDR in the nonorthogonal case. In fact, the FDR increases
with $k$ as predicted.  The values of FDR are substantially smaller
when $n=2p$ as compared to the case when $n=p$. This is naturally in
line with our prediction since $\E(X_i' X_j)^2 = 1/n$ for $i \neq j$
so that random vectors in dimension $n = 5,000$ statistically exhibit
higher sample correlations than vectors in a space of twice this size.

In closing, we have discussed the problem associated with the bias
induced by large regression coefficients. Looking at \eqref{eq:KKT2},
there are of course other sources of bias causing a variance inflation
such as those coefficients $\beta_j \neq 0$ with vanishing estimates,
i.e.,~$\hat \beta_j = 0$.


\subsection{Adjusting the regularizing sequence for SLOPE}
\label{sec:lambda}

Selecting a sequence $\{\lambda_i\}$ for use in SLOPE is
an interesting research topic that is beyond the scope of this work.
Having said this, an initial thought would be to work with the same
sequence as in an orthogonal design, namely, with $\lambda_i =
\LBH(i)$ as in Theorem \ref{teo:fdr_control1}. However, we have
demonstrated that this choice is too liberal and in this section, we
use our qualitative insights to propose an intuitive adjustment. Our
treatment here is informal.

\newcommand{\betaS}{\beta_{\mathcal{S}}}
\newcommand{\XS}{X_{\mathcal{S}}}
\newcommand{\lambdaS}{\lambda_{\mathcal{S}}}

Imagine we use $\LBH$ and that there are $k$ large coefficients. To
simplify notation, we suppose that $\beta_1 \ge \beta_2 \ge \ldots
\beta_k \gg 1$ and let $\mathcal{S}$ be the support set $\{1, \ldots,
k\}$. Assuming SLOPE correctly detects these variables and
correctly estimates the signs of the regression coefficients, the
estimate of the nonzero components is very roughly equal to
\[
(\XS' \XS)^{-1} (\XS' y - \lambdaS),  
\]
where $\lambdaS = (\LBH(1), \ldots, \LBH(k))'$ causing a bias
approximately equal to 
\[
\E \XS(\betaS - \hat{\beta}_{\mathcal{S}}) \approx \XS (\XS' \XS)^{-1}
\lambdaS. 
\]
We now return to \eqref{eq:KKT2} and ask about the size of the
variance inflation: in other words, what is the typical size of $ X'_i
\XS (\XS' \XS)^{-1} \lambdaS$? For a Gaussian design where the entries
of $X$ are i.i.d.~$\mathcal{N}(0,1)$, it is not hard to see that for
$i \notin \mathcal{S}$,
\[
\E (X'_i \XS (\XS' \XS)^{-1} \lambdaS)^2 = \lambdaS' \E(\XS' \XS)^{-1}
\lambdaS = \frac{\|\lambdaS\|_{\ell_2}^2}{n - |\mathcal{S}| - 1},
\]
where the last equality uses the fact that the expected value of an
inverse dimensional Wishart of dimension $k$ with $n$ degrees of
freedom is equal to $I_k/(n-k-1)$. 

\begin{figure}[h!]
 \centering
        \begin{subfigure}[b]{0.495\textwidth}
                \centering
         \includegraphics[width=\textwidth]{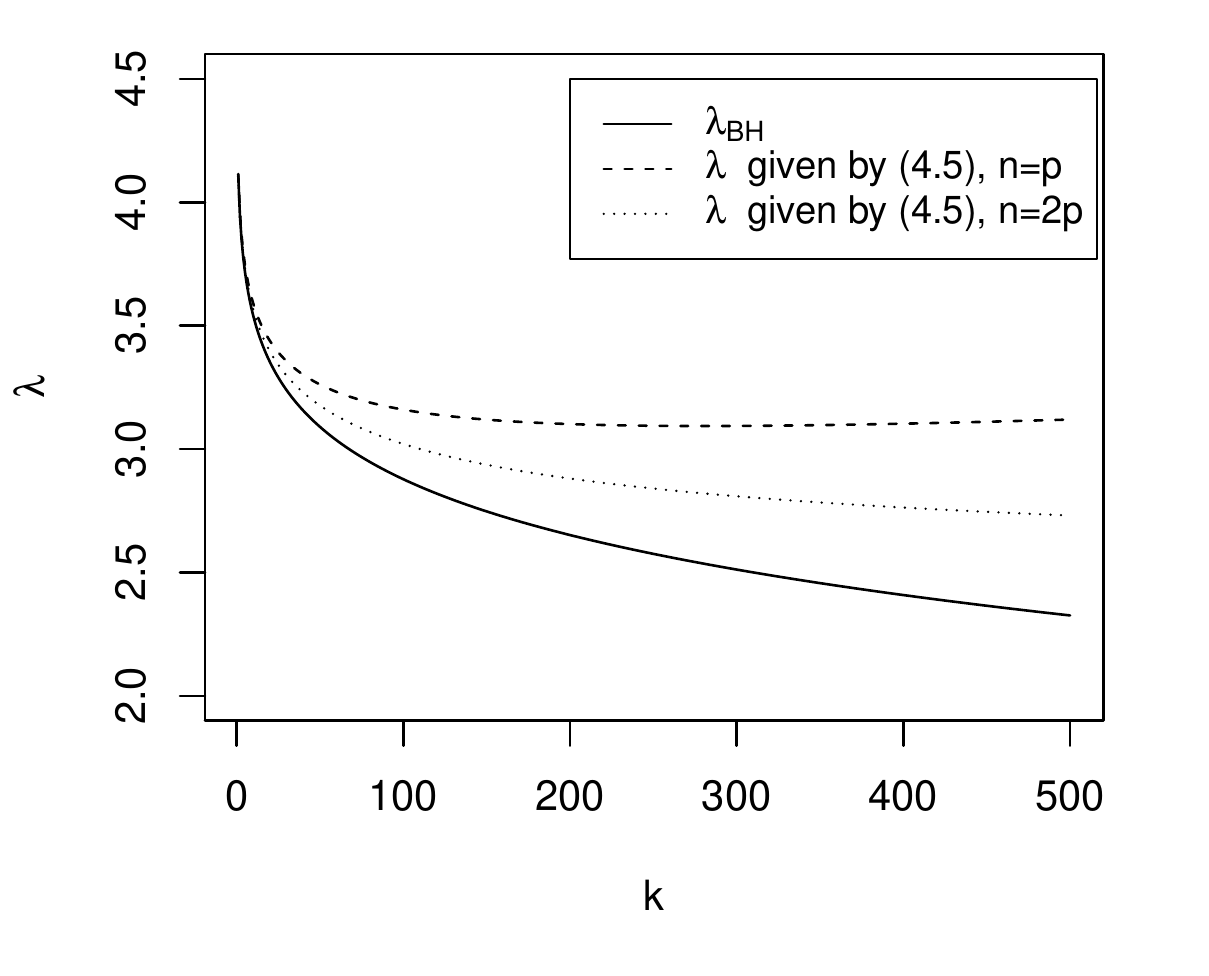}
 \caption{$q = 0.2$.}
\label{fig:Fig4a}
 \end{subfigure}
        \begin{subfigure}[b]{0.495\textwidth}
                \centering
          \includegraphics[width=\textwidth]{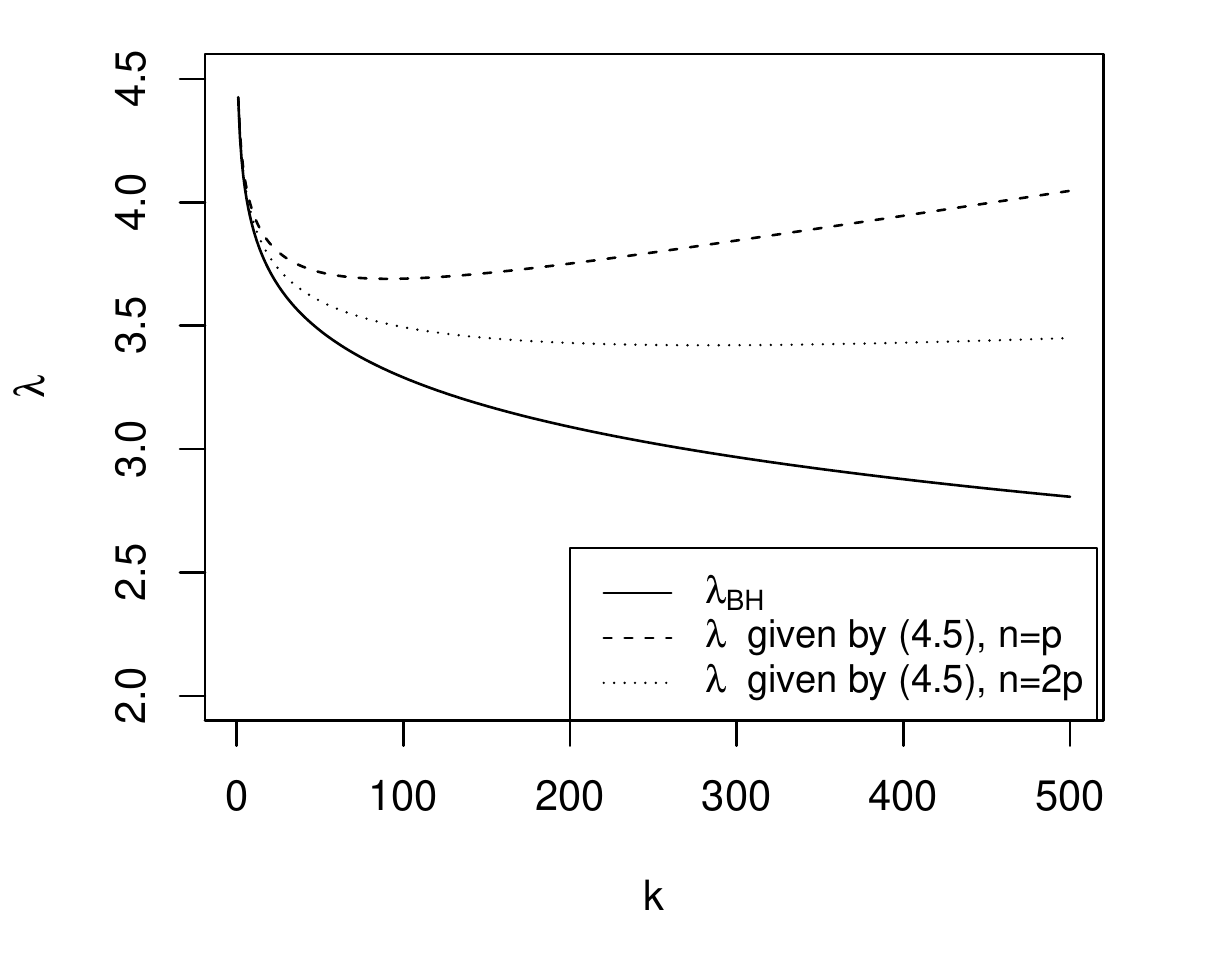}
 \caption{$q = 0.05$.}
\label{fig:Fig4b}
 \end{subfigure}
 \caption{Graphical representation of sequences $\{\lambda_i\}$ for
   $p=5000$. The solid line is $\LBH$, the dashed (resp.~dotted) line
   is $\lambda$ given by \eqref{eq:LOL} for $n = p$ (resp.~$n =
   2p$).}
 \label{fig:Fig4}
\end{figure}

This suggests a correction of the following form: we start with
$\lambda_1 = \LBH(1)$. At the next stage, however, we need to account
  for the slight increase in variance so that we do not want to use
  $\LBH(2)$ but rather
\[
\lambda_2 = \LBH(2) \sqrt{1 + \frac{\lambda_1^2}{n-2}}.
\]
Continuing, this gives 
\begin{equation}
  \label{eq:correction}
  \lambda_i = \LBH(i) \sqrt{1 + \frac{\sum_{j < i} \lambda_j^2}{n-i}}. 
\end{equation}
In our simulations, we shall use a variation on this idea and work
with $\lambda_1 = \LBH(1)$ and for $i > 1$,
\begin{equation}
  \label{eq:LOL}
  \lambda_i =  \LBH(i) \sqrt{1 + \frac{\sum_{j < i} \LBH^2(j)}{n-i}}. 
\end{equation}
In practice, the performance of this slightly less conservative
sequence does not differ from \eqref{eq:correction}. Figure
\ref{fig:Fig4} plots the adjusted values given by \eqref{eq:LOL}. As
is clear, these new values yield a procedure that is more conservative
than that based on $\LBH$.

For small values of $q$, the sequence $\lambda_i$ is no longer
decreasing. Rather it decreases until it reaches a minimum value and
then increases. It would not make sense to use such a sequence---note
that we would also lose convexity---and letting $k^\star = k(n,p,q)$
be the location of the global minimum, we shall work with
\begin{equation}
\label{EC}
\LBHc(i) = \begin{cases} \lambda_i, & i \le k^\star,\\
\lambda_{k^\star}, & i > k^\star, 
\end{cases}
\quad \text{with $\lambda_i$ as in \eqref{eq:LOL}}.
\end{equation}

\begin{figure}[h!]
        \centering
        \begin{subfigure}[b]{0.495\textwidth}
                \centering
                \includegraphics[width=\textwidth]{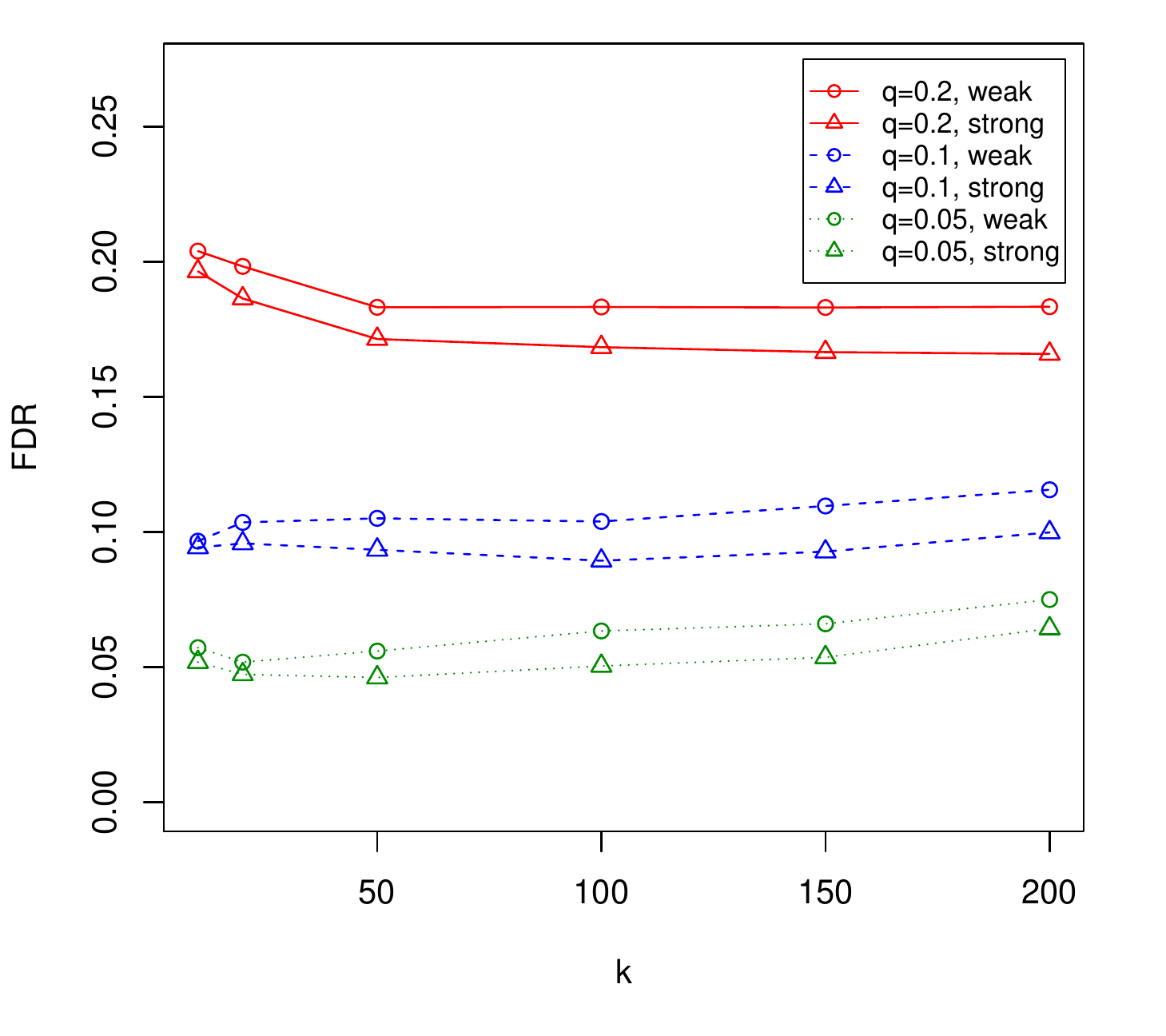}
                \caption{$n=p=5,000$.}
                \label{fig:Fig5a}
        \end{subfigure}
        \begin{subfigure}[b]{0.495\textwidth}
                \centering
                \includegraphics[width=\textwidth]{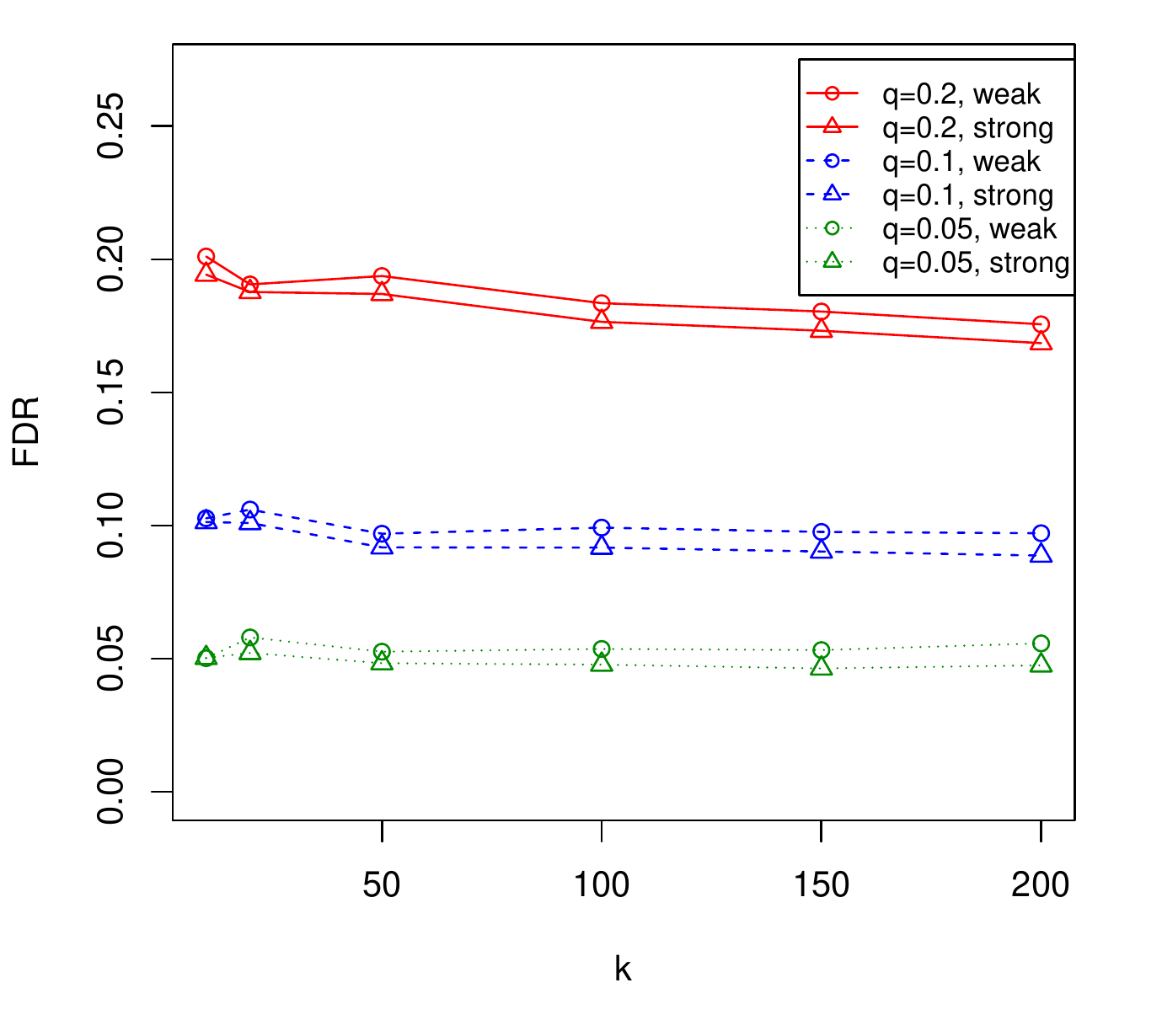}
\caption{$n=2p=10,000$.}
\label{fig:Fig5b}
        \end{subfigure}
        \caption{FDR of SLOPE with $\lambda_i = \LBHc(i)$
          as in \eqref{EC}.}
          \label{fig:Fig5}
\end{figure}

Working with $\LBHc$, Figure \ref{fig:Fig5} plots the observed FDR of
SLOPE in the same setup as in Figure \ref{fig:Fig3b}.  For
$n = p =5,000$, the values of the critical point $k^\star$ are $91$
for $q = 0.05$, $141$ for $q = 0.1$, and $279$ for $q = 0.2$. For
$n=2p=10,000$, they become $283$, $560$, and $2,\!976$,
respectively. It can be observed that SLOPE keeps the FDR
at a level close to the nominal level even after passing the critical
point. When $n = p = 5,000$ and $q = 0.05$, which corresponds to the
situation in which the critical point is earliest ($k^\star = 91$),
one can observe a slight increase in FDR above the nominal value when
$k$ ranges from $100$ to $200$.  It is also interesting to observe
that FDR control is more difficult when the coefficients have moderate
amplitudes rather than when they have large ones. Although this effect
is very small, an interpretation is that the correction is not strong
enough to account for the loss of power; that is, for not including a
large fraction of true regressors.

\begin{figure}[h!]
        \centering
        \begin{subfigure}[b]{0.495\textwidth}
                \centering
                \includegraphics[width=\textwidth]{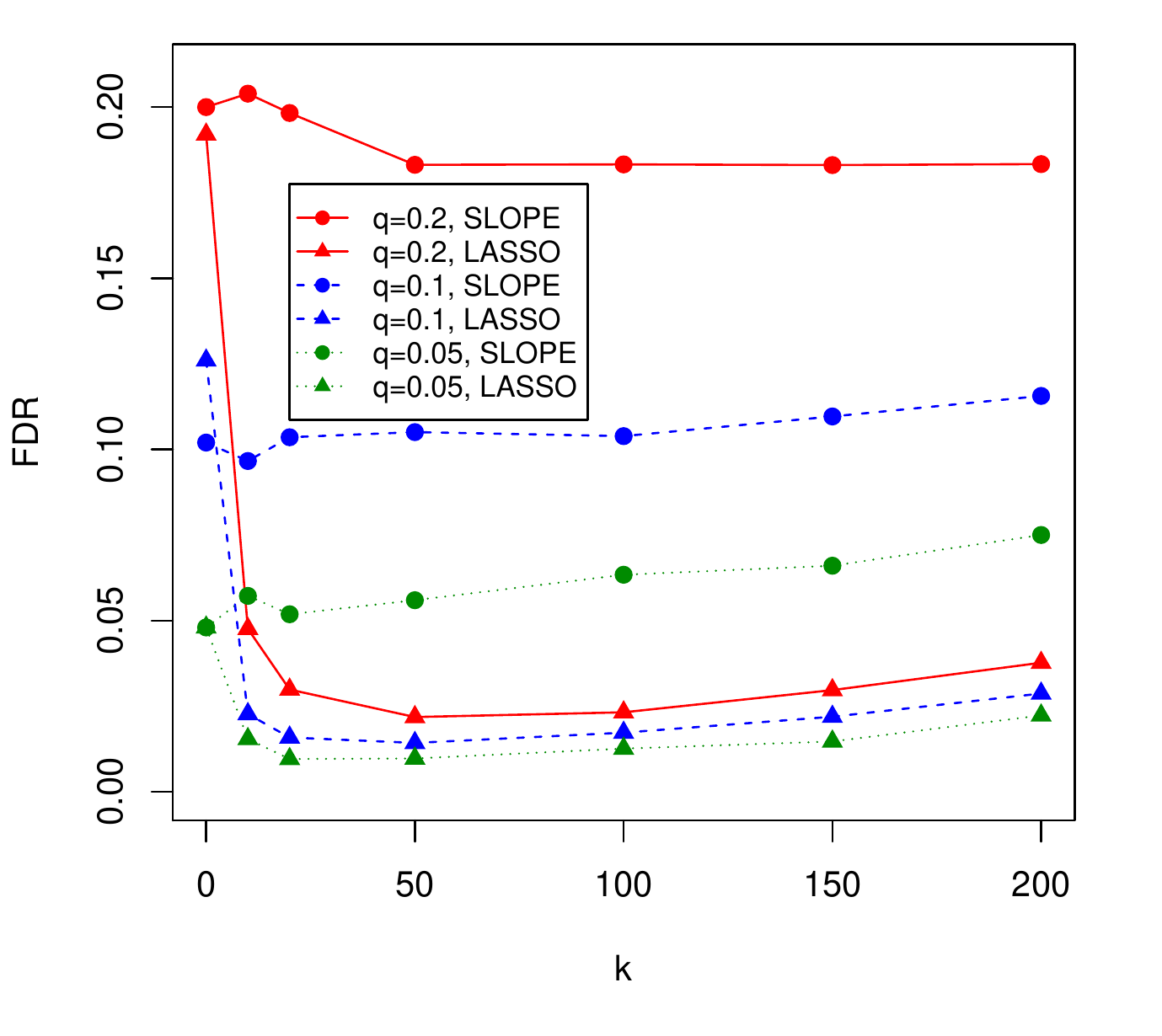}
                \label{fig:Fig6a}
        \end{subfigure}
        \begin{subfigure}[b]{0.495\textwidth}
                \centering
                \includegraphics[width=\textwidth]{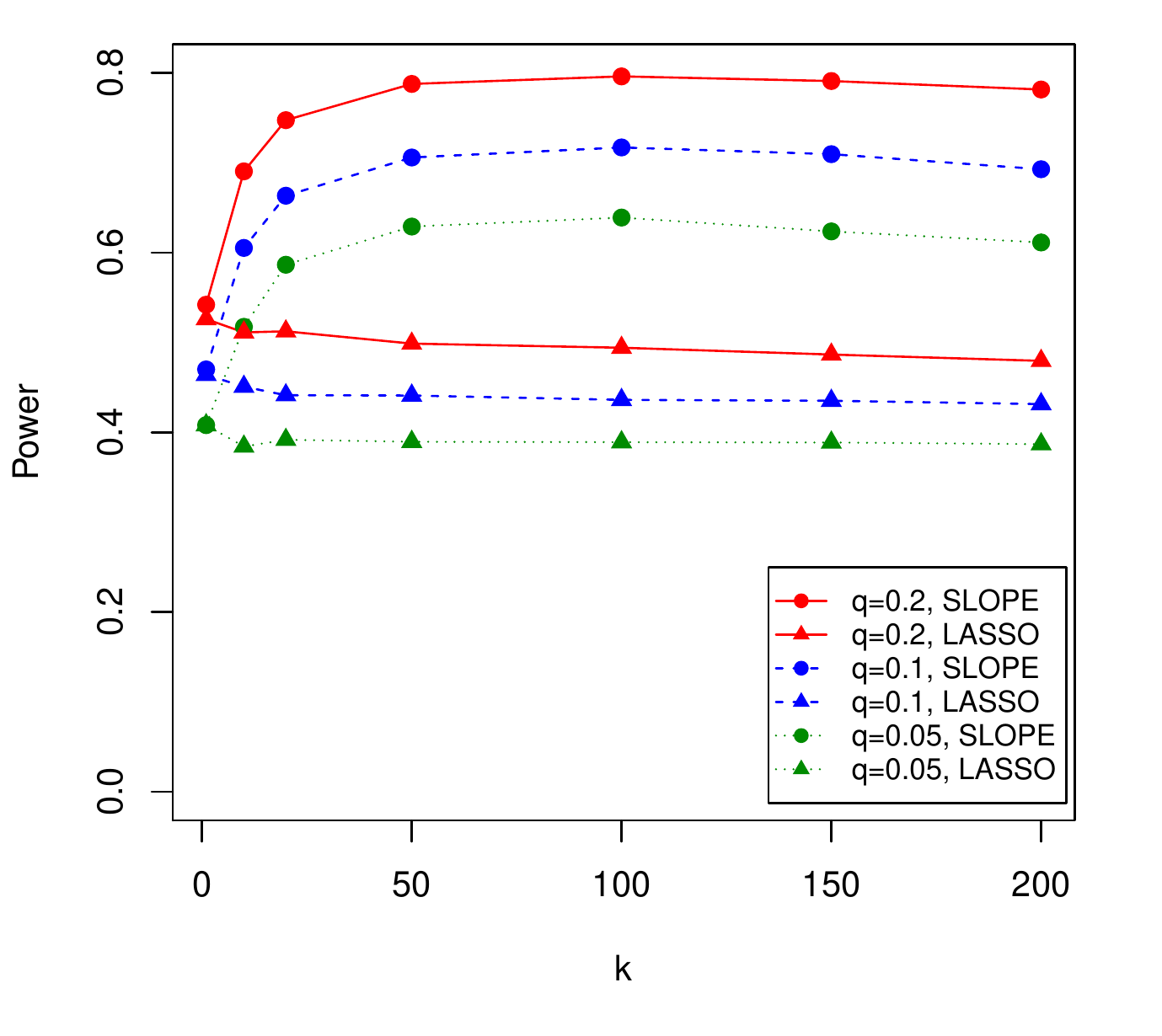}
\label{fig:Fig6b}
        \end{subfigure}
\vspace*{-1cm}
\caption{FDR and power of SLOPE and of the lasso with $\lambda =
  \LBH(1) = \LBHc(1)$; $n=p=5000$, $\beta_i=\sqrt{2\log p}$ for
  $i=1,\ldots,k$.}
          \label{fig:Fig6}
\end{figure}

Figure \ref{fig:Fig6} illustrates the advantage of using an initially
decreasing sequence of thresholds (BHq style) as compared to the
classical lasso with $\lambda =\LBHc(1)$ (Bonferroni style). The
setting of the experiment is the same as in Figure \ref{fig:Fig3b}
with $n=p=5000$ and weak signals $\beta_i=\sqrt{2\log p}$. It can be
observed that under the global null, both procedures work the same and
keep the FDR or FWER at the assumed level. As $k$ increases, however,
the FDR of SLOPE remains constant and close to the nominal level,
while that of the lasso rapidly decreases and starts increasing after
$k = 50$.  The gain for keeping FDR at the assumed level is a
substantial gain in power for small values of $k$. The power of the
lasso remains approximately constant for all $k\in\{1,\ldots, 200\}$,
while the power of SLOPE exhibits a rapid increase at small values of
$k$. The gain is already quite clear for $k=10$, where for $q=0.1$ the
power of SLOPE exceeds 60\%, whereas that of the lasso remains
constant at 45\%. For $k=100$, the power of SLOPE reaches 71\%, while
that of the lasso is still at 45\%.

%% file: numerical.tex
\section{Numerical Examples}
\label{sec:numerical}

\subsection{Multiple testing}
\label{sec:testing}

\subsubsection{Genome wide association studies}

\begin{figure}[h!]
  \centering
        \begin{subfigure}[b]{0.495\textwidth}
                \centering
                \includegraphics[width=\textwidth]{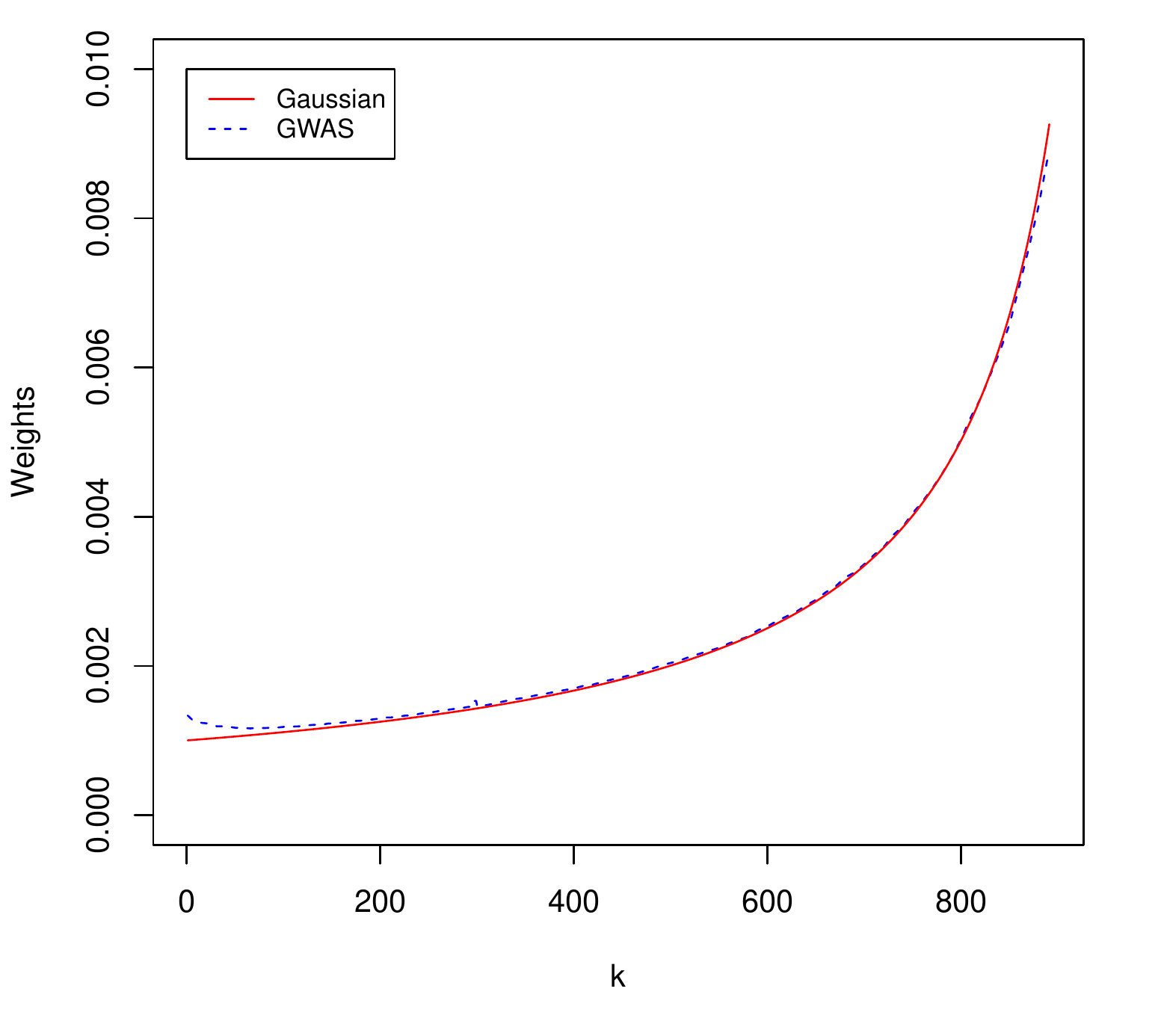}
                \subcaption{Weights $w_k$ as a function of $k$.}
               \label{fig:weights}
        \end{subfigure}
        \begin{subfigure}[b]{0.495\textwidth}
                \centering
                \includegraphics[width=\textwidth]{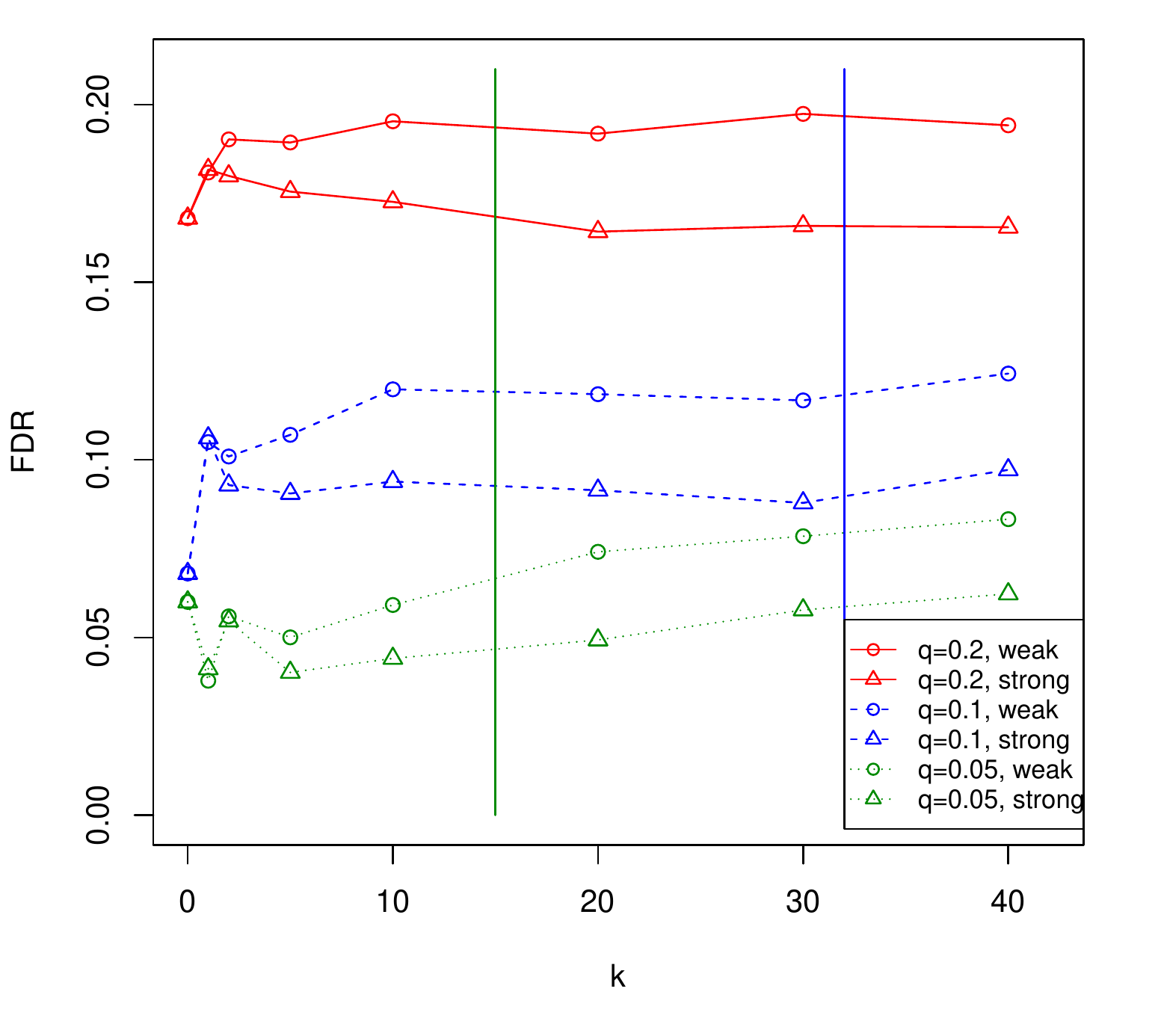}
                \subcaption{FDR as a function of $k$.}
                \label{fig:fdr_GWAS}
        \end{subfigure}
  \begin{subfigure}[b]{0.495\textwidth}
                \centering
                \includegraphics[width=\textwidth]{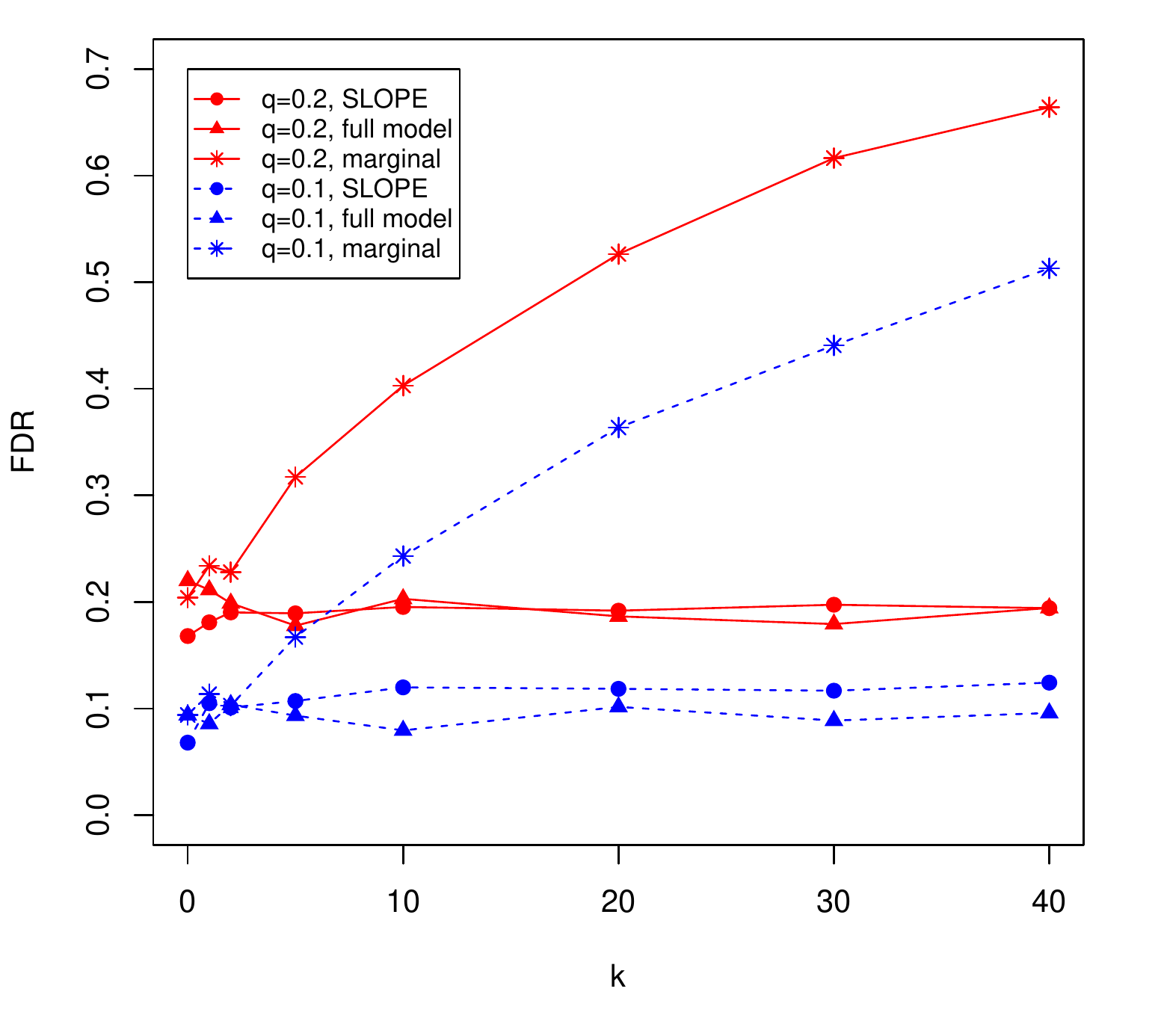}
                \subcaption{FDR as a function of $k$.}
                \label{fig:fdr_GWAS_comp}
         \end{subfigure}
        \begin{subfigure}[b]{0.495\textwidth}
                \centering
                \includegraphics[width=\textwidth]{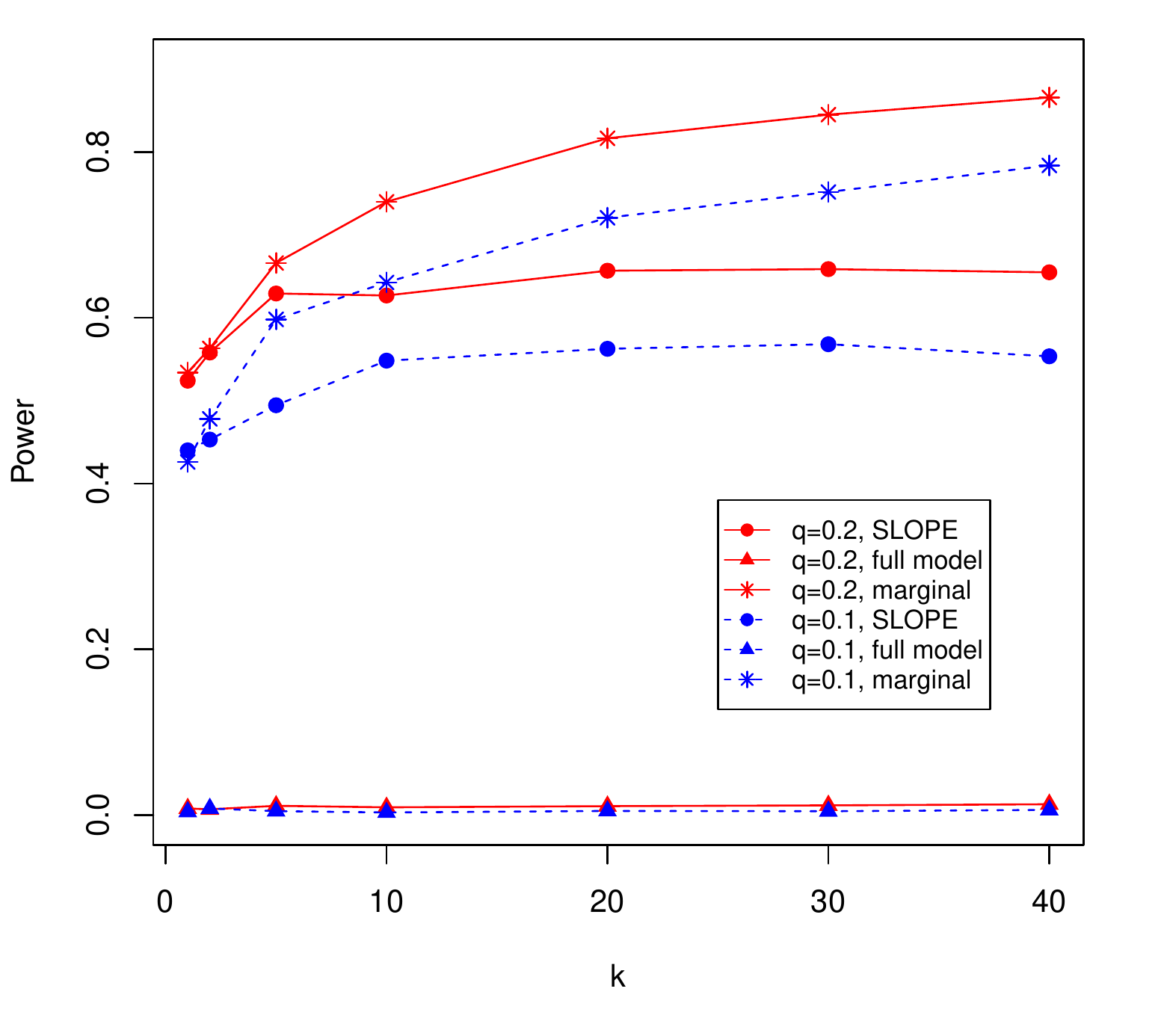}
                \subcaption{Power as a function of $k$.}
                \label{fig:power_GWAS_comp}
        \end{subfigure} 
        \caption{Simulation results for the GWAS inspired
          examples. The weights for SLOPE are shown in (a).  In (b),
          the vertical lines represent the locations of the critical
          points for $q = 0.05$ and $q = 0.1$. As mentioned in the
          text weak (resp.~strong) signals have amplitudes equal to
          $\sqrt{2\log p}$ (resp.~$5\sqrt{2\log p}$). Plots (c) and
          (d) show the FDR and power of the different methods
          operating at the nominal level $q = 0.1$ in the case of weak
          signals.}
\label{fig:GWAS}
\end{figure}

In this section we report the results of simulations inspired by the
genome-wide search for influential genes. In this case the regressor
variables can take on only three values according to the genotype of a
genetic marker. In details, $X=0$ if a given individual has two copies
of a reference allele at a given marker, $X=2$ if she has two copies
of a variant allele, and $X=1$ if this individual is heterozygous,
i.e.,~has one copy of a reference and one copy of a variant allele. For
this study we used simulated data relating to 1,000 individuals from
the admixture of the African-American (ASW) and European (CEU)
populations, based on the HapMap \cite{Hap} genotype data. The details
of the simulation of the admixture are described in \cite{BFST}. The
original data set contains 482,298 markers (locations on the genome)
in which genotypes at neighboring markers are usually strongly
correlated. To avoid ambiguities related to the definition of the true
and false positives we extensively pruned the design matrix, leaving
only 892 markers distributed over all chromosomes. In this final
data set the maximal pairwise correlation between those genotypes at
different marker locations is equal to 0.2. The columns of the design
matrix are further standardized, so that each variable has zero mean
and variance equal to one (in other words, the $\ell_2$ norm of each
column is equal to one).  The design matrix used in our simulation
study is available at
\url{http://www-stat.stanford.edu/~candes/SortedL1/Gen.txt}. Following
the arguments from Section \ref{sec:lambda}, the weights $w_k$ for
$\LBHc$ are 
\begin{equation}\label{Eq:ExpWk}
w_k = \frac{1}{k}\mathbb{E}\Vert (X_{\mathcal{S}}'X_\mathcal{S})^{-1}X'_\mathcal{S}X_i\Vert_2^2,
\end{equation}
where the expectation is taken over all subsets $\mathcal{S} \subseteq
\{1, \ldots, p\}$ with $\vert\mathcal{S}\vert = k$ and $i \notin
\mathcal{S}$. Here, we substitute this expectation with an average
over 5000 random samples when $k<300$ and 2500 random samples for
larger $k$. The weights are estimated for $k \in \{1,6, 11,
\ldots,296\}$ and $k\in\{300, 310, 320, \ldots,890\}$ and we use
linear interpolation to estimate the remaining values. As seen in
Figure \ref{fig:weights}, for small $k$ the estimated weights are
slightly larger than the corresponding weights for the Gaussian design
matrix. In the most relevant region where $k<50$ they slightly
decrease with $k$.  When using this selection of weights the critical
points $k^\star$ for $q \in \{0.05,0.1,0.2\}$ are equal to $15$, $32$
and $67$.

In our study, we compare the performance of SLOPE with two other
approaches that are sometimes used to control the false discovery
rate. The first approach is often used in real Genome Wide Association
Studies (GWAS) and operates as follows: first, carry out simple
regression (or marginal correlation) tests at each marker and then use
the Benjamini-Hochberg procedure to adjust for multiplicity (and
hopefully control the overall FDR). The second approach uses the
Benjamini-Hochberg procedure with $p$-values obtained by testing the
significance of individual predictors within the full regression model
with 892 regressors.  We assume Gaussian errors and according to
Theorem 1.3 in \cite{BY}, applying the BHq procedure with $q=
{q_0}/S_p$, $S_p = \sum_{i=1}^{p} {1}/{i}$, controls the FDR at level
$q_0$. In our context, the logarithmic factor makes the procedure too
conservative and in our comparison, we shall use the BHq procedure
with $q$ being equal to the FDR level we aim to obtain. The reader
will observe that in our experiments, this choice keeps the FDR around
or below the nominal level $q$ (had we opted for $q/S_p$, we would
have had an FDR well below the nominal level and essentially no
power).

The trait or response is simulated according to the linear model
\eqref{eq:linear}, where $z$ is a 1,000-dimensional vector with
i.i.d.~$\mathcal{N}(0,1)$ entries. The number $k$ of nonzero elements
in the regression coefficient vector $\beta$ varies between 0 and
40. For each $k$, we report the values of the FDR and power by
averaging false discovery proportions over 500 replicates. In each
replicate, we generate a new noise vector and a regression coefficient
vector at the appropriate sparsity level by selecting locations of the
nonzero coefficients uniformly at random. The amplitudes are all
equal to $5\sqrt{2\log p}$ (strong signals) or $\sqrt{2\log p}$ (weak
signals).

As observed in Figure \ref{fig:fdr_GWAS}, when the signals are strong
SLOPE keeps the FDR close to or below the nominal level whenever the
number of non-nulls is less or equal to the value $k(n,p,q)$ of the
critical point.  As in the case of Gaussian matrices, FDR control is
more difficult when signals have a moderate amplitude; even in this
case, however, the FDR is kept close to the nominal level as long as
$k$ is below the critical point.

Figures \ref{fig:fdr_GWAS_comp} and \ref{fig:power_GWAS_comp} compare
the FDR and the power of the three different approaches when effects
are weak ($\beta=\sqrt{2\log p}$).  In Figure \ref{fig:fdr_GWAS_comp}
we observe that the method based on marginal tests completely fails to
control FDR when the number of true regressors exceeds 10. For $k=10$
this method has a FDR two times larger than the nominal level. For
$k=40$, the FDR exceeds 50\% meaning that the `marginal method'
carried out at the nominal FDR level $q=0.1$ detects more false than
true regressors. This behavior can be easily explained by observing
that univariate models are not exact: in a univariate least-squares
model, we estimate the projection of the full model on a simpler model
with one regressor. Even a small sample correlation between the true
regressor $X_1$ and some other variable $X_2$ will make the expected
value of the response variable dependent on $X_2$.  In passing, this
phenomenon illustrates a basic problem related to the application of
marginal tests in the context of GWAS; marginal tests yield far too
many false regressors even in situations where correlations between
the columns of the design matrix are rather small.  The method based
on the tests within the full regression model behaves in a completely
different way. It controls the FDR at the assumed level but has no
detection power. This can be explained by observing that least squares
estimates of regression coefficients have a large standard deviation
when $n$ is comparable to $p$. In comparison, SLOPE
performs very well. When $k\leq 40$, the FDR is kept close to the
nominal level and the method has a large power, especially when one
considers that the magnitude of the simulated signals are comparable
to the upper quantiles of the noise distribution.

\subsubsection{Multiple mean testing from correlated statistics}

We now illustrate the properties of our method as applied to a
classical multiple testing problem with correlated test
statistics. Imagine that we perform $p = 1,000$ tests in each of 5
different laboratories and that there is a significant random
laboratory effect.  The test statistics can be modeled as
\[
y_{i,j} = \mu_i + \tau_j + z_{i,j}, \quad 1 \le i \le 1000, \,\, 1 \le j
\le 5,
\]
where the laboratory effects $\tau_j$ are i.i.d.~mean-zero Gaussian
random variables as are the errors $z_{i,j}$. The random lab effects
are independent from the $z$'s.  We wish to test whether $H_j: \mu_j =
0$ versus a two-sided alternative. In order to do this, imagine
averaging the scores over all five labs, which gives
\[
\bar{y}_i = \mu_i + \bar{\tau} + \bar{z}_i, \quad 1 \le i \le n.
\]
We assume that things are normalized so that
$\operatorname{Var}(\bar{y}_i) = 1$, the vector $\bar{y} \sim
\mathcal{N}(0,\Sigma)$ where $\Sigma_{i,i} = 1$ and $\Sigma_{i,j} =
\rho > 0$ for $i \neq j$. Below, we shall work with $\rho = 0.5$.

Our problem is to test the means of a multivariate Gaussian vector
with equicorrelated entries.  One possible approach is to use the
Benjamini-Hochberg procedure, that is we order $|\bar{y}|_{(1)} \ge
|\bar{y}|_{(2)} \ge \ldots \ge |\bar{y}|_{(p)}$ and apply the step-up
procedure with critical values equal to $\Phi^{-1}(1-iq/2p)$. Because
the statistics are correlated we would use a conservative level equal
to $q/S_p$ as suggested in \cite[Theorem 1.3]{BY}. However, this is
really too conservative and, therefore, in our comparisons we selected
$q$ to be the FDR level we wish to obtain. (The reader will again
observe that in our experiments, this choice keeps the FDR below the
nominal level $q$. Had we opted for $q/S_p$, we would have had an FDR
well below the nominal level and essentially no power).

Another possible approach is to `whiten the noise' and express our
multiple testing problem in the form of a regression equation
\begin{equation}\label{test_reg}
  \tilde y =\Sigma^{-1/2} \bar y = \Sigma^{-1/2} \mu + \epsilon,
\end{equation}
where $\epsilon \sim \mathcal{N}(0,I_p)$, and use model selection
tools for identifying the nonzero elements of $\mu$.

Interestingly, while the matrix $\Sigma$ is far from being diagonal,
$\Sigma^{-1/2}$ is diagonally dominant. Specifically,
$\Sigma^{-1/2}_{i,i}=1.4128$ and $\Sigma^{-1/2}_{i,j}=-0.0014$ for $i
\neq j$.  This makes our multiple testing problem well suited for the
application of SLOPE with the original
$\lambda_{\text{BH}}$ values. Hence, we work with \eqref{test_reg} in
which we normalize $\Sigma^{-1/2}$ as to have unit norm columns.

For the purpose of this study we simulate a sequence of sparse
multiple testing problems, with the number $k$ of nonzero $\mu_i$
varying between 0 and 80. With $\Sigma^{-1/2}$ normalized as above,
all the nonzero means are equal to $\sqrt{2 \log p}$ so that $\tilde y
\sim \mathcal{N}(\tilde \Sigma^{-1/2} \mu, I_p)$ in which $\tilde
\Sigma^{-1/2}$ has unit normed columns. The magnitude of true effects
was chosen so as to obtain a moderate power of their detection.  For
each sparsity level $k$, the results are averaged over 500 independent
replicates.

\begin{figure}[h!]
  \centering
        \begin{subfigure}[b]{0.495\textwidth}
                \centering
                \includegraphics[width=\textwidth]{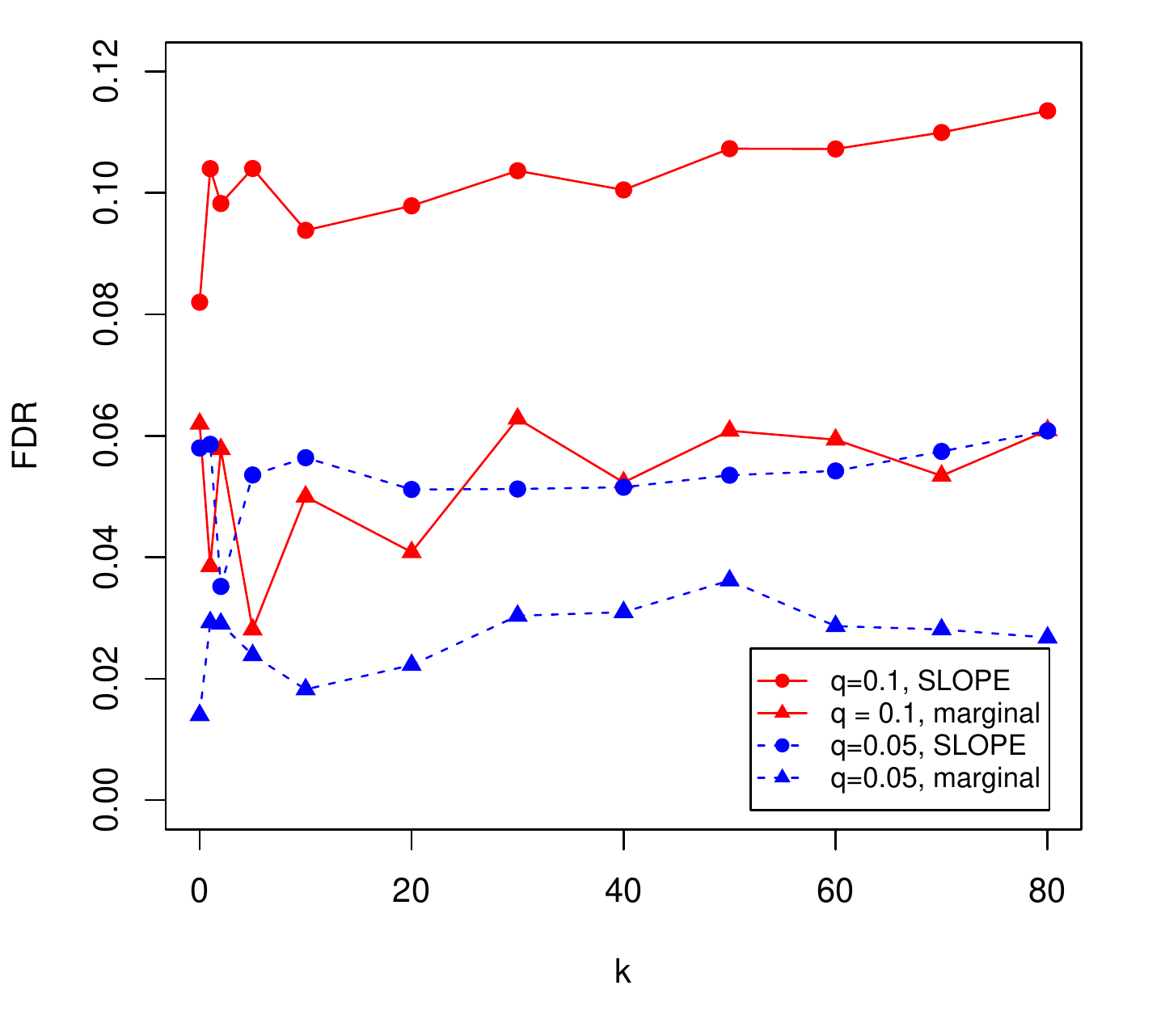}
        \end{subfigure}
        \begin{subfigure}[b]{0.495\textwidth}
                \centering
                \includegraphics[width=\textwidth]{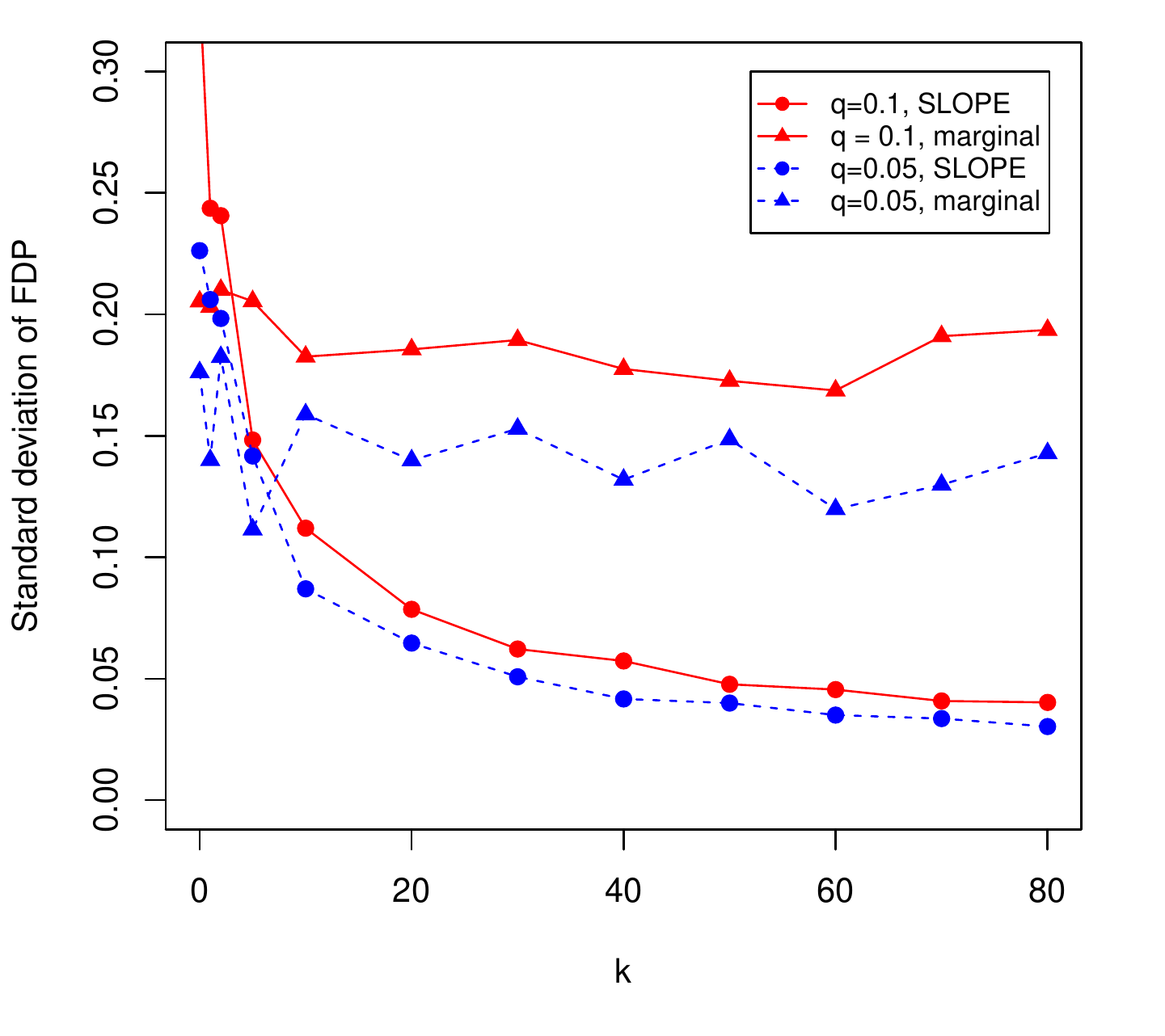}
        \end{subfigure}
  \begin{subfigure}[b]{0.495\textwidth}
                \centering
                \includegraphics[width=\textwidth]{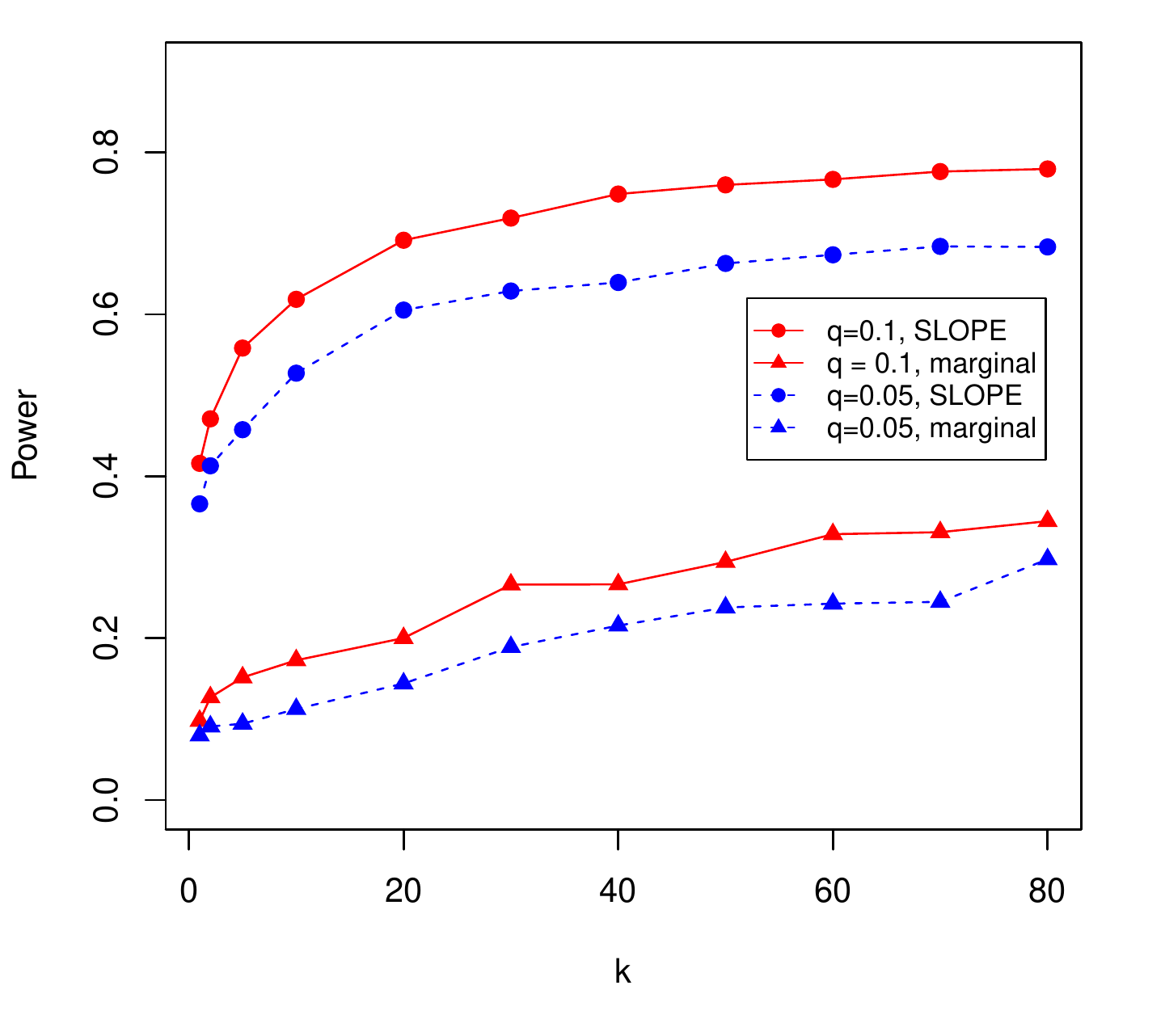}
        \end{subfigure}
        \caption{Simulation results for testing multiple means from
          correlated statistics. (a) FDR levels for SLOPE
          and marginal tests as a function of $k$. (b) Variability of
          the false discovery proportions for both methods. (c) Power
          plot.}
        \label{fig:anova1}
\end{figure}

\begin{figure}[h!]
  \centering
        \begin{subfigure}[b]{0.495\textwidth}
                \centering
                \includegraphics[width=\textwidth]{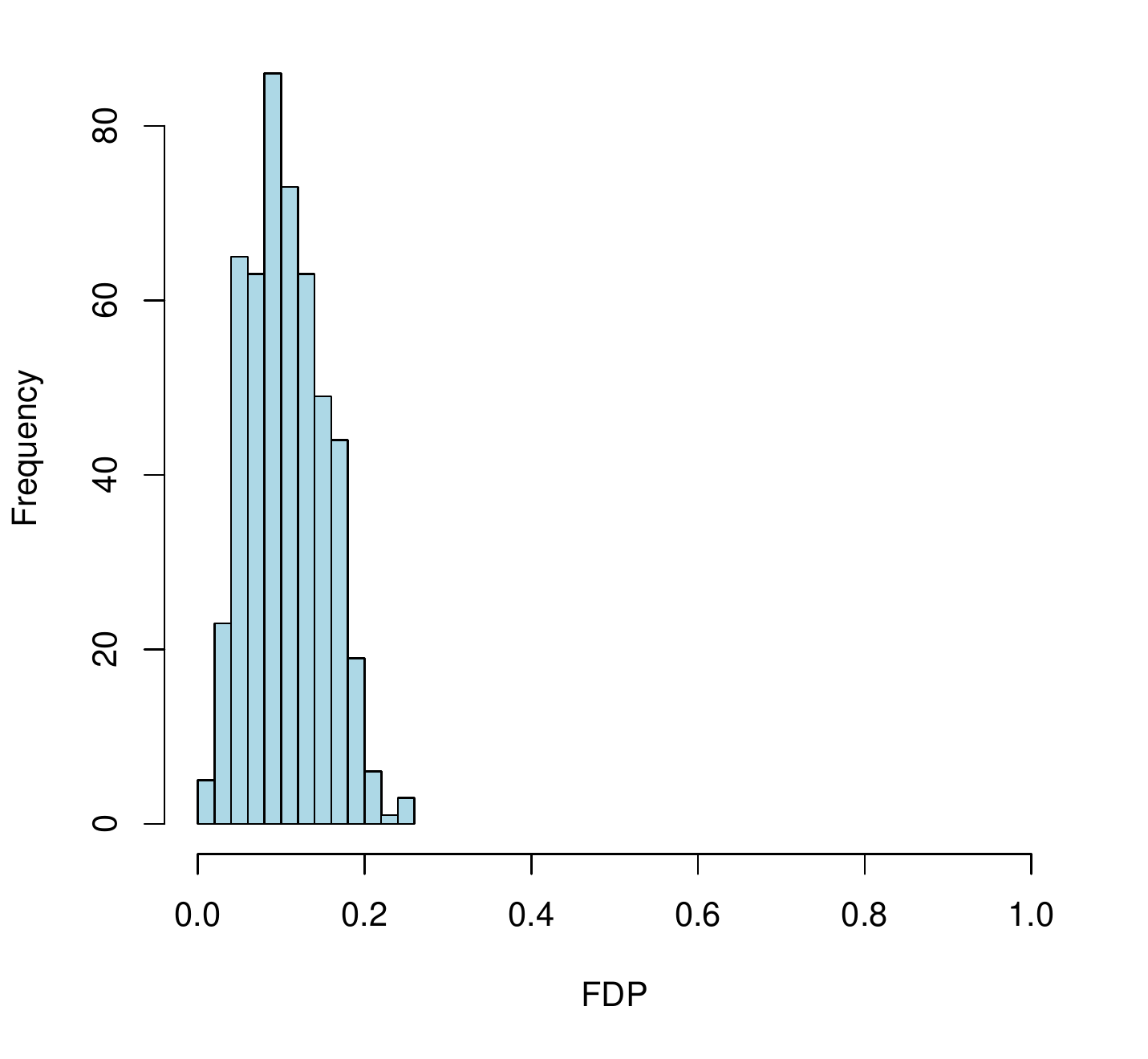}
                \subcaption{Observed FDP for SLOPE.}
                \label{fig:Fig9a}
        \end{subfigure}
        \begin{subfigure}[b]{0.495\textwidth}
                \centering
                \includegraphics[width=\textwidth]{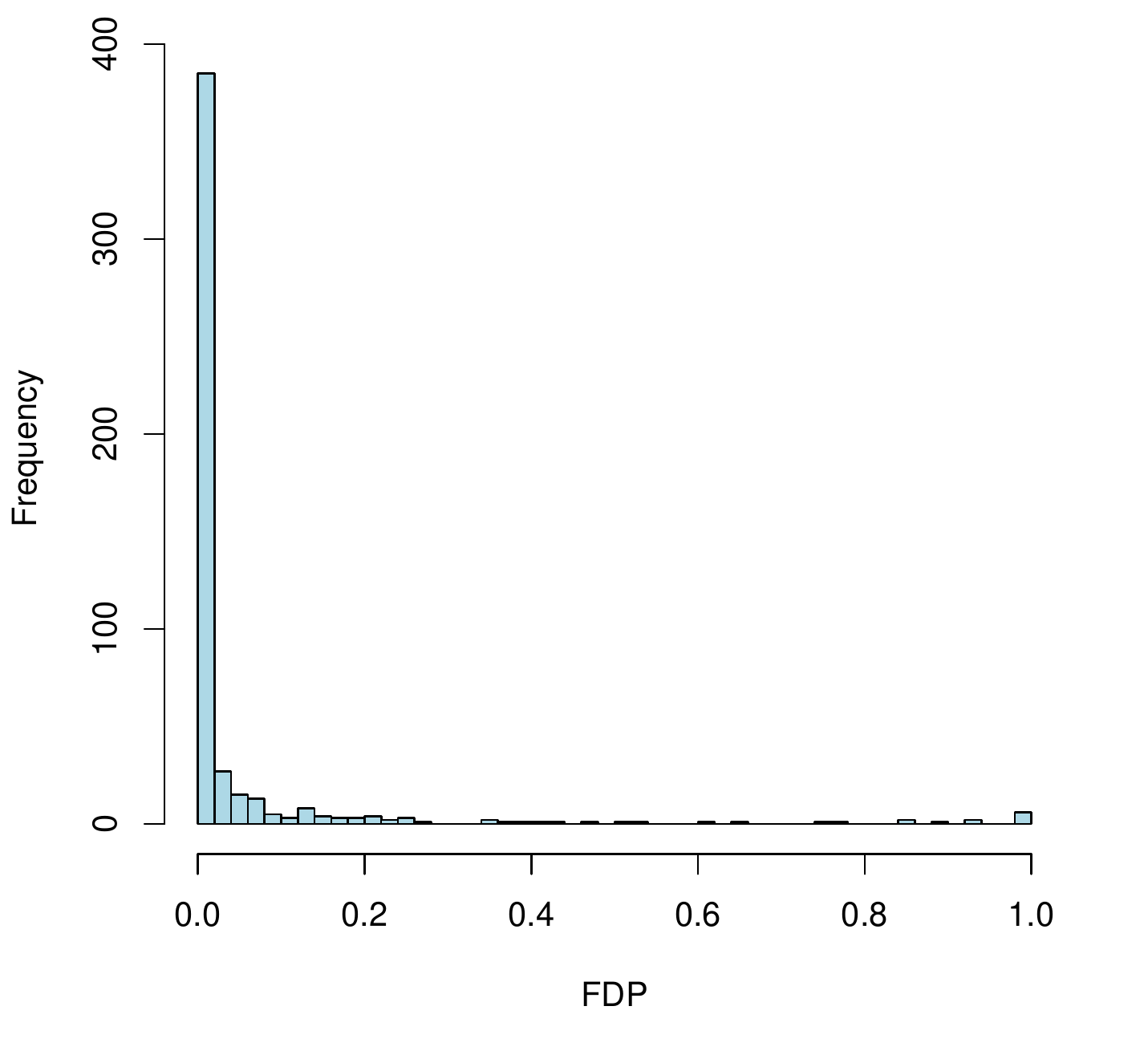}
                \subcaption{Observed FDP for marginal test.}
                \label{fig:Fig9b}
        \end{subfigure}
  \begin{subfigure}[b]{0.495\textwidth}
                \centering
                \includegraphics[width=\textwidth]{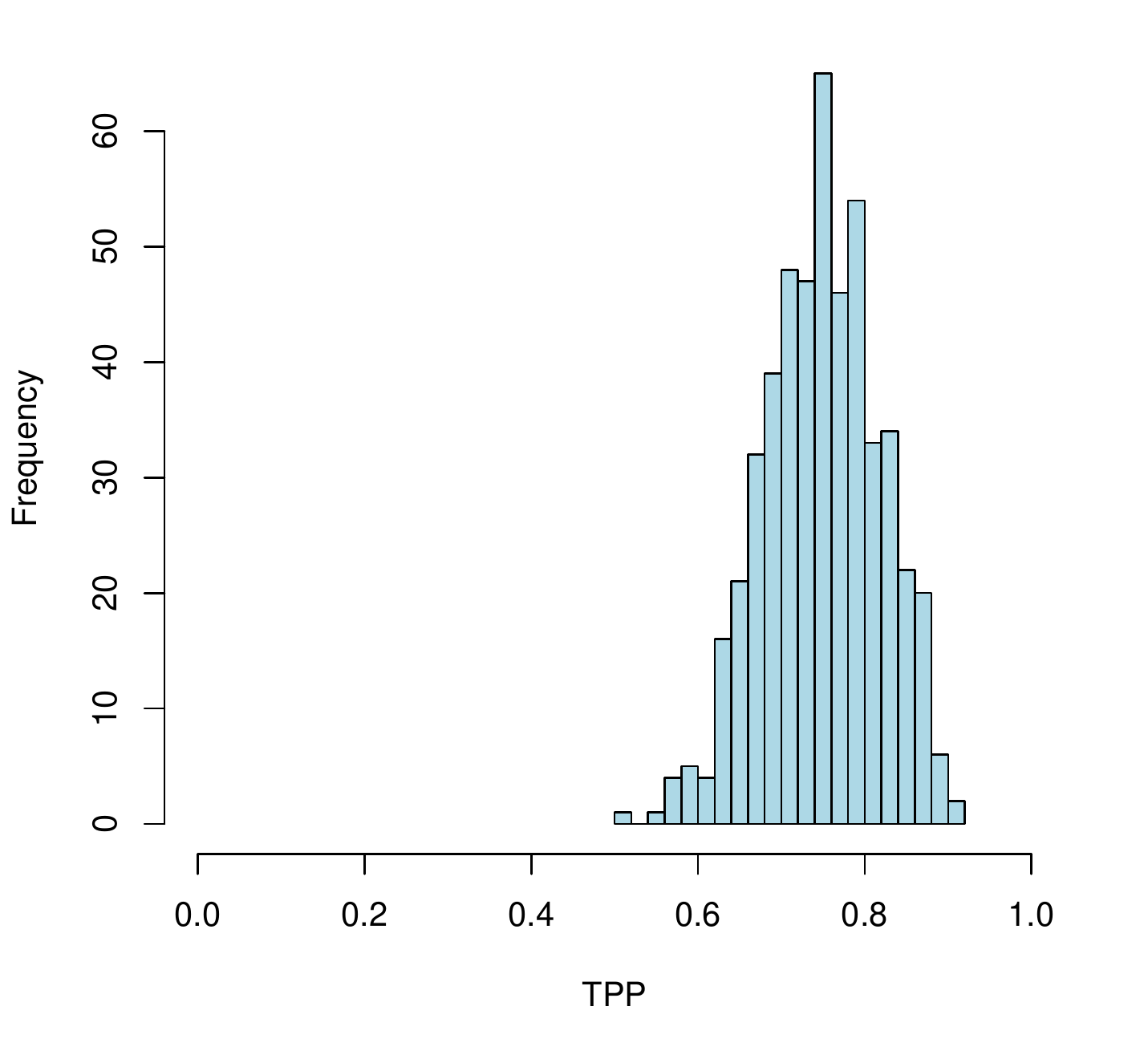}
                \subcaption{Observed TPP for SLOPE.}
                \label{fig:Fig9c}
        \end{subfigure}
\begin{subfigure}[b]{0.495\textwidth}
                \centering
                \includegraphics[width=\textwidth]{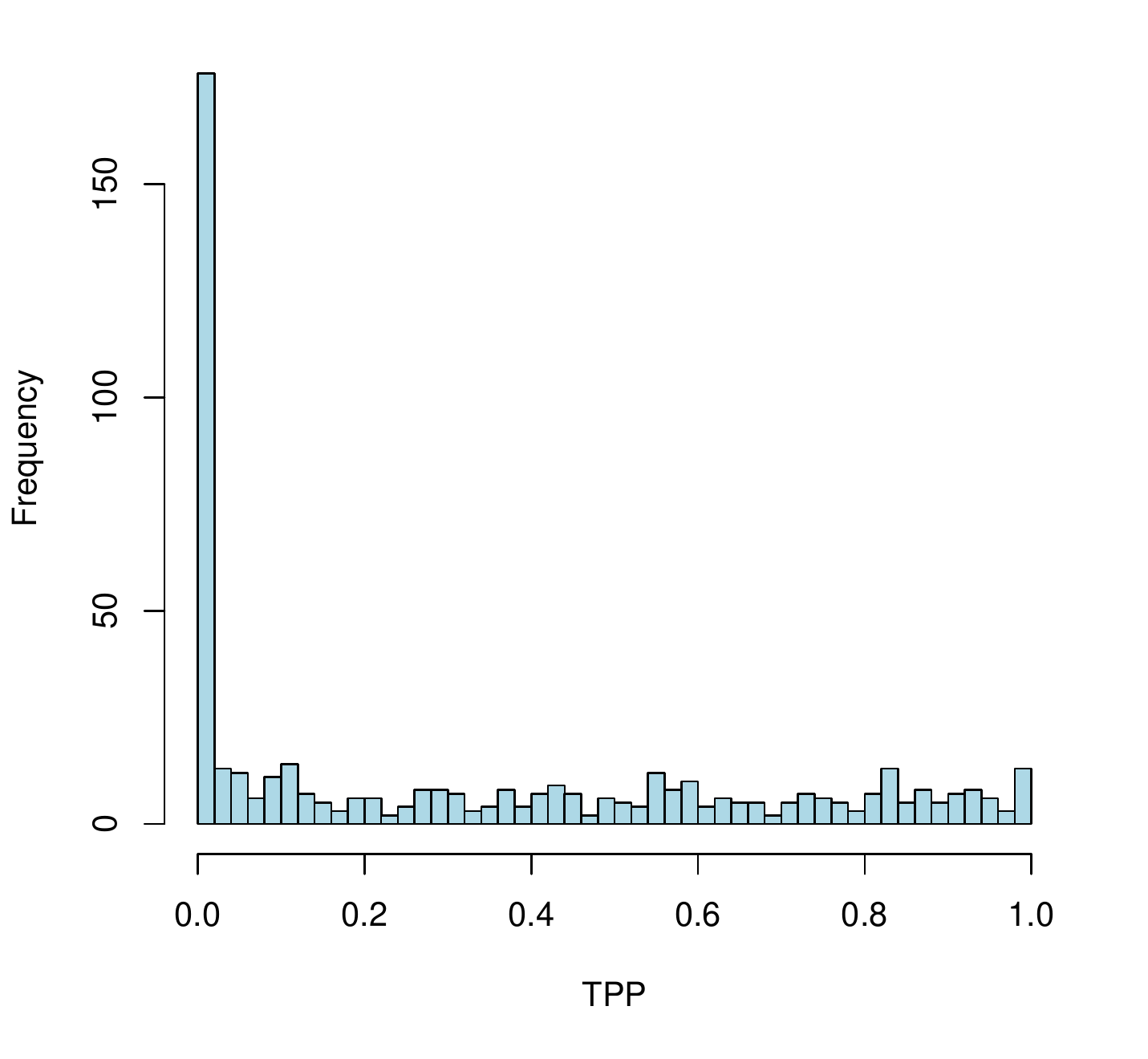}
                \subcaption{Observed TPP for marginal test.}
                \label{fig:Fig9d}
        \end{subfigure}
        \caption{Testing example with $q = 0.1$ and $k = 50$.}
        \label{fig:anova2}
\end{figure}

In Figures \ref{fig:anova1} and \ref{fig:anova2} we present the
summary of our simulation study, which reveals that SLOPE solves the
multiple testing problem in a much better fashion than the BHq
procedure applied to marginal test statistics. In our setting, the BHq
procedure is too conservative as it keeps the FDR below the nominal
level. Moreover, as observed in Figure \ref{fig:anova2}, when $q=0.1$
and $k=50$, in approximately 75\% of the cases the observed False
Dicovery Proportion (FDP) is equal to 0, while in the remaining 25\%
of the cases, it takes values which are distributed over the whole
interval (0,1). FDP taking the value zero in a majority of the cases
is not welcome since it is related to a low power. To be sure, in
approximately 35\% of all cases BHq did not make any rejections
(i.e.,~$R=0$). Conditional on $R>0$, the mean of FDP is equal to 0.22
with a standard deviation of 0.28, which clearly shows that the
observed FDP is typically far away from the nominal value of $q=0.1$.
Because of this behavior, the applicability of the BHq procedure under
this correlation structure is in question since for any given
realization, the FDP may be very different from its targeted expected
value. In comparison to BHq, SLOPE offers a more predictable FDR and
substantially larger and more predictable True Positive Proportion
(TPP, fraction of correctly identified true signals), compare the
spread of the histograms in Figure \ref{fig:anova2}.

\subsection{High-dimensional compressive sensing examples}
\label{sec:estimation}

The experiments presented so far all used an $n\times p$ matrix $X$
with $n \approx p$. In this section we consider matrices with $n \ll
p$. Such matrices are used in compressive sensing
\cite{CAN2006RTa,CAN2006Ta,DON2006c} to acquire linear measurements of
signals $\beta$ that are either sparse or can be sparsely represented
in a suitable basis. In particular, we observe $y = X\beta + z$, where
$z$ is an additive noise term. We assume that $z \sim
\mathcal{N}(0,\sigma^2I)$, and that $\sigma$ is known.  Despite the
underdetermined nature of the problem, there is now a large volume of
theoretical work in the field of compressed sensing that shows that
accurate reconstructions of $\beta$ can nevertheless be obtained using
$\ell_1$-based algorithms such as the lasso.


The goal of our simulations is to compare the performance of the lasso
and SLOPE in this setting, and to evaluate their sensitivity with
respect to the regularizing parameters $\lambda$ and $q$,
respectively.

\subsubsection{Sparse signals}\label{Sec:SparseSignals}

As a first experiment we generate an $n\times p$ matrix $X$ by taking
the $p\times p$ matrix corresponding to the one-dimensional discrete
cosine transformation (DCT-II)\footnote{Please see \cite{Mallat} for a
  formal definition of DCT-II.} for signals of length $p$, and
randomly selecting $n$ different rows in such a way that the first row
is always included. We then scale the matrix by a factor $\sqrt{p/n}$
to obtain approximately unit-norm columns. Note that including the
first row of the DCT matrix, all of whose entries are identical,
ensures that we have information about the mean of the
signal. Throughout this section we use a fixed instance of $X$ with $n
= p/2$ and $p = 262,\!144$.

The signals $\beta$ are constructed as follows. First a random support
set $\mathcal{S} \subset \{1, \ldots, p\}$ of cardinality $k$ is drawn
uniformly at random. We then set the off-support entries to zero,
i.e., $\beta_i = 0$ for all $i \not\in\mathcal{S}$, and choose the
remaining $k$ entries $\beta_i$ with $i \in \mathcal{S}$ according to
one of six signal classes:
\begin{enumerate}
\item Random Gaussian entries: $\beta_i \sim \mathcal{N}(0,\sigma^2)$ with
$\sigma = 2\sqrt{2\log p}$.
\item Random Gaussian entries: $\beta_i \sim \mathcal{N}(0,\sigma^2)$ with
  $\sigma = 3\sqrt{2\log p}$.
\item Constant values: $\beta_i = 1.2\sqrt{2\log{p}}$.
\item Linearly decreasing from $1.2\sqrt{2\log p}$ to $0.6\sqrt{2\log p}$.
\item Linearly decreasing from $1.5\sqrt{2\log p}$ to $0.5\sqrt{2\log p}$.
\item Linearly decreasing from $4.5\sqrt{2\log p}$ to $1.5\sqrt{2\log p}$.
\end{enumerate}
In the last three classes, the linear ranges are randomly permuted
before assigning their values to the respective entries in $\beta$. As
a final signal class, $\beta$ is chosen to be a random permutation of
the following dense vector $v$:
\begin{enumerate}
\setcounter{enumi}{6}
\item Exponentially decaying entries: $v_i = 1.2\sqrt{2\log{p}} \left(i/k\right)^{-1.2}$. 
\end{enumerate}
We consider nine sparsity levels $k$: 1, 5, 10, 50, 100, 500, 1000,
1\%, and 5\%, where the percentages are relative to $p$ (thus giving
$k=2,621$ and $k=13,017$, respectively for the last two choices).  The
noise vector $z$ is always generated from the multivariate normal
$\mathcal{N}(0,1)$ distribution.

The $k$ largest entries in $\beta$ are designed to lie around the
critical level with respect to the noise. Especially for highly sparse
$\beta$, this means that the power and FDR obtained with any of the
methods under consideration depends strongly on the exact combination
of $\beta$ and $z$. To avoid large fluctuations in individual results,
we therefore report the result as averaged over 100 random signal and
noise instances for $k \leq 50$, and over 50 instances for $k=100$ and
$k=500$. For larger $k$ individual instances were found to be sufficient.

The mean square error of the estimated regression coefficients,
$\hat\beta$ is defined as
\[
\mathrm{MSE} = \frac{1}{p}\sum_{i=1}^p \E (\hat\beta_i - \beta_i)^2.  
\]
%
In our numerical experiments we work with estimated MSE values in
which the expectation is replaced by the mean over a number of
realizations of $\beta$ and the corresponding $\hat{\beta}$.  For
convenience we refer to this as the MSE, but it should be kept in mind
that this is only an estimation of the actual MSE.

\begin{figure}[htb]
\centering
\begin{tabular}{cc}
\includegraphics[width=0.42\textwidth]{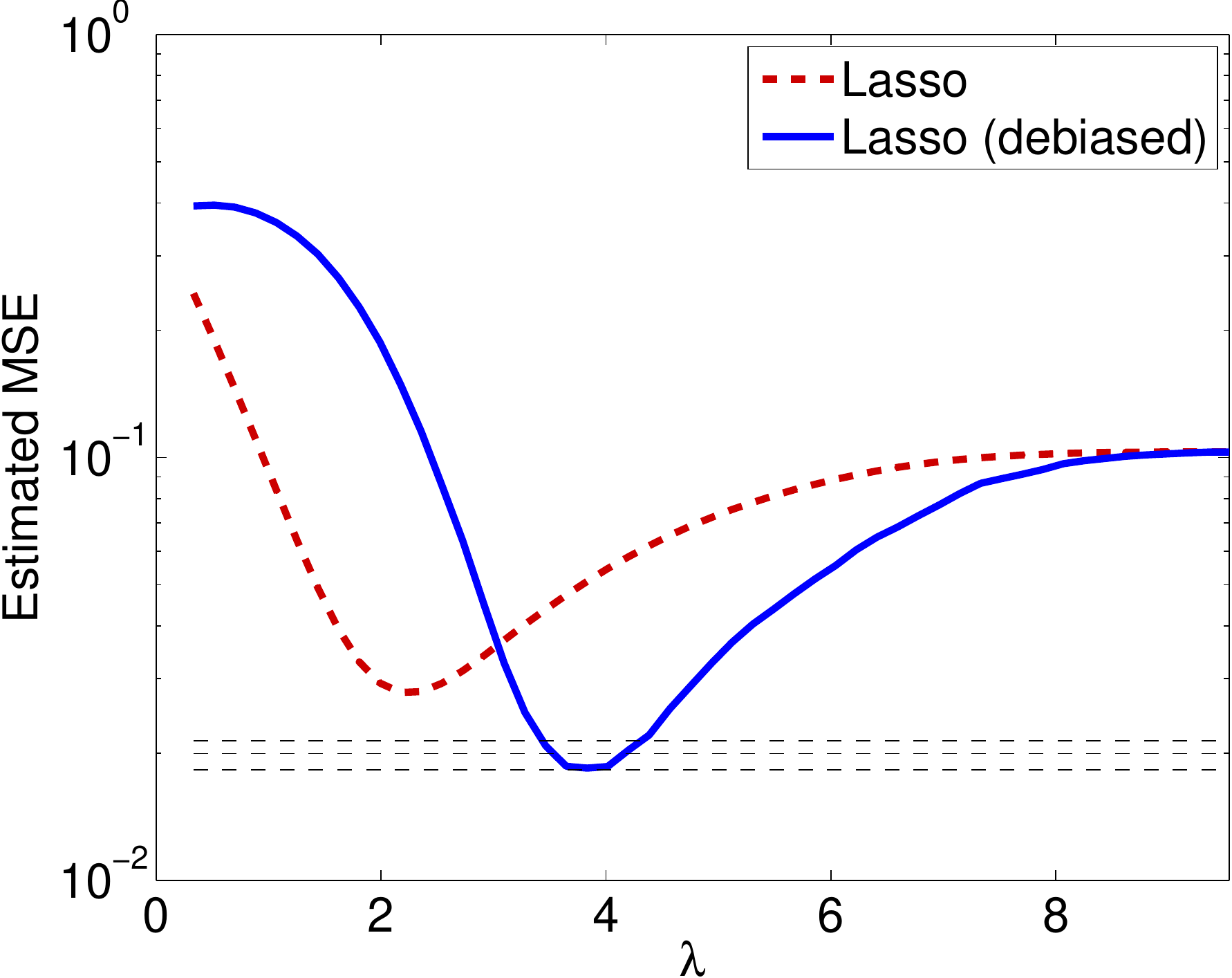} &
\includegraphics[width=0.42\textwidth]{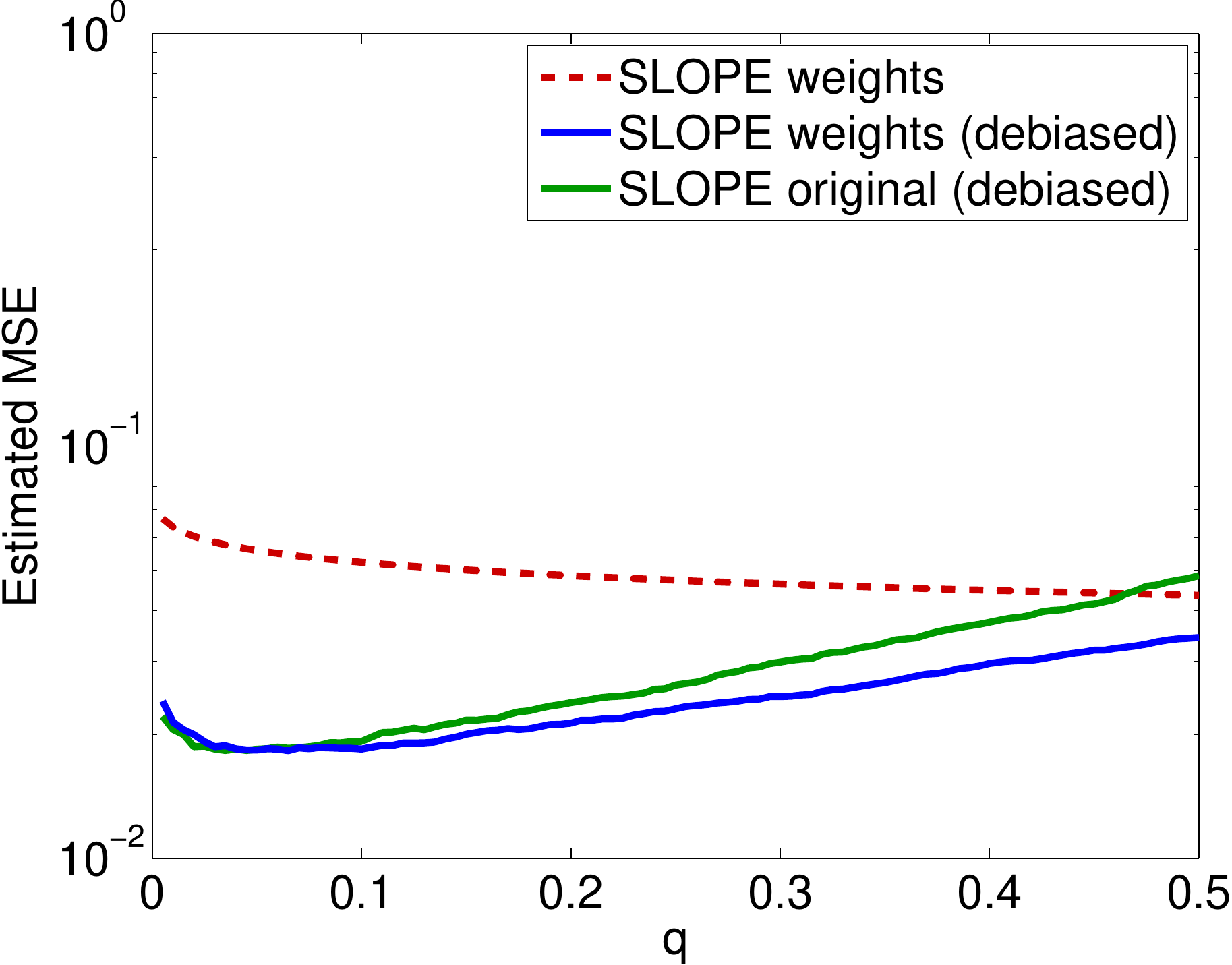} \\
({\bf{a}}) & ({\bf{b}}) \\[3pt]
\includegraphics[width=0.42\textwidth]{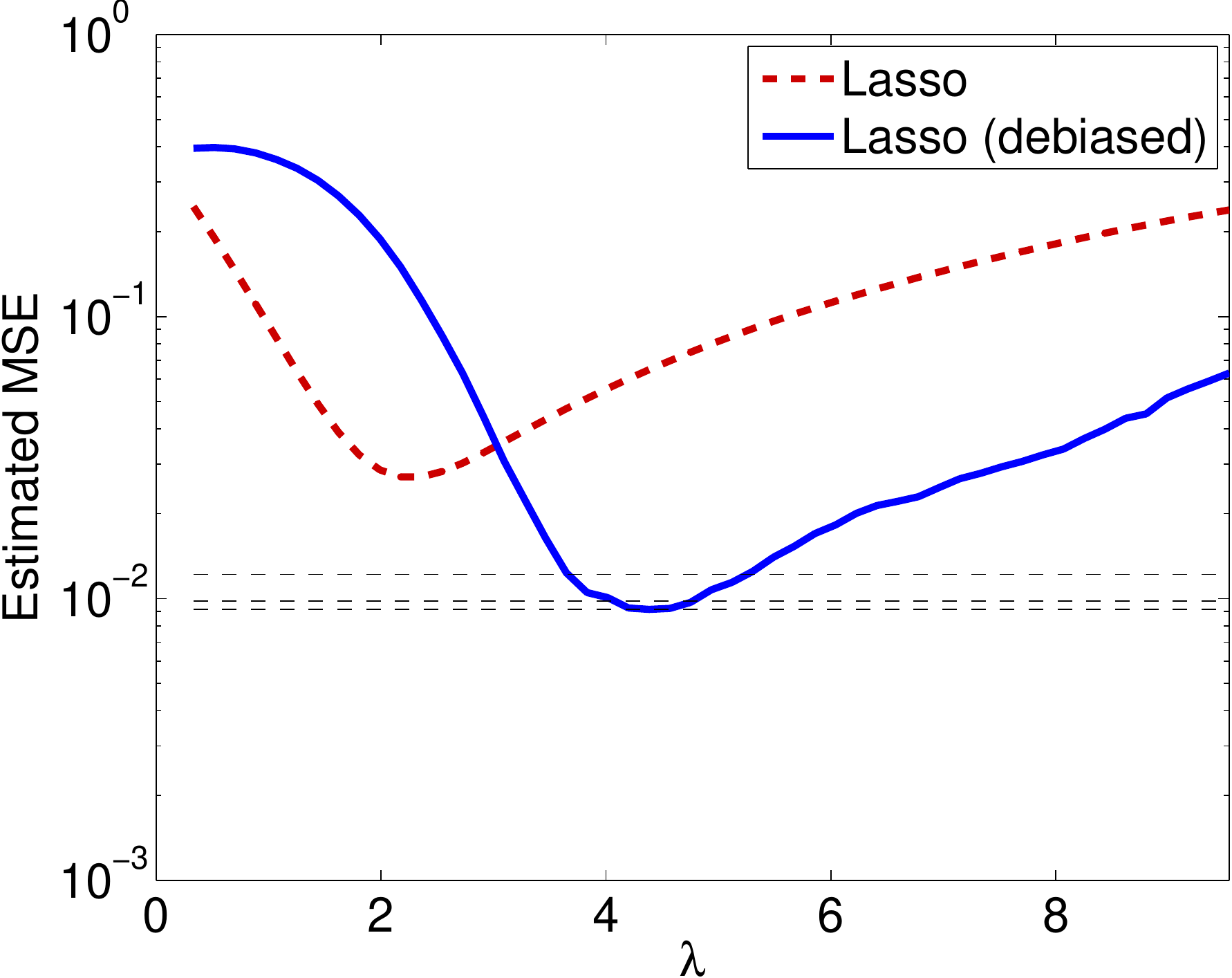} &
\includegraphics[width=0.42\textwidth]{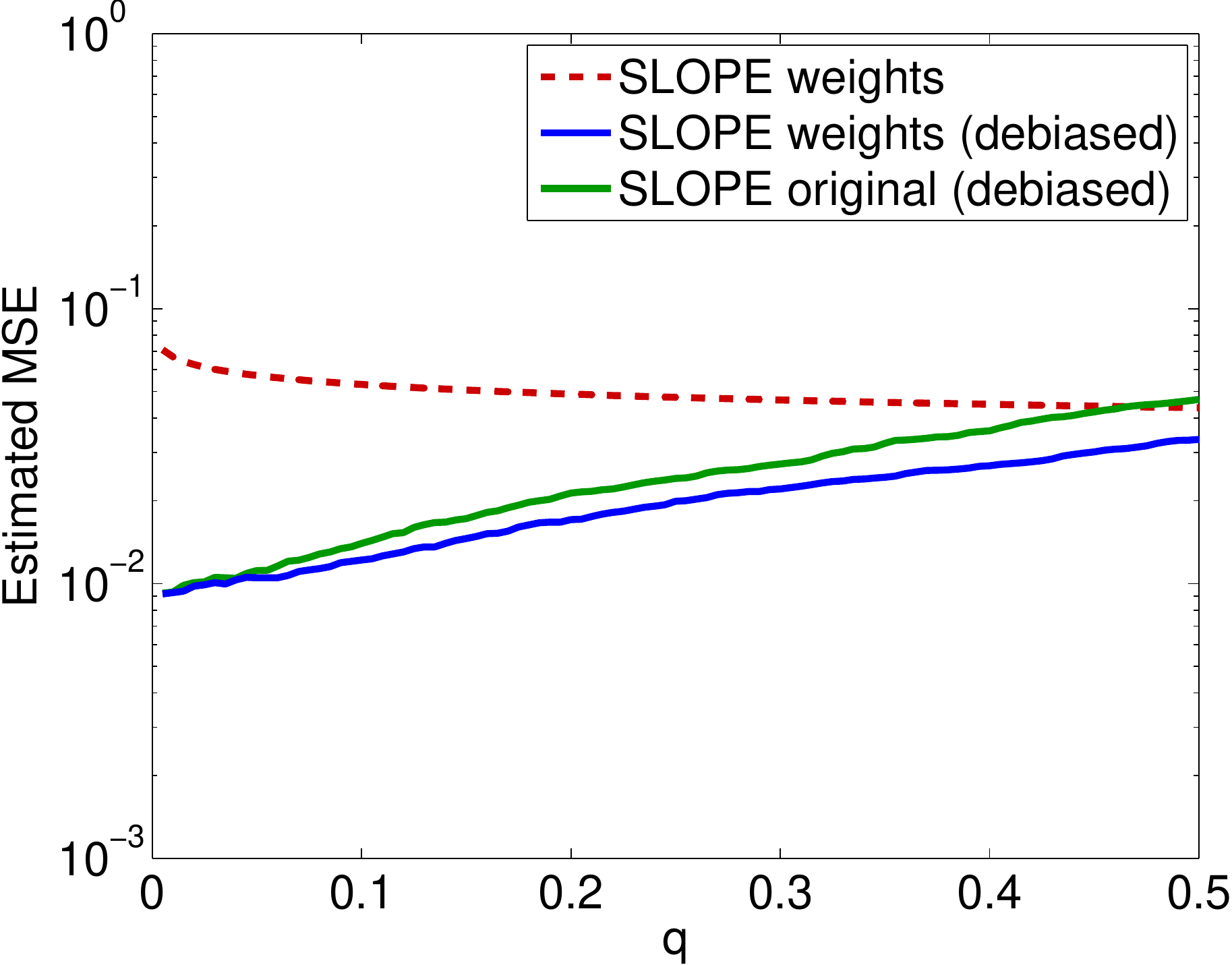} \\
({\bf{c}}) & ({\bf{d}}) \\[3pt]
\includegraphics[width=0.42\textwidth]{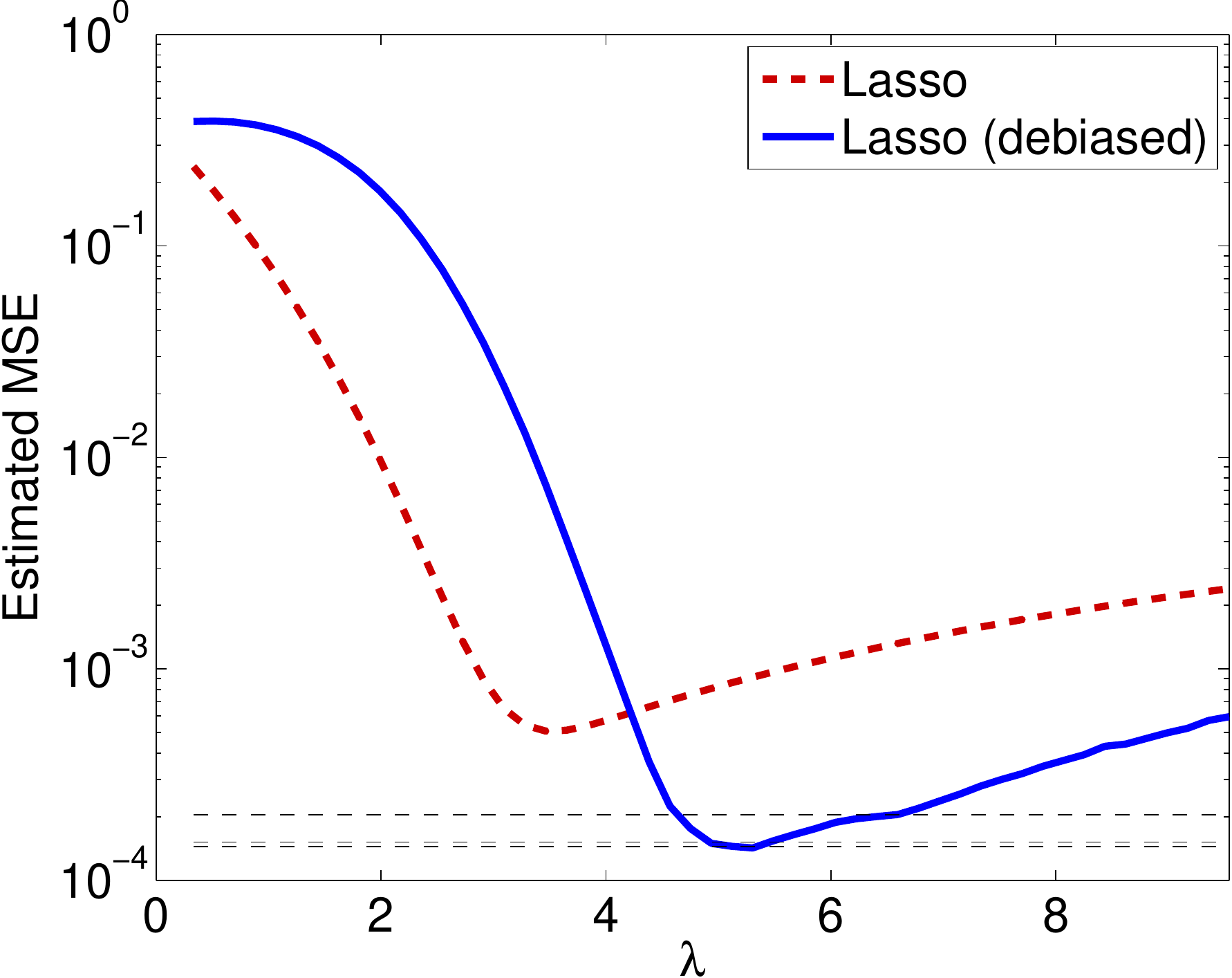} &
\includegraphics[width=0.42\textwidth]{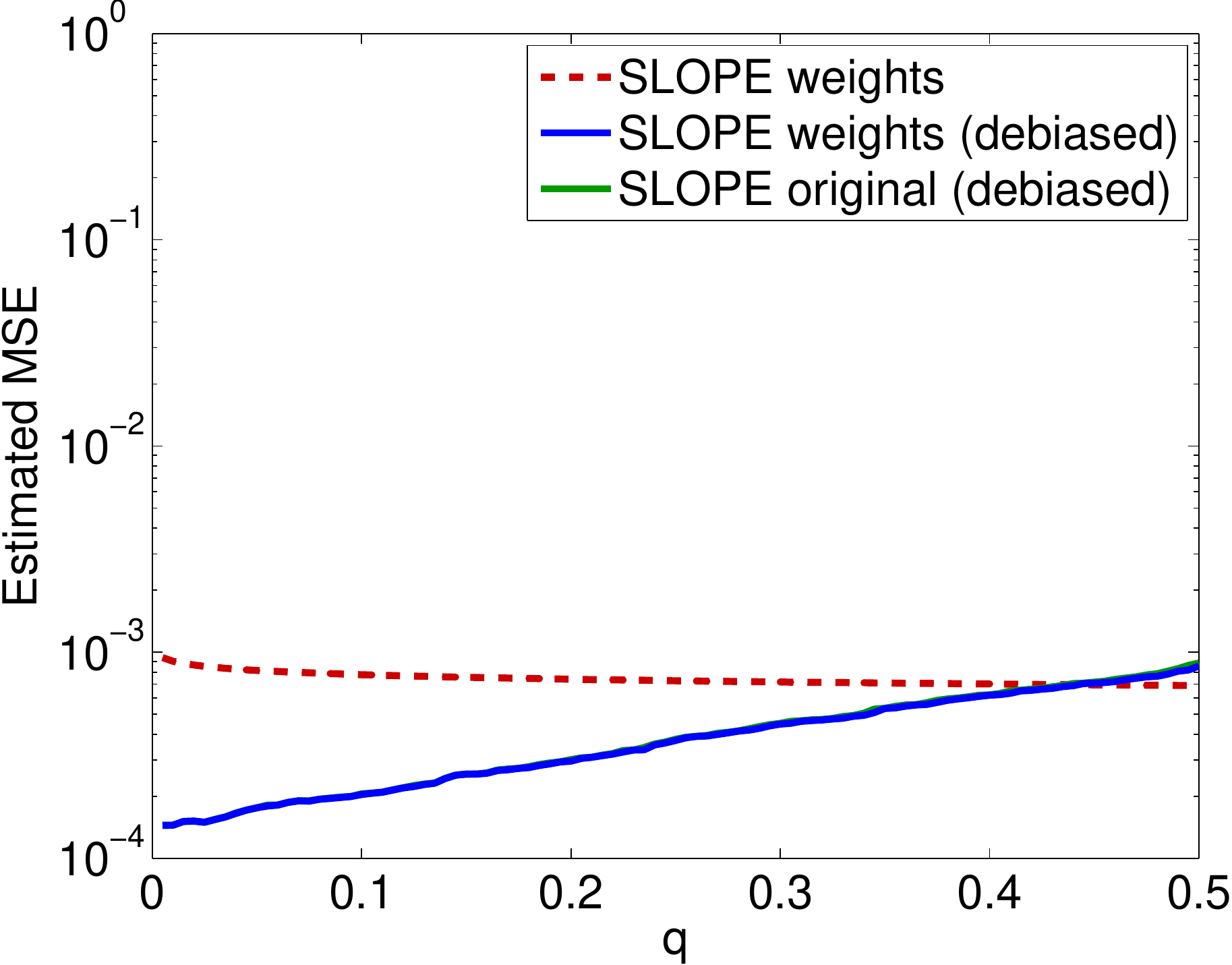} \\
({\bf{e}}) & ({\bf{f}})
\end{tabular}
\caption{Estimated mean square error value for $\hat{\beta}$ before
  and after debiasing for (a,b) signal class $5$, $k=1000$, (c,d)
  signal class $2$, $k=1000$, and (e,f) signal class $2$, $k=10$. The
  plots on the left show the results obtained using the  lasso
  algorithm, along with horizontal lines indicating the performance of
  SLOPE with from bottom to top: the optimal
  value of $q$; $q=0.02$; and the worst value of $q$ in the interval
  $0.01$ to $0.1$. The plots on the right show the (debiased) results
  of SLOPE with and without adaptive weights.
}\label{Fig:MSEExample}
\end{figure}

In preliminary experiments we observed that under sparse scenarios the
bias due to the shrinkage of regression coefficients has a
deteriorating influence on the MSE of both the lasso and SLOPE
estimates. We therefore suggest to use a debiasing step in which the
results from the lasso or SLOPE are used only to identify the nonzero
regression coefficients. The final estimation of their values is then
performed with the classical method of least squares on that
support. Having too many incorrect entries in the support distorts the
estimates of the true nonzero regression coefficients and it is thus
important to maintain a low FDR. At the same time we need a large
power: omitting entries from the support not only leads to zero
estimates for those coefficients, but also distorts the estimates of
those coefficients belonging to the
support. Figure~\ref{Fig:MSEExample} shows the effect the debiasing
step has on the (estimated) MSE. (In the consecutive graphs we report
only the results for the debiased version of the lasso and SLOPE). In
all but the least sparse cases, the minimum MSE with debiasing is
always lower than the best MSE obtained without it. The figure also
shows the dependency of the MSE on $\lambda$ for the lasso and $q$ for
SLOPE. The optimal choice of $\lambda$ for the lasso was found to
shift considerably over the various problem settings. By contrast, the
optimal $q$ for SLOPE remained fairly constant throughout. Moreover,
in many cases, using a fixed value of $q=0.02$ gives MSE values that
are close to, and sometimes even below, the lowest level obtained
using the lasso.

\begin{figure}
\centering
\begin{tabular}{cc}
\includegraphics[width=0.45\textwidth]{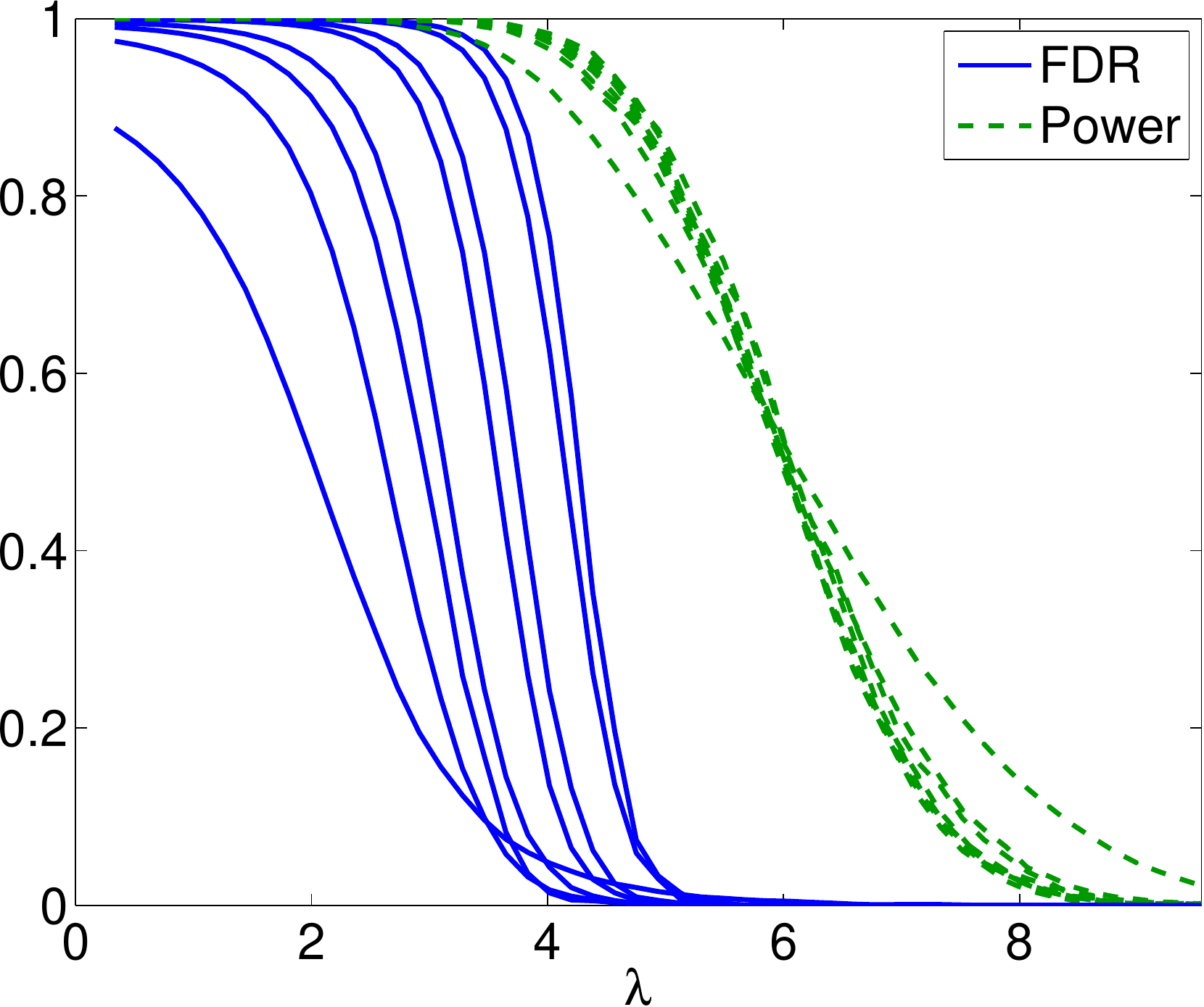}&
\includegraphics[width=0.45\textwidth]{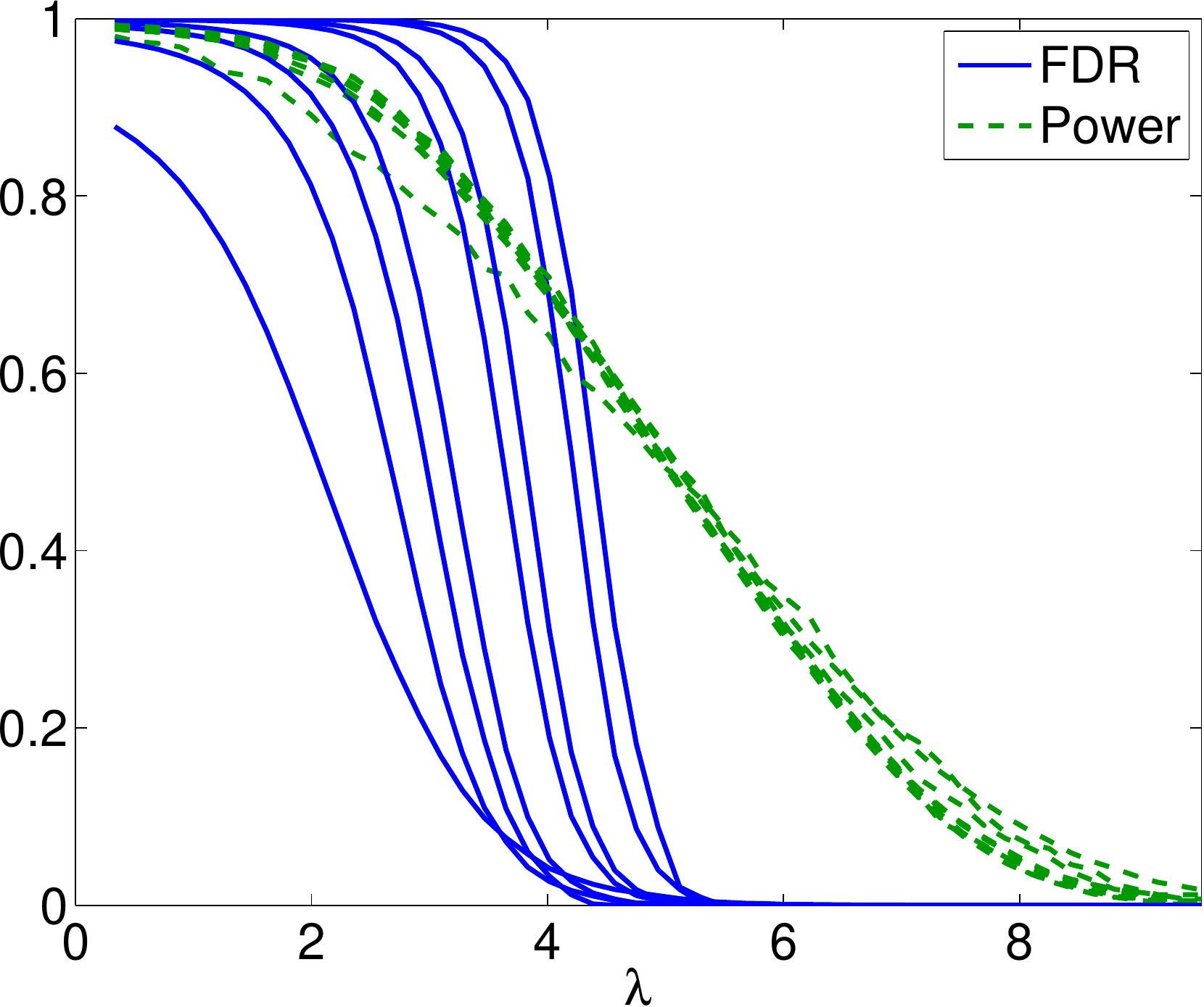}\\
({\bf{a}}) & ({\bf{b}})
\end{tabular}
\caption{FDR and power results obtained using  lasso for
  different sparsity levels and (a) signals of class 3, and (b) signal
  of class 5. The FDR curves from the right to the left correspond to
  decreasing sparsity (more nonzero entries in $\beta$): $k=1, 5, 10,
  50, 100, 500, 1000, 2621$, and $13,017$.}\label{Fig:LassoProblem}
\end{figure}

Figure~\ref{Fig:LassoProblem} shows the power and FDR results obtained
for signals of classes 3 and 5 using the  lasso with various
$\lambda$ values. The different FDR curves in the plot correspond to
increasing levels of sparsity (fewer nonzeros) from left to right. For
fixed $\lambda$ values, the power for a fixed signal class is nearly
independent of the sparsity level. The FDR, on the other hand,
increases significantly with sparsity.

\begin{figure}
\centering
\begin{tabular}{cc}
\includegraphics[width=0.45\textwidth]{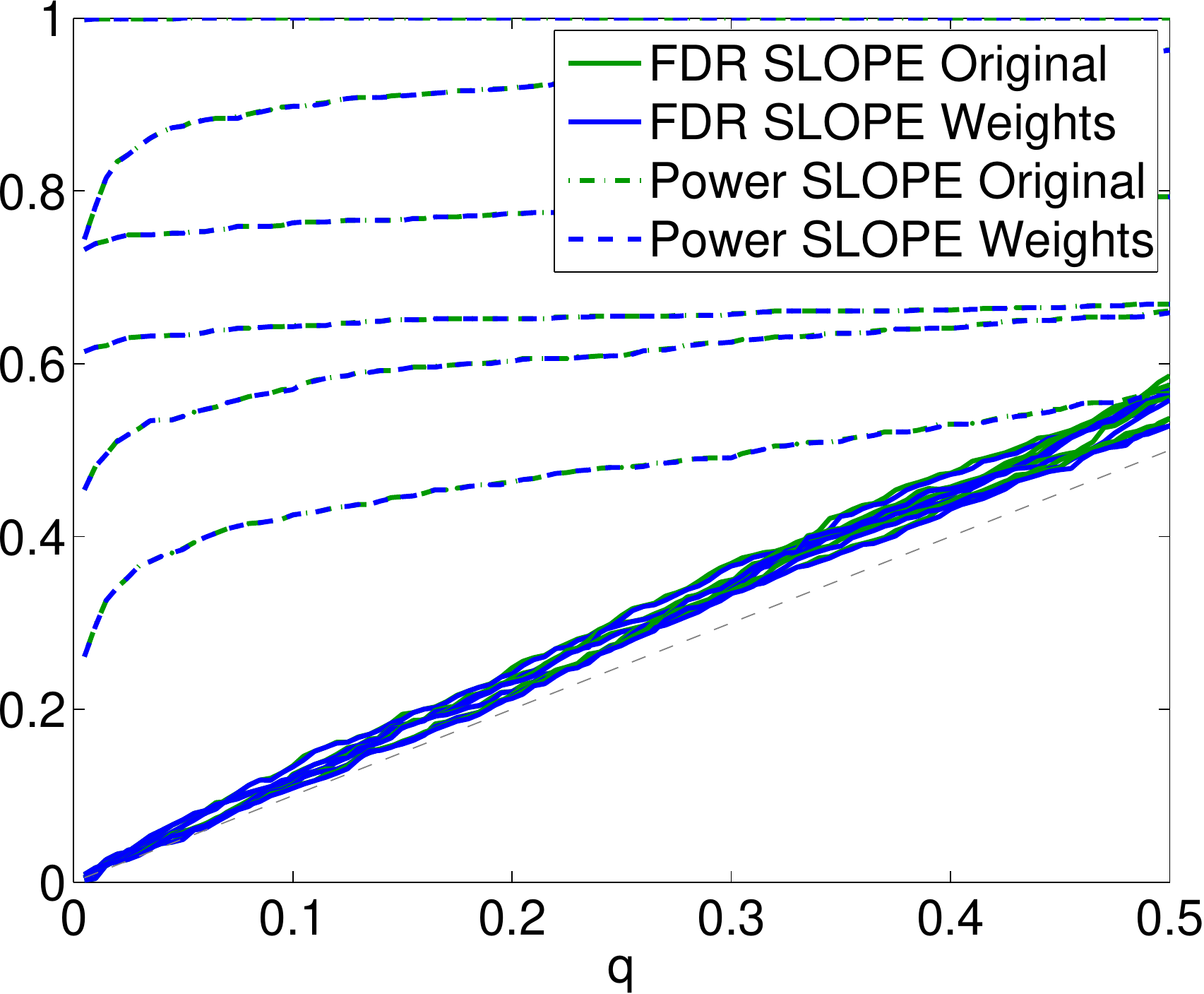} &
\includegraphics[width=0.45\textwidth]{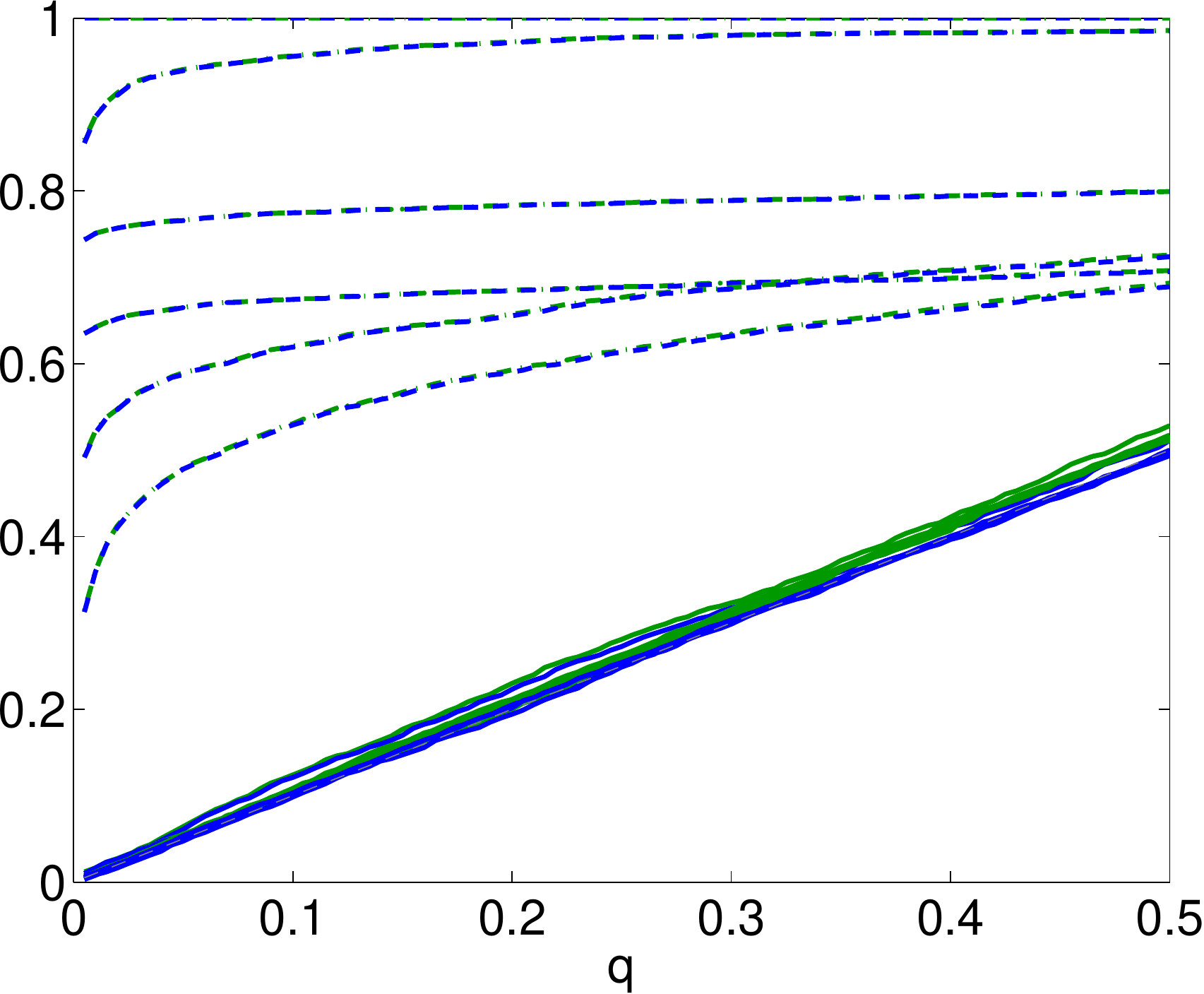} \\
({\bf{a}}) $k = 10$ & ({\bf{b}}) $k = 50$\\
\includegraphics[width=0.45\textwidth]{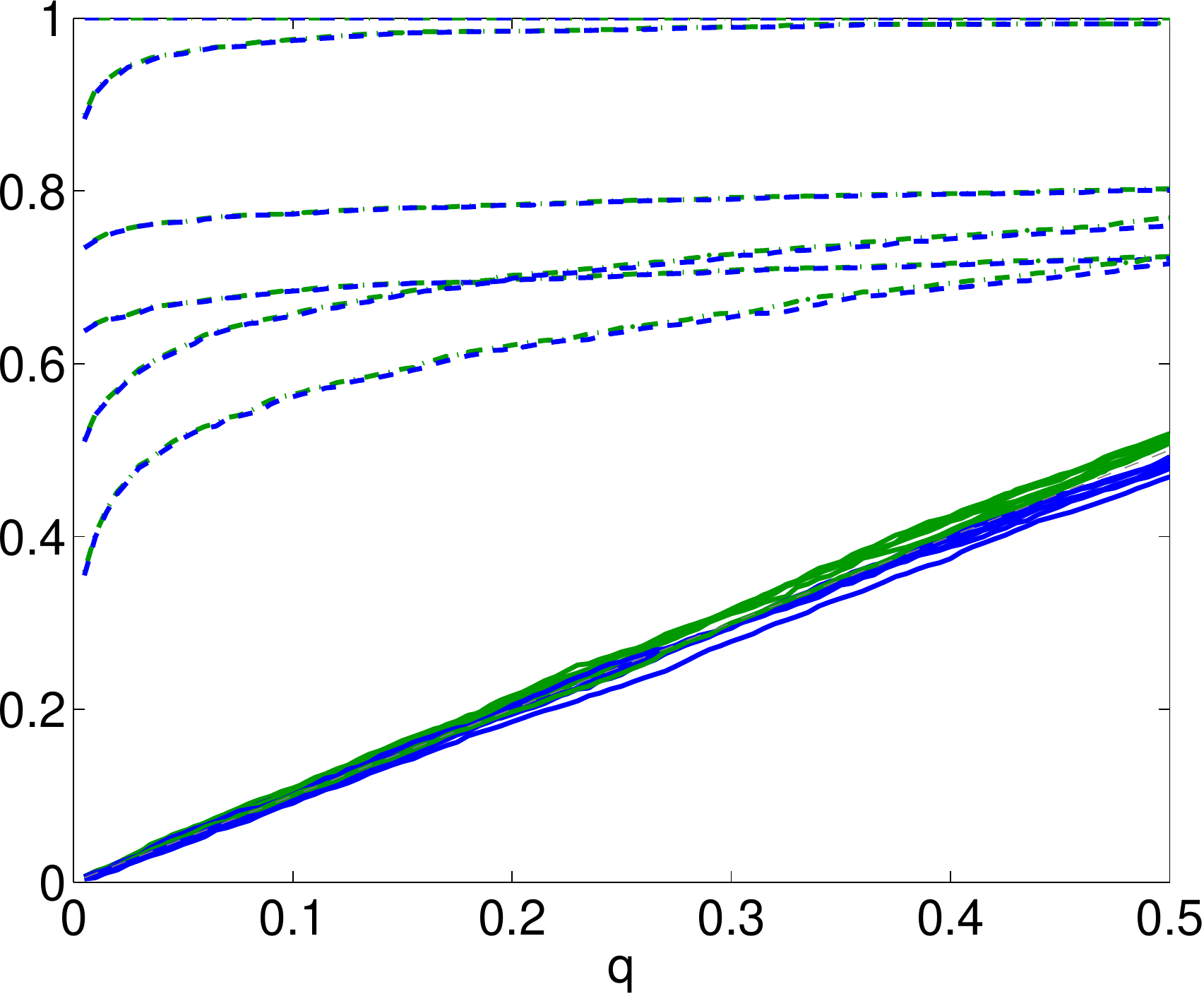} &
\includegraphics[width=0.45\textwidth]{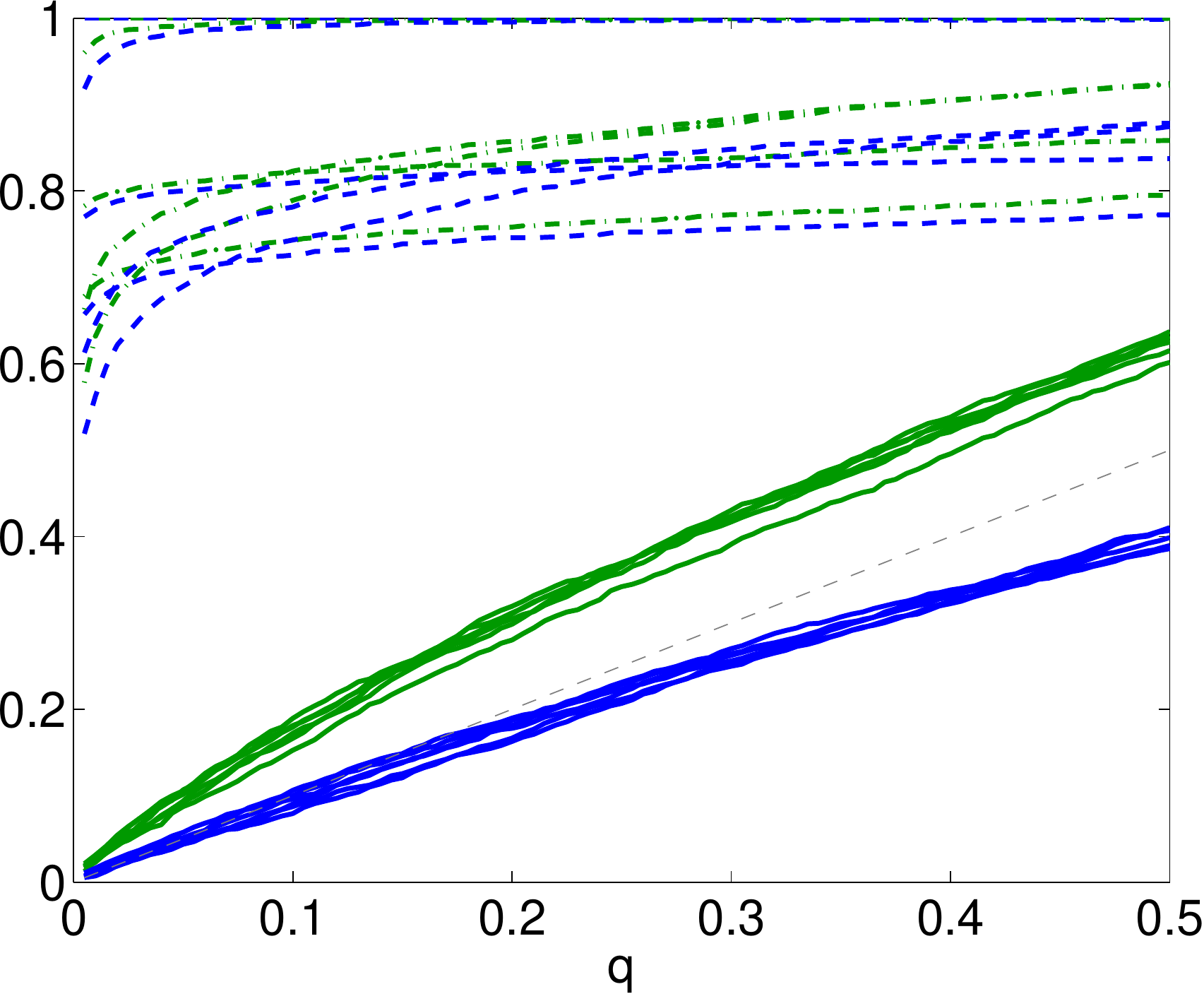} \\
({\bf{c}}) $k=100$ & ({\bf{d}}) $k=1\%$
\end{tabular}
\caption{Power and FDR results obtained using SLOPE. Each
  plots shows the results for all seven signal classes for a fixed
  sparsity level. At $q=0.1$ the lines from top to bottom correspond
  to classes 4,5,1,2,3, and 6 for plots  (a)--(c), and to 1,4,5,2,3, and 6
  for plot (d). The dashed gray line indicates the nominal FDR
  level.}\label{Fig:OLSparsity}
\end{figure}

The FDR and power results obtained with SLOPE are shown in
Figure~\ref{Fig:OLSparsity}. Here, the results are grouped by sparsity
level, and the different curves in each plot correspond to the seven
different signal classes. The FDR curves for each sparsity level are
very similar. The power levels differ quite a bit depending on how
`difficult' the signal is to recover. In plots (a)--(c) the results
between SLOPE with fixed and adapted weights (see
Section~\ref{sec:lambda}) are almost
identical. In plot (d), however, the FDR of SLOPE with
adapted weights is much lower, at the expense of some of the
power. Importantly, it can be seen that for a wide range of sparsity
levels, the adaptive version keeps FDR very close to the nominal level
indicated by the gray dashed line.

We mentioned above that the optimal parameter choice for SLOPE is much
more stable than that for the  lasso. Because the optimal value
depends on the problem it would be much more convenient to work with a
fixed $\lambda$ or $q$. In the following experiment we fix the
parameter for each method and then determine over all problem
instances the maximum ratio between the optimal (two-norm) misfit over
all parameters (for both lasso and SLOPE) and the misfit obtained with
the given parameter choice. Table~\ref{Table:DctMinimax} summarizes
the results. It can be seen that the optimal parameter choice for a
series of signals from the different classes generated with a fixed
sparsity level changes rapidly for lasso (between $\lambda
= 3.46$ and $\lambda = 5.49$), whereas it remains fairly constant for
SLOPE (between $q=0.01$ and $q=0.04$. Importantly, the maximum
relative misfit for lasso over all problem instances varies
tremendously for each of the $\lambda$ values in the given range. For
SLOPE, the same maximum relative misfit changes slightly over the
given parameter range for $q$, but far more moderately. Finally, the
results obtained with the parameter that minimizes the maximum
relative misfit over all problem instances are much better for SLOPE.

\begin{table}
\centering
\input{./tables/TableDctDiffExt.tex}
\caption{Entries in the table give the maximum ratio $(||\hat{b}-b||_2
  /  ||\hat{b}_{\mathrm{opt}} - b||_2) - 1$ in percent over all signal
  types with the given sparsity. Here $b$ is an instance of $\beta$;
  $\hat{b}$ is the corresponding estimate obtained using lasso and SLOPE 
  with the given parameter value; and
  $\hat{b}_{\mathrm{opt}}$ is the estimate with the lowest two-norm
  misfit over all parameters and both methods. The entries in red show
  the lowest ratio obtained with either method for the given sparsity
  level. The entries in blue correspond to the parameter choice that
  gives the lowest maximum ratio over all problems.}\label{Table:DctMinimax}
\end{table}

\subsubsection{Phantom}\label{Sec:Phantom}

As a more concrete application of the methods, we now discuss the
results for the reconstruction of phantom data from subsampled
discrete-cosine measurements. Denoting by $x \in \mathbb{R}^p$ the
vectorized version of a phantom with $p$ pixel values, we observe $y =
RDx$, where $R$ is a restriction operator and $D$ is a two-dimensional
discrete-cosine transformation\footnote{In medical imaging
  measurements would correspond to restricted Fourier measurements. We
  use the discrete-cosine transform to avoid complex numbers,
  although, with minor modifications to the prox function, the solver
  can readily deal with complex numbers.}. Direct reconstruction of
the signal from $y$ is difficult because $x$ is not sparse. However,
when expressed in a suitable wavelet basis the coefficients become
highly compressible. Using the two-dimensional Haar wavelet
transformation $H$, we can write $\beta = Hx$, or by orthogonality of
$H$, $x = H^T\beta$. Combining this we get to the desired setting $y =
X\beta$ with $X := RDH^T$, and $\beta$ approximately sparse. Once the
estimated $\hat{\beta}$ is known we can obtain the corresponding
signal estimation using $\hat{x} = H^T\hat{\beta}$. Note that due to
orthogonality of $H$ we have that $\Vert{x - \hat{x}}\Vert_2 =
\Vert{\beta - \hat{\beta}}\Vert_2$, and it therefore suffices to
report only the MSE values for $\hat{\beta}$.

\begin{figure}
\centering
\begin{tabular}{cc}
\includegraphics[height=5.5cm]{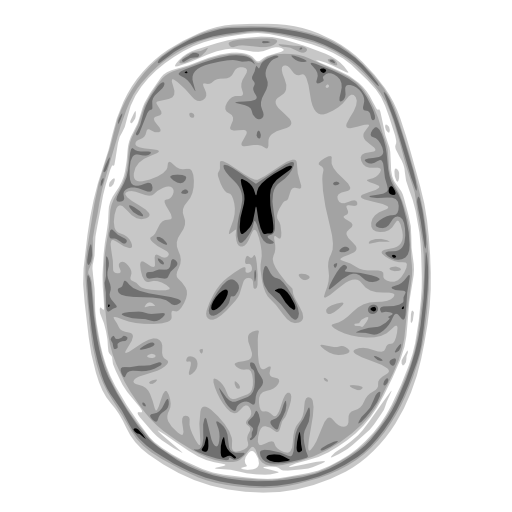} & 
\includegraphics[height=5.5cm]{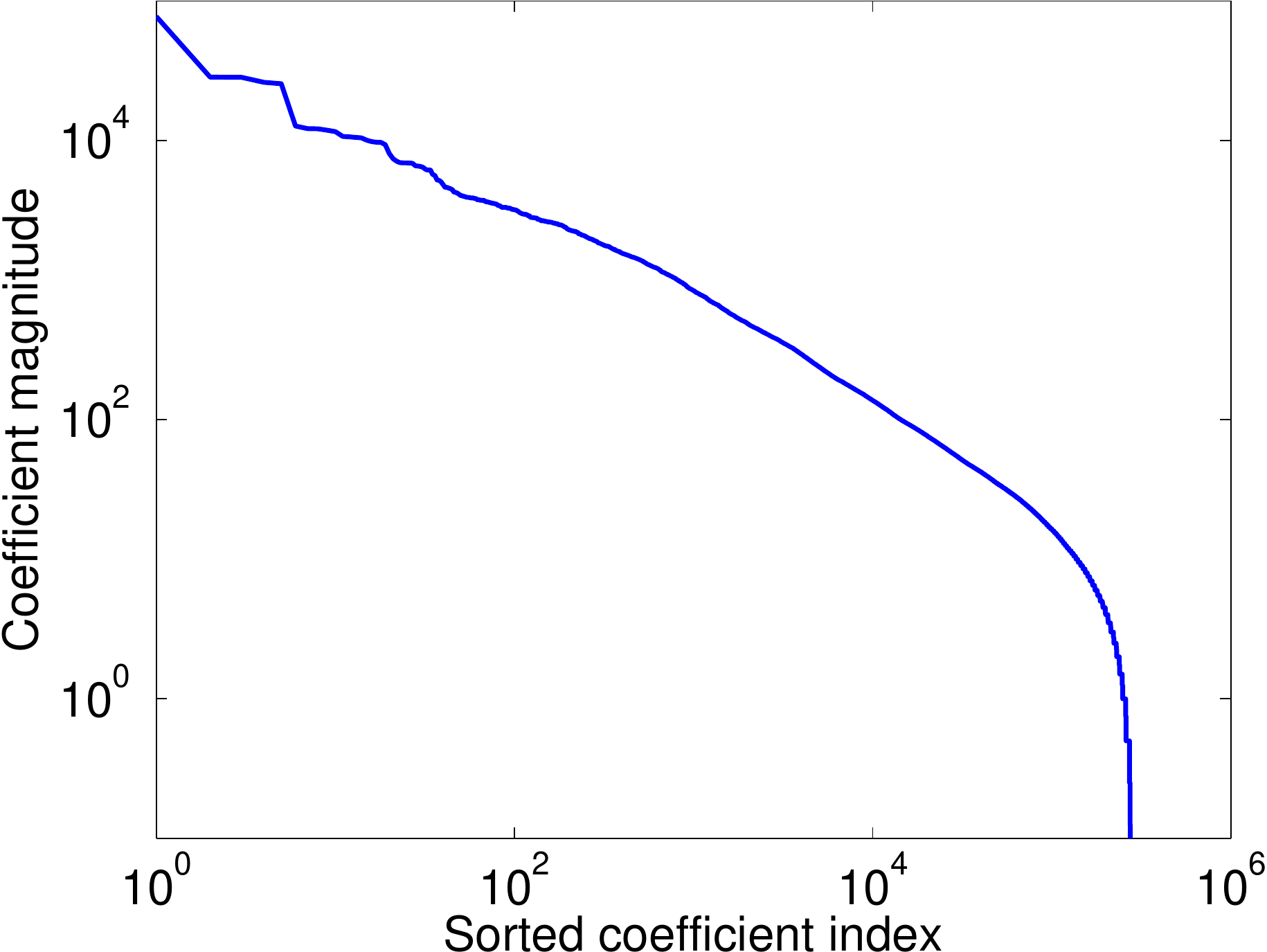}\\
({\bf{a}}) & ({\bf{b}})
\end{tabular}
\caption{Illustration of (a) a discretized version of the continuous
  phantom data, shown with inverted grayscale, and (b) the magnitudes
  of the Haar wavelet coefficients in decreasing order.
}\label{Fig:EPFLPhantom}
\end{figure}

For our experiments we generated a test phantom by discretizing the
analytical phantoms described in Guerquin-Kern, et
al.~\cite{EPFLPhantom} and available on the accompanying website. The
resulting $2048\times 2048$ image, illustrated in
Figure~\ref{Fig:EPFLPhantom}, is then vectorized to obtain $x$ with $p
= 2048^2$. We obtain $\beta$ using the two-dimensional Haar
transformation implemented in the Rice wavelet toolbox
\cite{Software:RWT} and interfaced though the Spot linear operator
toolbox \cite{Software:Spot}. As summarized in
Table~\ref{Table:PhantomSettings}, we consider two subsampling levels:
one with $n = p/2$, the other with $n =p/5$.  For each of the
subsampling levels we choose a number of different noise levels based
on a target sparsity level $k$. In particular, we choose $k = \lceil
\gamma p/10^4\rceil$, and then set $\sigma =
\vert\beta\vert_{[k]}/(1.2\sqrt{2\log p})$. The values of $\gamma$ and
the resulting signal-to-noise ratios ($\norm{X\beta}_2 /
\norm{X\beta-y}$) for each problem setting are listed in
Table~\ref{Table:PhantomSettings}.

\begin{table}
\centering
\begin{tabular}{lcccccccc}
\hline
Problem & 1 & 2 & 3 & 4 & 5 & 6 & 7 & 8 \\
\hline
$n/p$ & 0.5 & 0.5 & 0.5 & 0.5 & 0.2 & 0.2 & 0.2 & 0.2 \\
$\gamma$ & 30 & 20 & 10 & 5 & 500 & 50 & 20 & 10 \\
SNR & 5.97 & 4.25 &  2.39 &  1.43 & 411.7 & 19.5 &  9.10 &  5.12 \\
\hline
\end{tabular}
\caption{Different problem configurations for phantom reconstruction
  with $p = 2048^2$. The number of observations $n$ is prescribed by
  the given ratios $n/p$. The noise level $\sigma$ is set to
  $\vert\beta\vert_{[k]} / (1.2\sqrt{2\log{p}})$ with $k = \lceil
  \gamma p/10^{4}\rceil$.}\label{Table:PhantomSettings}
\end{table}

As in the previous section we are interested in the sensitivity of the
results with respect to the parameter choice, as well as in the
determination of a parameter value that does well over a wide range of
problems. We therefore started by running the lasso and SLOPE
algorithms with various choices of $\lambda$ and $q$,
respectively. The resulting relative two-norm misfit values are
summarized in Table~\ref{Table:PhantomMSE}. The minimum MSE value
obtained using the  lasso are typically slightly smaller than
those obtained using SLOPE, except for problems 3 and 4. On the other
hand, the relative misfit values obtained over the given parameter
ranges vary more than those for SLOPE.

\begin{table}
\centering
\input{tables/TablePhantomMSE}
\caption{Relative misfit values $||\hat{b} - b||_2 / ||b||_2$ (scaled
  by 100 for display purposes) for different
  phantom test problem instances with $b = \beta$ and its
  corresponding estimate $\hat{b}$, using lasso and SLOPE. 
  The highlighted entries indicate the lowest misfit
  values obtained on a given problem using each of the two methods. Not
  shown is the minimum relative misfit of $1.971$ for problem 5 obtained
  using the  lasso using $\lambda =
  6.0$.}\label{Table:PhantomMSE}
\end{table}

Table~\ref{Table:PhantomDiff} shows the percentage difference of each
estimate to the lowest two-norm misfit obtained with either of the two
methods for each problem. The results obtained with the parameters
that minimize the maximum difference over all problem instances are
highlighted. In this case, using the given parameters for each methods
gives a similar maximum difference of approximately $1.4\%$. This
value is again very sensitive to the choice of $\lambda$ for the
 lasso but remains much lower for the given parameter choices
for SLOPE. More interestingly perhaps is to compare the
results in Tables~\ref{Table:DctMinimax} and \ref{Table:PhantomDiff}.
Choosing the best fixed value for the lasso obtained for the
DCT problems ($\lambda = 4.94$) and applying this to the phantom
problem gives a maximum difference close to $10\%$. This effect is
even more pronounced when using the best parameter choice ($\lambda =
4.00$) from the phantom experiment and applying it to the DCT
problems. In that case it leads to deviations of up to $800\%$
relative to the best. For SLOPE the differences are minor:
applying the DCT-optimal parameter $q=0.02$ gives a maximum deviation
of $1.474\%$ instead of the best $1.410\%$ on the phantom problems. Vice
versa, applying the optimal parameter for the phantom, $q=0.03$ to the
DCT problems gives a maximum deviation of $31.2\%$ compared to
$24.0\%$ for the best.

\begin{table}
\centering
\input{tables/TablePhantomDiffExt}
\caption{
  Entries in the table give the maximum ratio $(||\hat{b}-b||_2
  /  ||\hat{b}_{\mathrm{opt}} - b||_2) - 1$ in percent for each
  problem and parameter setting. Here $b$ is an instance of $\beta$;
  $\hat{b}$ is the corresponding estimate obtained using lasso or SLOPE 
  with the given parameter value; and
  $\hat{b}_{\mathrm{opt}}$ is the estimate with the lowest two-norm
  misfit over all parameters and both methods for a given problem. The
  entries in blue correspond to the parameter choice that gives the
  lowest maximum ratio over all problem
  instances.}\label{Table:PhantomDiff}
\end{table}

%
%

\subsubsection{Weights}

When using SLOPE with the adaptive procedure \eqref{EC}
described in Section~\ref{sec:lambda}, we need to evaluate
\eqref{Eq:ExpWk}.
For most matrices $X$, there is no closed-form solution to the above
expectation, and numerical evaluation would be prohibitively
expensive. However, we can get good approximations $\hat{w}_k$ of
$w_k$ values by means of Monte-Carlo simulations. Even so, it remains
computationally expensive to evaluate $w_k$ for all values $k \in
[1,\min(n,p-1)]$. Our approach is to take 21 equidistant $k$ values
within this interval, including the end points, and evaluate
$\hat{w}_k$ at those points. The remaining weights are obtained using
linear interpolation between the known values. For the first interval
this linear approximation was found to be too crude, and we therefore
sample $\hat{w}_k$ at an addition 19 equispaced points within the
first interval, thus giving a total of 40 sampled points. The next
question is how many random trials to take for each
$\hat{w}_k$. Preliminary experiments showed that the variance in the
sampled values reduces with increasing $k$, while the computation cost
of evaluating one term in the expectation in \eqref{Eq:ExpWk} grows
substantially. To avoid choosing a fixed number of samples and risk
inaccurate estimations of $\hat{w}_k$ for small $k$, or waste time
when evaluating $\hat{w}_k$ for large $k$, we used a dynamic sampling
scheme. Starting with a small number of samples, we obtain an
intermediate estimate $\hat{w}_k$. We then compute a reference
approximation using the same number of samples, record the difference
between the two values and update $\hat{w}_k$ to the average of the
two. If the original difference was sufficiently small we terminate
the algorithm, otherwise we double the number of samples and compute a
new reference approximation, which is then again first compared to the
current $\hat{w}_k$, and then merge into it, and so on. This process
continues until the algorithm either terminates naturally, or when a
pre-defined maximum sample size is reached.

\begin{figure}
\centering
\begin{tabular}{ccc}
\includegraphics[width=0.315\textwidth]{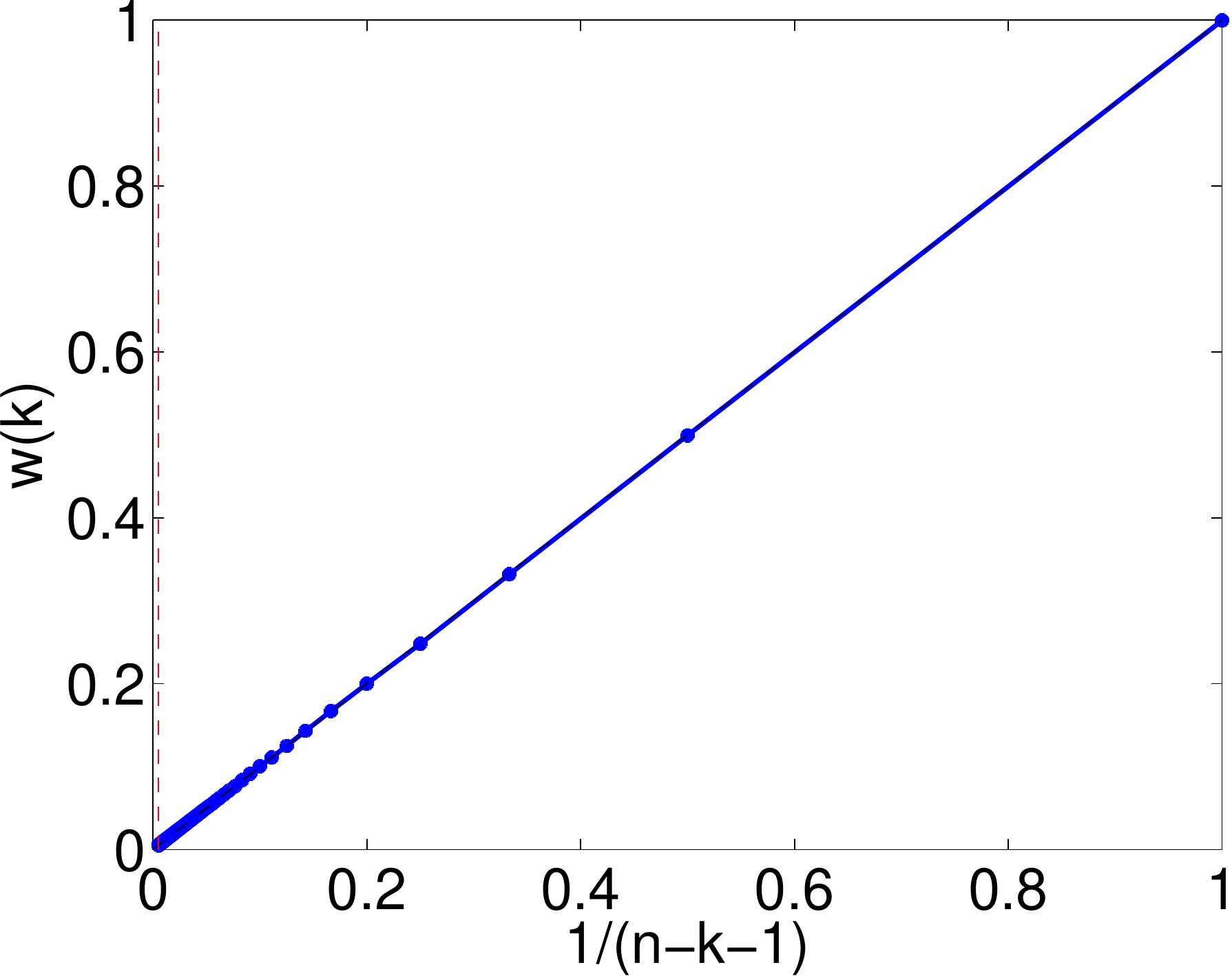} &
\includegraphics[width=0.315\textwidth]{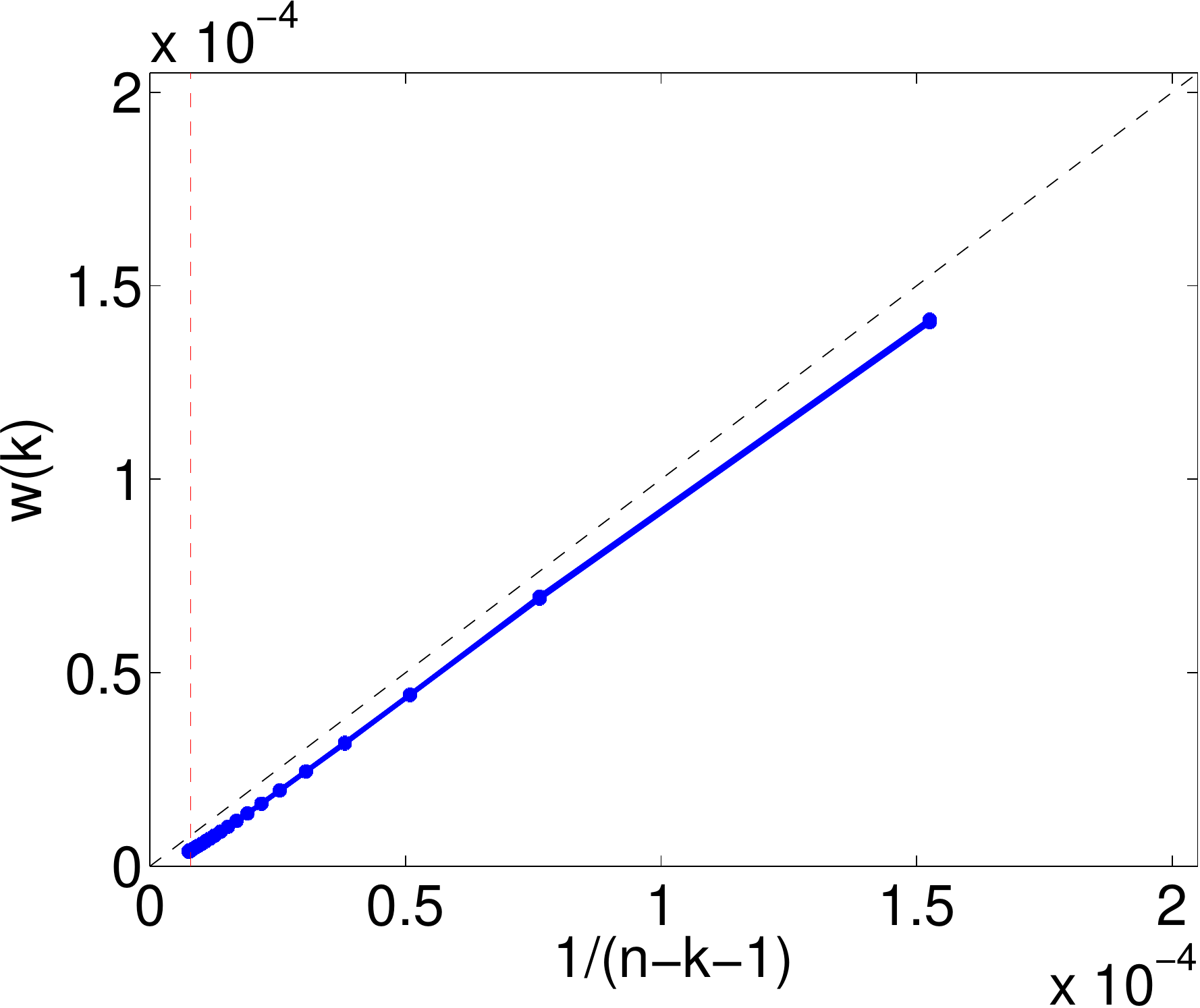} &
\includegraphics[width=0.315\textwidth]{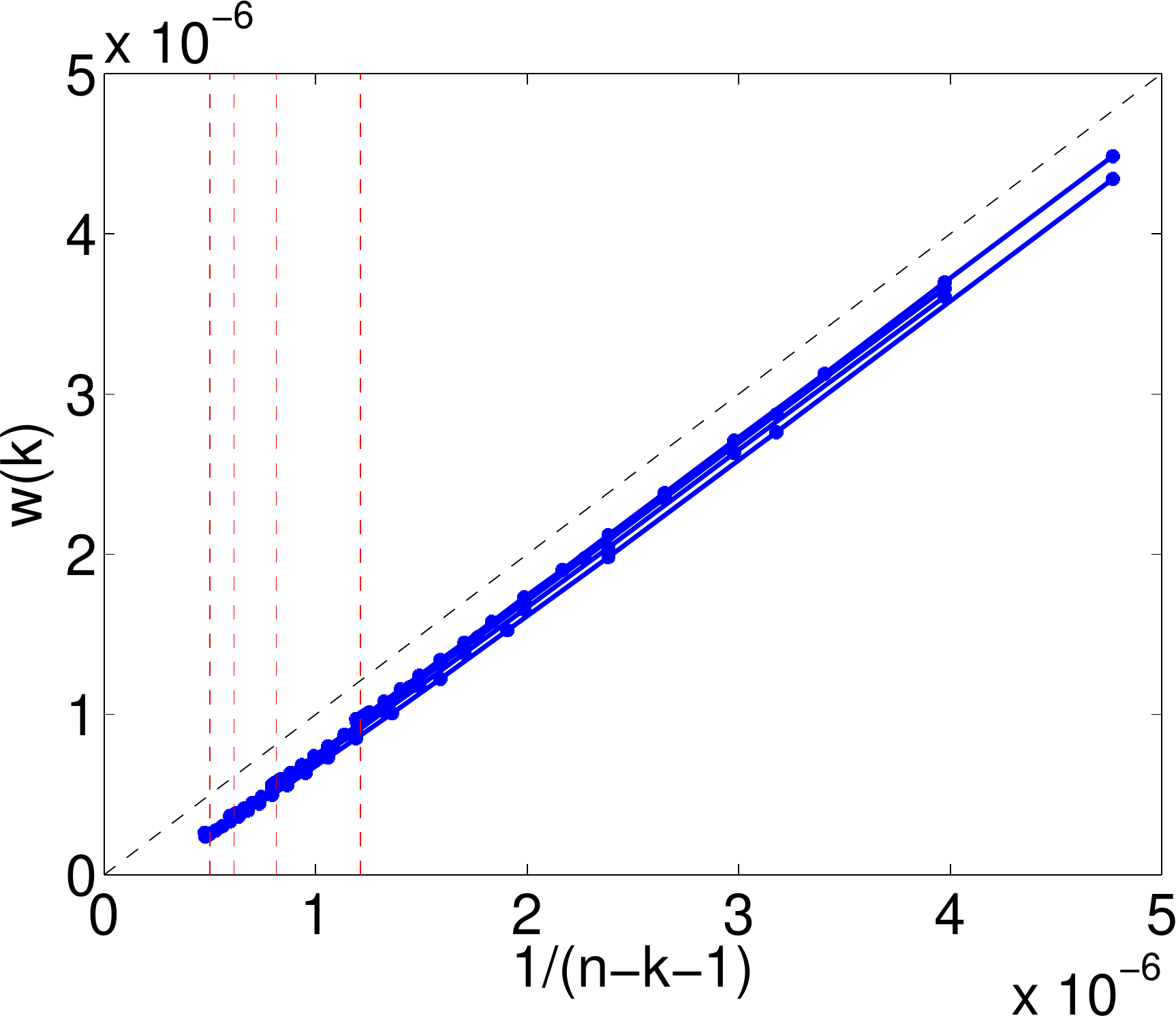}\\
({\bf{a}}) Gaussian & ({\bf{b}}) DCT & ({\bf{c}}) DCT and Haar
\end{tabular}
\caption{Weights $w(k)$ plotted against $1 / (n-k-1)$ for (a) a
  $200\times 600$ matrix with entries sampled i.i.d.\ from the normal
  distribution; (b) five instances of the restricted DCT matrix
  described in Section~\ref{Sec:Phantom}; and (c) matrices as
  described in Section~\ref{Sec:Phantom} for $n/p$ ranging from 0.2 to
  0.5. The vertical red lines indicate the weights needed to evaluate
  $\LBHc$ for $q=0.1$.}\label{Fig:MatrixWeights}
\end{figure}

In Figure~\ref{Fig:MatrixWeights} we plot the resulting weights
obtained for three different types of matrices against $1 / (n - k -
1)$. The results in plot (a) are obtained for a $200\times 600$ matrix
with entries sampled i.i.d.\ from the normal distribution, and with
$\hat{w}_k$ for all $k\in[1,200]$ based on 10,000 samples. The blue
line exactly follows the theoretical line for $w_k = 1/(n-k-1)$, as
derived in Section~\ref{sec:lambda}. Plots (b) show the results
obtained for five instances of the the restricted DCT operator
described in Section~\ref{Sec:SparseSignals}. We omit the value of
$\hat{w}_k$ for $k=n$, which would otherwise distort the figure due to
the choice of axes. In plot (c) we plot the weights for the matrices
used in Section~\ref{Sec:Phantom}, along with additional instances
with $n/p = 0.3$ and $n/p = 0.4$. The results in plots (b) and (c) are
slightly below the reference line. The vertical dashed red lines in
the plots indicate the maximum value of $k$ needed to evaluate $\LBHc$
with $q = 0.1$. In the DCT case we only need up to $k=5439$, and for
DCT and Haar we need $k=14,322$ for $n/p = 0.2$ and $k=95,472$ for
$n/p = 0.5$. In both cases this $k$ falls well within the first
interval of the coarsely sampled grid. For a fixed $q$ it may
therefore be possible to much further reduce the computational cost by
evaluating successive $\hat{w}_k$ for increasing $k$ until the
critical value is reached.

%% file: tables/TableDctDiffExt.tex
\begin{tabular}{llrrrrrrrrrr}
\multicolumn{2}{r}{$k$}& \multicolumn{1}{c}{1}& \multicolumn{1}{c}{5}& \multicolumn{1}{c}{10}& \multicolumn{1}{c}{50}& \multicolumn{1}{c}{100}& \multicolumn{1}{c}{500}& \multicolumn{1}{c}{1000}& \multicolumn{1}{c}{1\%}& \multicolumn{1}{c}{5\%}& Max. \\
\hline
\multirow{10}{*}{\rotatebox{90}{\begin{minipage}{2cm}\begin{center}Lasso\end{center}\end{minipage}}}
 &    $\lambda=3.28$  &    3100.9  &    2118.3  &    1698.5  &    705.5  &    477.9  &    183.8  &    123.0  &    62.1  &    24.8  & 3100.9 \\
 &    $\lambda=3.46$  &    2316.3  &    1571.0  &    1256.0  &    510.8  &    342.4  &    126.9  &    83.7  &    40.2  & {\color{darkred}{18.1}} & 2316.3 \\
 &    $\lambda=3.83$  &    1199.0  &    790.8  &    636.2  &    240.9  &    147.4  &    51.8  &    33.2  &    16.3  &    18.7  & 1199.0 \\
 &    $\lambda=4.02$  &    827.5  &    540.7  &    430.2  &    154.3  &    90.7  &    28.3  &    18.1  & {\color{darkred}{11.5}} &    25.4  & 827.5 \\
 &    $\lambda=4.20$  &    546.8  &    352.5  &    281.5  &    91.7  &    52.2  & {\color{darkred}{14.4}} & {\color{darkred}{12.4}} &    19.1  &    38.2  & 546.8 \\
 &    $\lambda=4.57$  &    200.9  &    116.2  &    94.4  &    24.2  & {\color{darkred}{12.9}} &    23.1  &    36.4  &    45.7  &    64.0  & 200.9 \\
 &    $\lambda=4.75$  &    113.1  &    56.4  &    48.6  & {\color{darkred}{11.7}} &    21.5  &    37.2  &    57.9  &    65.6  &    77.3  & 113.1 \\
 &    $\lambda=4.94$  & {\color{blue}{53.5}} & {\color{blue}{33.9}} & {\color{blue}{22.0}} & {\color{blue}{23.7}} & {\color{blue}{39.9}} & {\color{blue}{53.8}} & {\color{blue}{75.6}} & {\color{blue}{84.0}} & {\color{blue}{90.9}} & 90.9 \\
 &    $\lambda=5.31$  &    20.0  & {\color{darkred}{28.3}} &    27.4  &    58.7  &    77.2  &    92.0  &    128.0  &    122.0  &    116.6  & 128.0 \\
 &    $\lambda=5.49$  & {\color{darkred}{11.3}} &    39.4  &    38.7  &    74.5  &    94.3  &    111.9  &    149.2  &    141.2  &    127.9  & 149.2 \\
\hline
\multirow{7}{*}{\rotatebox{90}{\begin{minipage}{2cm}\begin{center}SLOPE\end{center}\end{minipage}}}
 &    $q=0.01$  &    14.9  &    27.1  & {\color{darkred}{17.3}} & {\color{darkred}{13.1}} & {\color{darkred}{14.3}} & {\color{darkred}{14.7}} &    16.8  &    20.3  &    38.6  & 38.6 \\
 &    $q=0.02$  & {\color{blue}{11.5}} & {\color{blue}{17.8}} & {\color{blue}{24.0}} & {\color{blue}{20.1}} & {\color{blue}{17.8}} & {\color{blue}{15.4}} & {\color{blue}{14.2}} & {\color{blue}{11.6}} & {\color{blue}{24.0}} & 24.0 \\
 &    $q=0.03$  &    17.9  &    29.8  &    31.2  &    28.0  &    24.6  &    22.1  &    23.7  &    15.2  &    19.7  & 31.2 \\
 &    $q=0.04$  &    17.9  &    36.6  &    45.0  &    36.4  &    32.9  &    29.2  &    29.5  &    19.1  & {\color{darkred}{16.5}} & 45.0 \\
 &    $q=0.10$  &    47.9  &    66.6  &    89.6  &    77.5  &    69.0  &    63.3  &    54.8  &    38.9  &    19.3  & 89.6 \\
 &    $q=0.12$  &    52.7  &    82.2  &    101.0  &    89.1  &    78.5  &    72.2  &    65.4  &    46.0  &    22.2  & 101.0 \\
 &    $q=0.15$  &    78.9  &    102.0  &    123.7  &    106.7  &    94.4  &    85.4  &    77.2  &    54.8  &    25.8  & 123.7 \\
\hline
\end{tabular}

%% file: tables/TablePhantomMSE.tex
\begin{tabular}{llrrrrrrrr}
\multicolumn{2}{l}{Problem idx.}& \multicolumn{1}{c}{1}& \multicolumn{1}{c}{2}& \multicolumn{1}{c}{3}& \multicolumn{1}{c}{4}& \multicolumn{1}{c}{5}& \multicolumn{1}{c}{6}& \multicolumn{1}{c}{7}& \multicolumn{1}{c}{8} \\
\hline
\multirow{7}{*}{\rotatebox{90}{\begin{minipage}{2cm}\begin{center}Lasso\end{center}\end{minipage}}}
 & $\lambda=3.50$ &    8.589  &    10.544  &    14.627  &    19.474  &    2.018  &    7.821  &    11.742  &    15.848 \\
 & $\lambda=3.75$ & {\color{blue}{8.570}} & {\color{blue}{10.415}} &    14.158  &    18.324  &    2.010  &    7.723  &    11.390  &    15.098 \\
 & $\lambda=4.00$ &    8.667  &    10.513  & {\color{blue}{14.101}} & {\color{blue}{17.884}} &    2.000  &    7.670  &    11.198  &    14.720 \\
 & $\lambda=4.25$ &    8.851  &    10.683  &    14.251  &    17.995  &    1.992  & {\color{blue}{7.669}} & {\color{blue}{11.163}} & {\color{blue}{14.635}}\\
 & $\lambda=4.50$ &    9.067  &    10.904  &    14.525  &    18.222  &    1.985  &    7.694  &    11.234  &    14.764 \\
 & $\lambda=4.75$ &    9.300  &    11.140  &    14.820  &    18.614  &    1.980  &    7.769  &    11.384  &    14.966 \\
 & $\lambda=5.00$ &    9.536  &    11.388  &    15.092  &    19.042  &    1.979  &    7.884  &    11.592  &    15.202 \\
\hline
\multirow{7}{*}{\rotatebox{90}{\begin{minipage}{2cm}\begin{center}SLOPE\end{center}\end{minipage}}}
 & $q=0.01$ &    8.837  &    10.702  &    14.364  &    18.211  & {\color{blue}{1.988}} &    7.724  &    11.289  &    14.836 \\
 & $q=0.02$ &    8.694  &    10.569  &    14.201  &    18.037  &    1.994  &    7.679  &    11.195  &    14.711 \\
 & $q=0.03$ &    8.627  &    10.495  &    14.118  &    17.948  &    1.999  &    7.678  & {\color{blue}{11.178}} &    14.658 \\
 & $q=0.04$ &    8.593  &    10.449  &    14.100  &    17.876  &    2.002  & {\color{blue}{7.677}} &    11.179  & {\color{blue}{14.651}}\\
 & $q=0.05$ & {\color{blue}{8.580}} & {\color{blue}{10.430}} & {\color{blue}{14.089}} & {\color{blue}{17.876}} &    2.006  &    7.678  &    11.190  &    14.670 \\
 & $q=0.10$ &    8.598  &    10.487  &    14.188  &    18.061  &    2.017  &    7.695  &    11.258  &    14.815 \\
 & $q=0.12$ &    8.635  &    10.542  &    14.274  &    18.160  &    2.021  &    7.707  &    11.304  &    14.916 \\
\hline
\end{tabular}

%% file: tables/TablePhantomDiffExt.tex
\begin{tabular}{llrrrrrrrrr}
  \multicolumn{2}{l}{Problem idx.}& \multicolumn{1}{c}{1}& \multicolumn{1}{c}{2}& \multicolumn{1}{c}{3}& \multicolumn{1}{c}{4}& \multicolumn{1}{c}{5}& \multicolumn{1}{c}{6}& \multicolumn{1}{c}{7}& \multicolumn{1}{c}{8}& Max. \\
  \hline
  \multirow{7}{*}{\rotatebox{90}{\begin{minipage}{2cm}\begin{center}Lasso\end{center}\end{minipage}}}
  &    $\lambda=3.50$  &    0.232  &    1.237  &    3.817  &    8.940  &    2.378  &    1.989  &    5.190  &    8.287  & 8.940 \\
  &    $\lambda=3.75$  &    0.000  &    0.000  &    0.489  &    2.504  &    1.952  &    0.702  &    2.041  &    3.163  & 3.163 \\
  &    $\lambda=4.00$  & {\color{blue}{1.140}} & {\color{blue}{0.938}} & {\color{blue}{0.085}} & {\color{blue}{0.045}} & {\color{blue}{1.475}} & {\color{blue}{0.011}} & {\color{blue}{0.315}} & {\color{blue}{0.581}} & 1.475 \\
  &    $\lambda=4.25$  &    3.284  &    2.574  &    1.148  &    0.667  &    1.038  &    0.000  &    0.000  &    0.000  & 3.284 \\
  &    $\lambda=4.50$  &    5.806  &    4.690  &    3.091  &    1.936  &    0.715  &    0.330  &    0.638  &    0.879  & 5.806 \\
  &    $\lambda=4.75$  &    8.527  &    6.956  &    5.189  &    4.131  &    0.443  &    1.300  &    1.985  &    2.259  & 8.527 \\
  &    $\lambda=5.00$  &    11.281  &    9.340  &    7.118  &    6.525  &    0.372  &    2.804  &    3.850  &    3.869  & 11.281 \\
  \hline
  \multirow{7}{*}{\rotatebox{90}{\begin{minipage}{2cm}\begin{center}SLOPE\end{center}\end{minipage}}}
 &    $q=0.01$  &    3.123  &    2.758  &    1.949  &    1.873  &    0.866  &    0.722  &    1.132  &    1.370  & 3.123 \\
 &    $q=0.02$  &    1.449  &    1.474  &    0.795  &    0.903  &    1.134  &    0.130  &    0.293  &    0.519  & 1.474 \\
 &    $q=0.03$  & {\color{blue}{0.668}} & {\color{blue}{0.770}} & {\color{blue}{0.202}} & {\color{blue}{0.406}} & {\color{blue}{1.410}} & {\color{blue}{0.126}} & {\color{blue}{0.138}} & {\color{blue}{0.153}} & 1.410 \\
 &    $q=0.04$  &    0.270  &    0.327  &    0.074  &    0.001  &    1.585  &    0.110  &    0.145  &    0.106  & 1.585 \\
 &    $q=0.05$  &    0.117  &    0.140  &    0.000  &    0.000  &    1.743  &    0.124  &    0.242  &    0.239  & 1.743 \\
 &    $q=0.10$  &    0.326  &    0.687  &    0.697  &    1.033  &    2.346  &    0.338  &    0.859  &    1.231  & 2.346 \\
 &    $q=0.12$  &    0.767  &    1.221  &    1.309  &    1.587  &    2.507  &    0.493  &    1.270  &    1.920  & 2.507 \\
\hline
\end{tabular}

%% file: discussion.tex
\section{Discussion}
\label{sec:discussion}

In this paper, we have shown that the sorted $\ell_1$ norm may be
useful in statistical applications both for multiple hypothesis
testing and for parameter estimation. In particular, we have
demonstrated that the sorted $\ell_1$ norm can be optimized
efficiently and established the correctness of SLOPE for
FDR control under orthogonal designs. We also demonstrated via
simulation studies that in some settings, SLOPE can keep
the FDR at a reasonable level while offering increased power. Finally,
SLOPE can be used to obtain accurate estimates in sparse
or nearly sparse regression problems in the high-dimensional regime.

Our work suggests further research and we list a few open problems we
find stimulating. First, our methods assume that we have knowledge of
the noise standard deviation and it would be interesting to have
access to methods that would not require this knowledge. A tantalizing
perspective would be to design joint optimization schemes to
simultaneously estimate the regression coefficients via the sorted
$\ell_1$ norm and the noise level as in \cite{ScaledLasso} for the
lasso. Second, just as in the BHq procedure, where the test statistics
are compared with fixed critical values, we have only considered in
this paper fixed values of the regularizing sequence
$\{\lambda_i\}$. It would be interesting to know whether it is
possible to select such parameters in a data-driven fashion as to
achieve desirable statistical properties. For the simpler lasso
problem for instance, an important question is whether it is possible
to select $\lambda$ on the lasso path as to control the FDR, see
\cite{TibsTaylor} for contemporary research related to this
issue. Finally, we have demonstrated the limited ability of the lasso
and SLOPE to control the FDR in general. It would be of great interest
to know what kinds of positive theoretical results can be obtained in
perhaps restricted settings.


%% file: appendix.tex
\section{Proofs of Intermediate Results}
\label{sec:appendix}

\subsection{Proof of Proposition \ref{prop:orderedl1}}

\begin{lemma}
\label{lem:wh}
For all $w$ and $h$ in $\R^n$, we have
\[
\sum_i w_i h_i \le \sum_i |w|_{(i)} |h|_{(i)}. 
\]
\end{lemma}
\begin{proof}
Since
\[
\sum_i w_i h_i \le \sum_i |w_i| |h_i|, 
\]
it is sufficient to prove the claim for $w, h \ge 0$, which we now
assume. Without loss of generality, suppose that $w_1 \ge w_2 \ge
\ldots \ge w_n \ge 0$. Then 
\begin{align*}
  \sum_i w_i h_i    
& = w_n (h_1 + \ldots + h_n) + (w_{n-1} - w_n) (h_1 + \ldots + h_{n-1}) + \ldots + (w_1 - w_2) h_1\\
  & \le w_n (h_{(1)} + \ldots + h_{(n)}) + (w_{n-1} - w_n) (h_{(1)} +
  \ldots + h_{(n-1)}) + \ldots + (w_1 - w_2) h_{(1)}\\
  & = w_n h_{(n)} + w_{n-1} h_{(n-1)} + \ldots + w_1 h_{(1)} \\
  & = \sum_i w_{(i)} h_{(i)},
\end{align*}
which proves the claim. 
\end{proof}

We are in position to prove Proposition \ref{prop:orderedl1}. Suppose
$w \in C_\lambda$, then  Lemma \ref{lem:wh} gives 
\begin{multline*}
  \sum_i  w_i b_i 
   \le \sum_i |w|_{(i)} |b|_{(i)}
   = |b|_{(p)} (|w|_{(1)} + \ldots + |w|_{(p)})\\ + (|b|_{(p-1)} -
  |b|_{(p)}) (|w|_{(1)} + \ldots + |w|_{(p-1)}) + \ldots + (|b|_{(1)}
  -
  |b|_{(2)}) |w|_{(1)}. 
\end{multline*}
Next, membership to $C_\lambda$ allows to bound the right-hand side as 
\[
|b|_{(p)} (\lambda_1 + \ldots + \lambda_{p}) + (|b|_{(p-1)} -
  |b|_{(p)}) (\lambda_1 + \ldots + \lambda_{p-1}) + \ldots +
  (|b|_{(1)} -
  |b|_{(2)}) \lambda_{1} = \sum_i \lambda_i |b|_{(i)}.
\]
This shows that $\sup_{w \in C_\lambda} \, \<z, w\> \le
J_\lambda(b)$. For the equality, assume without loss of generality
that $|b_1| \ge |b_2| \ge \ldots \ge |b_p|$ and take $w_i = \lambda_i
\sgn(b_i)$. Then it is clear that $\<w, b\> = \sum_i \lambda_i |b_i| =
J_\lambda(b)$.

\subsection{Proof of Proposition \ref{prop:BHq-ol}}

Set $y = \tilde y$ for notational convenience, and following Section
\ref{sec:prelim}, assume without loss that $y_1 \ge y_2 \ge \ldots \ge
y_p \ge 0$.  Consider a feasible input $x_1 \ge \ldots \ge x_p$ such
that $x_{\iSU+1} > 0$ and define $x'$ as 
\[
x'_i = \begin{cases} x_i & i \le \iSU\\
0 & i > \iSU.
\end{cases}
\]
Then the difference in the objective functional is equal to 
\[
f(x) - f(x') = \sum_{i > \iSU} \Bigl\{\half ((y_i - x_i)^2 - y_i^2) + \lambda_i x_i\Bigr\} = \sum_{i > \iSU} \Bigl\{\half x_i^2 + (\lambda_i - y_i) x_i\Bigr\}.
\]
Since by definition $y_i \le \lambda_i$ for $i > \iSU$, we see that
$f(x) - f(x') > 0$. This shows that $i^\star \le \iSU$.

We now argue that $\i^\star \ge \iSD$. Take $x$ feasible as before obeying
$x_1 \ge \ldots x_{i_0} > 0$ and $x_{i_0 +1} = \ldots = x_p = 0$ where
$i_0 < \iSD$.
Define $x'$ as
\[
x'_i = \begin{cases} x_i & i \le i_0\\
h & i_0 < i \le \iSD\\
0 & i > \iSD,
\end{cases}
\]
where $0 < h \le x_{i_0}$ is an appropriately small scalar. Observe
that $x'$ is feasible (i.e.~is nonnegative and nonincreasing). The
difference in the objective functional is equal to
\[
f(x) - f(x') = \sum_{i_0 < i \le \iSD} \Bigl\{\half (y_i^2 -
(y_i-h)^2) - \lambda_i h\Bigr\} = \sum_{i_0 < i \le \iSD} \Bigl\{(y_i -
\lambda_i) h - \half h^2\Bigr\}.
\]
Since by definition $y_i > \lambda_i$ for $i \le \iSD$, we see that
taking $h$ small enough gives $f(x) - f(x') > 0$, concluding the
proof.

\section{Asymptotic Predictions of FDR Levels}
\label{sec:Montanari}

Figures \ref{fig:Fig1} and \ref{fig:minimax} in Section
\ref{sec:fdr_gauss_design} display asymptotic predictions of FDR
levels and it is now time to explain where they come from. These
predictions can be heuristically derived by extending results from
\cite{Rangan09} and \cite{BM12}. In a nutshell, \cite{Rangan09}
applies the replica method from statistical physics for computing the
asymptotic performance of the lasso with Gaussian designs. In
contrast, \cite{BM12} rigorously proves asymptotic properties of the
lasso solution in the same setting by studying an iterative scheme
called approximate message passing (AMP) \cite{DMM09} inspired by
belief-propagation on dense graphical models. Below, we follow
\cite{BM12}.

\subsection{Asymptotic properties of the lasso}

Our predictions use the main result from \cite[Theorem 1.5]{BM12} that
we shall state below with a bit less of generality. We begin with two
definitions. First, a function $\varphi:
\mathbb{R}^2\rightarrow\mathbb{R}$ is said to be pseudo-Lipschitz if
there is a numerical constant $L$ such that for all $x,
y\in\mathbb{R}^2$,
\[
|\varphi(x)-\varphi(y)|\le L(1+\|x\|_{\ell_2}+\|y\|_{\ell_2})\|x-y\|_{\ell_2}. 
\]
Second, for any $\delta>0$, we let $\alpha_{\text{min}} =
\alpha_{\text{min}}(\delta)$ be the unique solution to
\[
2(1+\alpha^2)\Phi(-\alpha)-2\alpha\phi(\alpha)-\delta=0
\] 
if $\delta\le 1$, and 0 otherwise.
\begin{theorem}\cite[Theorem 1.5]{BM12}\label{amp-thm}
  Consider the linear model with i.i.d.~$\mathcal{N}(0,1)$ errors in
  which $X$ is an $n\times p$ matrix with i.i.d.~$\mathcal{N}(0,1/n)$
  entries. Suppose that the $\beta_i$'s are i.i.d.~random variables,
  independent of $X$, and with positive variance (below, $\Theta$ is a
  random variable distributed as $\beta_i$). Then for any
  pseudo-Lipschitz function $\varphi$, the lasso solution $\hat\beta$
  to \eqref{eq:lasso} with fixed $\lambda$ obeys
  \begin{equation}
    \label{eq:BM12}
    \frac{1}{p}\sum_{i=1}^p\varphi(\hat\beta_i,\beta_i) \, \, \longrightarrow \,\, \mathbb{E}\varphi(\eta_{\alpha\tau}(\Theta + \tau Z), \Theta), 
\end{equation}
where the convergence holds in probability as $p, n\rightarrow \infty$
in such a way that ${n}/{p}\rightarrow \delta$. Above, $Z \sim
\mathcal{N}(0,1)$ independent of $\Theta$, and $\tau>0,
\alpha>\alpha_{\text{\em min}}(\delta)$ are the unique solutions to
\begin{equation}
\label{eq:amp-bm}
\begin{aligned}
\tau^2 &= 1 + \frac{1}{\delta}\mathbb{E}\Big{(}\eta_{\alpha\tau}(\Theta + \tau Z) - \Theta\Big{)}^2,\\
\lambda &= \Big{(}1 - \frac{1}{\delta}\mathbb{P}(|\Theta + \tau Z| > \alpha\tau)\Big{)}\alpha\tau.\\
\end{aligned}
\end{equation}
\end{theorem}
The predicted curves plotted in Figures \ref{fig:Fig1} and
\ref{fig:minimax} are obtained by using this result together with some
heuristic arguments.
Setting
\[
\varphi_V(x, y) = 1(x\ne 0) 1(y=0) 
\quad \text{and} \quad 
\varphi_R(x, y) = 1(x\ne 0), 
\]
the number of false discoveries $V$ and of discoveries $R$ take the form 
\[
V = \sum_{i=1}^p \varphi_V(\hat\beta_i, \beta_i), \qquad R =
\sum_{i=1}^p \varphi_R(\hat\beta_i, \beta_i).
\]
Ignoring the fact that $\varphi_V$ and $\varphi_R$ are not pseudo-Lipschitz,
applying Theorem \ref{amp-thm} gives
\begin{equation}\label{eq:amp-fdr}
\begin{array}{lll}
 V/p & \goto & \mathbb{P}(\Theta=0)\P(|Z| > \alpha)\\
  R/p & \goto & \mathbb{P}(|\Theta+\tau Z|>\alpha\tau)
\end{array}
\quad \Longrightarrow \quad 
\text{FDP} \goto \frac{\mathbb{P}(\Theta=0)\P(|Z| > \alpha)}{\mathbb{P}(|\Theta+\tau Z|>\alpha\tau)}.
\end{equation}
Moreover, this also suggests that power also converges in probability
to
\begin{equation}\label{eq:amp-power}
  \text{Power} \goto \frac{\mathbb{P}(\Theta\ne 0, |\Theta+\tau Z|>\alpha\tau)}{\mathbb{P}(\Theta\ne 0)} =\mathbb{P}(|\Theta+\tau Z|>\alpha\tau|\Theta\ne 0).
\end{equation}
In a supplementary note \cite{supp}, we prove that even though
$\varphi_V$ and $\varphi_R$ are discontinuous so that Theorem \ref{amp-thm}
does not apply, the conclusions above are mathematically valid.


\subsection{Predictions in Figure \ref{fig:Fig1}}

In Figure \ref{fig:Fig1}, $n=p=1000$ and $\beta_1,\ldots,\beta_k$ are
i.i.d.~copies of $3\lambda Z$, $Z \sim \mathcal{N}(0, 1)$. We recall
that $\lambda=3.717 =\lambda_{BH}(1)$ with $q\approx 0.207$. To apply
Theorem \ref{amp-thm}, we let $\Theta$ be the mixture $\mathcal{N}(0,
(3\lambda)^2) + k/p \delta_0$.  Solving \eqref{eq:amp-bm} for $\alpha,
\tau$ gives a prediction of FDR and power according to
\eqref{eq:amp-fdr} and \eqref{eq:amp-power}. These are the prediction
curves plotted in the figure.

\subsection{Predicting the FDR in the high-SNR regime}

We now wish to explain the predicted curves in Figure
\ref{fig:minimax}. In the high-SNR regime, $\lambda$ is as large as we
want, but whatever its value, the true regression coefficients are
orders of magnitude greater. To connect this with Theorem
\ref{amp-thm}, imagine that the $\beta_i$'s are independent copies of
\[
\Theta = \begin{cases} M, & \text{with prob. } \epsilon,\\
  0, & \text{with prob. } 1 - \epsilon
\end{cases}
\]
for some $M>0$ and $\epsilon\in (0,1)$. If we denote by
$\text{FDR}(M,\lambda; \epsilon)$ the limiting FDR given by our
application of Theorem \ref{amp-thm} as above, the predicted curves
plotted in Figure \ref{fig:minimax} are
\[
\lim_{\lambda \goto \infty} \lim_{M \goto \infty} \text{FDR}(M,\lambda; \epsilon), 
\]
with $\epsilon = k/p$. One can compute this limit directly but this is
a little tedious. We present instead an intuitive method which gets to
the results quickly.


\textbf{Full asymptotic power.} As the name suggests, this is a
situation in which all the true regressors are eventually detected as
$M \goto \infty$. This happens in the case where $\delta \ge 1$ and
when $\delta < 1$ but $\epsilon$ is below the transition curve
$\epsilon^\star(\delta)$, which is the weak transition threshold
discussed in the compressive sensing literature, see
\cite{DonTan05}. (This curve has several different interpretations,
see e.g.~\cite{DMM09} or \cite{DJM13}.) In this case $\tau/M \goto 0$
as $M \goto \infty$, and it follows from \eqref{eq:amp-bm} that in the
limit $M \goto \infty$, we have
\begin{equation}\label{eq:amp-full}
\begin{aligned}
  \tau^2 &= 1 + \frac{(1-\epsilon)\tau^2}{\delta}\mathbb{E}\eta_{\alpha}(Z)^2 + \frac{\epsilon\tau^2(1+\alpha^2)}{\delta},\\
  \lambda &= \Big{(}1 - \frac{1}{\delta}\mathbb{P}(|\Theta + \tau Z| >
  \alpha\tau)\Big{)}\alpha\tau.
\end{aligned}
\end{equation}
We now take a limit as $\lambda \rightarrow \infty$. We can show that
$\tau \goto \infty$ while $\alpha$ converges to a positive constant
$\alpha^\star$. By \eqref{eq:amp-full}, $\alpha^{\star}$ is a solution
to
\begin{equation}\label{eq:amp-minimax-full}
2(1-\epsilon)((1+\alpha^2)\Phi(-\alpha)-\alpha\phi(\alpha)) + \epsilon(1+\alpha^2) = \delta.
\end{equation}
This equation admits one positive solution if $\delta\geq 1$, or if
$\delta < 1$ and $\epsilon < \epsilon^\star(\delta)\in (0,1)$.
When $\delta < 1$ and $\epsilon > \epsilon^\star(\delta)$,
\eqref{eq:amp-minimax-full} has two positive roots and only the
largest of these two, which we denote by $\alpha^\star$, is larger
than $\alpha_{\text{min}}$. These arguments give
\begin{equation*}
\begin{array}{lll}
  V/p & \goto & (1-\epsilon)\P(|Z| > \alpha^\star)\\
  R/p & \goto & \epsilon + (1-\epsilon) \P(|Z| > \alpha^\star)
\end{array}
\quad \Longrightarrow \quad 
\text{FDP} \goto \frac{(1-\epsilon)\P(|Z| > \alpha^\star)}{\epsilon + (1-\epsilon)\P(|Z| > \alpha^\star)}. 
\end{equation*}

\textbf{Limited asymptotic power.} We now consider $\delta < 1$ and
$\epsilon > \epsilon^\star(\delta)$, a setting in which we do not have
full asymptotic power. In this case, the quantity $1 -
{\delta}^{-1}\mathbb{P}(|\Theta + \tau Z| > \alpha\tau$) in
\eqref{eq:amp-bm} vanishes as $M\rightarrow\infty$. To study the
behavior in this case, we postulate that $M/\tau \goto \gamma^\star$
for some constant $\gamma^\star > 0$ whenever $M,
\lambda\rightarrow\infty$ with $\lambda/M \goto 0$. Then it follows
from the first equation in \eqref{eq:amp-bm} that
\begin{equation}
\label{eq:amp-minimax-nonfull}
2(1-\epsilon)((1+\alpha^2)\Phi(-\alpha)-\alpha\phi(\alpha)) + \epsilon\mathbb{E}\Big{(}\eta_{\alpha}(\gamma+Z)-\gamma\Big{)}^2=\delta. 
\end{equation}
We also have a second equation that comes from the fact that the
lasso solution has as many nonzero coefficients as there are rows in
$X$. In other words $R/p \goto \delta < 1$, which reads
\begin{equation}
\label{eq:amp-minimax-nonfull2}
2(1-\epsilon)\Phi(-\alpha)+\epsilon[\Phi(-\alpha-\gamma)+\Phi(-\alpha+\gamma)]
= \delta.
\end{equation}
Up to multiplication by a factor of $p$, the first (resp.~second)
term in the left-hand side is the asymptotic number of false
(resp.~correct) discoveries.  Letting $\alpha^\star$ and
$\gamma^\star$ be the solution to the system
\eqref{eq:amp-minimax-nonfull}--\eqref{eq:amp-minimax-nonfull2}, this
gives
\begin{equation*}
\begin{array}{lll}
  V/p & \goto &  2(1-\epsilon)\Phi(-\alpha^\star)\\
  R/p & \goto & \delta
\end{array}
\quad \Longrightarrow \quad 
\text{FDP} \goto  \frac{2(1-\epsilon)\Phi(-\alpha^\star)}{\delta}. 
\end{equation*}
Also, we have seen that 
\[
\text{Power}\, \goto \,
\Phi(-\alpha^\star-\gamma^\star)+\Phi(-\alpha^\star+\gamma^\star).
\]
This concludes the derivation of our expressions for the FDP (and
power) in the high signal-to-noise regime. The accuracy of these
predictions is quite visible in Figure \ref{fig:minimax}.